\documentclass[prb,twocolumn,superscriptaddress,amssymb]{revtex4-1}
\usepackage[utf8]{inputenc}
\usepackage[T1]{fontenc}
\usepackage{amsmath}
\usepackage{graphicx}
\usepackage{dcolumn}
\usepackage{bm}
\usepackage{epsfig}
\usepackage{float} 
\usepackage{color}
\usepackage{tikz}
\usepackage{subcaption}
\usepackage{verbatim}
\usepackage[normalem]{ulem}
\usepackage{multirow}
\usepackage[colorlinks=true,urlcolor=blue,linkcolor=blue,
            citecolor=blue]{hyperref}
\newcommand \beq{\begin{equation}}
\newcommand \eeq{\end{equation}}

\newcommand \exc{\texttt{EXC}}
\newcommand \exciting{\texttt{Exciting}}

\def\simge{\mathrel{%
       \rlap{\raise 0.511ex \hbox{$>$}}{\lower 0.511ex \hbox{$\sim$}}}}
\def\simle{\mathrel{
       \rlap{\raise 0.511ex \hbox{$<$}}{\lower 0.511ex \hbox{$\sim$}}}}
\def\beq {\begin{equation}}
\def\eeq {\end{equation}}
\def\w {\omega}
\def\bfq {\mathbf{q}}

\def\bfG {\mathbf{G}}
\def\bfk {\mathbf{k}}
\def\bfr {\mathbf{r}}

\newcommand{\bra}[1]{\langle #1|}
\newcommand{\ket}[1]{|#1\rangle}

\newcommand{\rv}{\mathbf{r}}
\newcommand{\kv}{\mathbf{k}}
\newcommand{\qv}{\mathbf{q}}

\newcommand{\rhot}{\tilde{\rho}}
\newcommand{\degree}{^\circ}
\newcommand{\alo}{$\alpha$-Al$_2$O$_3$}

\begin{document}

\title{Pseudopotential Bethe-Salpeter calculations for shallow-core x-ray absorption near-edge structures: excitonic effects 
in {\alo}}

\newcommand{\lsi}{LSI, CNRS, CEA/DRF/IRAMIS, \'Ecole Polytechnique, Institut Polytechnique de Paris, F-91120 Palaiseau, France}
\newcommand{\etsf}{European Theoretical Spectroscopy Facility (ETSF)}
\newcommand{\soleil}{Synchrotron SOLEIL, L'Orme des Merisiers, Saint-Aubin, BP 48, F-91192 Gif-sur-Yvette, France}

\date{\today}

\author{M. Laura Urquiza}
\affiliation{\lsi}
\affiliation{\etsf}

\author{Matteo Gatti}
\affiliation{\lsi}
\affiliation{\etsf}
\affiliation{\soleil}

\author{Francesco Sottile}
\affiliation{\lsi}
\affiliation{\etsf}

\begin{abstract}
We present an ab initio description of optical and shallow-core x-ray absorption spectroscopies in a unified formalism based on the pseudopotential plane-wave method at the level of the Bethe-Salpeter equation (BSE) within Green's functions theory. 
We show that norm-conserving pseudopotentials are reliable and accurate not only for valence, but also for semicore electron excitations. 
In order to validate our approach, we compare BSE absorption spectra obtained with two different codes:  the pseudopotential-based code {\exc} and the  all-electron full-potential code {\exciting}. 
We take corundum {\alo} as an example, being a prototypical material that presents strong electron-hole interactions for both valence and core electron excitations. 
We analyze in detail the optical absorption spectrum as well as the Al L$_1$ and L$_{2,3}$ edges in terms of anisotropy, crystal local fields, interference and excitonic effects. 
We perform a thorough inspection of the origin and localization of the lowest-energy excitons, and conclude highlighting the purely electronic
character off the pre-edge of L$_1$ and the dichroic nature of the optical and L$_{23}$ spectra.
\end{abstract}

\maketitle


\section{\label{sec:introduction} Introduction}

X-ray absorption spectroscopy (XAS) and optical absorption are complementary techniques to determine materials properties.
In optical absorption, valence electrons are excited into unoccupied conduction states across the band gap (or the Fermi energy in metals). Their excitations determine the color (or the transparency) of materials and are crucial to many  materials properties and functionalities, spanning from optoelectronics to solar energy conversion and storage.  
In XAS, promoted to unoccupied conduction bands are instead core electrons, tightly bound to the nuclei.
X-ray absorption near-edge structures (XANES), also known as near-edge X-ray absorption fine structure (NEXAFS), being element specific, is a probe of the atomic environment, giving structural and chemical information\cite{vanBokhoven2016}.
In the simplest independent-particle picture, XANES spectra are proportional to the unoccupied density of states, projected on the absorbing atom and the angular momentum component that is selected by dipole selection rules, whereas optical spectra can be interpreted on the basis of the joint density of states of valence and conduction bands.
In both spectroscopies, the interaction between the excited electron and the hole left behind can strongly alter this independent-particle picture.
Indeed, the electron-hole attraction can give rise to excitons, i.e  bound electron-hole pairs, leading to a transfer of spectral weight to lower energies in the spectra, including the formation of sharp peaks at their onset.

Given the importance of XANES spectroscopy, several theoretical methods have been developed to interpret the measured spectra in solids, taking care of  core-hole effects at different levels of approximation\cite{deGroot2021}.
The most efficient approaches are, on one side, multiple scattering methods\cite{Fujikawa1983,Tyson1992,Ahlers1998,Rehr2000,Rehr2009,Rehr2010}, and, on the other side,  multiplet models\cite{deGroot2005,DeGroot2008,Haverkort2012}.
While the former usually neglect the electronic interactions, 
the latter are often semiempirical (i.e.,  not entirely parameter-free) and generally neglect solid-state effects, 
being 
a many-body solution of finite-cluster models. 
Since the excitations of the core electrons are localised at the absorbing atoms, 
delta-self-consistent-field ($\Delta$SCF) methods can be also employed, nowadays usually within first-principles density-functional theory\cite{Mo_2000,Gougoussis2009,Taillefumier2002,Bunau2013,Mazevet2010,Hetenyi2004,Prendergast2006,Gao_2009,Prentice2020}.
The core-excited atom is treated as an impurity in a supercell approach, and the presence of the core hole is taken into account in different ways, from 
the Z+1 approximation\cite{Hjalmarson1981,Lie1999} (the absorbing atom is assumed to have one additional nuclear charge),  
to the half core-hole approximation \cite{Triguero1998,Klein_2021} (also known as Slater's transition-state method) or the full core-hole approximation (the electron removed from the core is put at lowest conduction band, or ionized).
Alternatively, XANES excitation spectra can be directly obtained within linear-response theory\cite{Rehr2005,Liang2017}, which is the standard approach for valence excitations and optical spectra as well\cite{Onida2002}. 
In this case, two possible options are time-dependent density-functional theory\cite{Besley_2009,Bunau2012,Bunau2012b} (TDDFT) and the Bethe-Salpeter equation\cite{Strinati1982,Strinati1984,Shirley1998,Carlisle1999,Shirley2000} (BSE) of Green's function theory\cite{Martin2016,Bechstedt2014}.
Since TDDFT lacks of efficient approximations for describing accurately excitonic effects in solids\cite{Botti_2007}, the BSE, even though computationally more expensive, is usually more reliable\cite{Onida2002}. In the present work, the solution of the BSE will therefore be also our preferred choice to simulate valence and shallow-core excitation spectra within the same formalism.

In the simulation of core excitation spectra, the intuitive technique to represent the single-particle wave functions are  all-electron methods. They explicitly deal with core electrons in extended materials by partitioning the space into interstitial and muffin-tin (MT) regions, where wave functions are described differently according to their localisation degree\cite{Wills2010,Andersen_1975,Sjostedt2000,Madsen2001}. 
Instead, methods that are based on plane-wave expansions cannot deal explicitly with the quickly oscillatory behavior of core electrons, tightly localised near the nuclei, which are instead generally taken into account effectively through the design of suitable pseudopotentials\cite{Payne_1992}. Plane-wave methods are computationally cheaper and new theoretical developments are easier to implement in plane-wave computer codes.
Moreover, the separation between core electrons, kept frozen, and valence electrons, treated explicitly, is often not rigid. Between valence and deep core electrons, there are often also shallow-core (or semicore) electrons, which in the pseudopotential framework can be in principle also treated as valence electrons, although at a price of higher computational cost.
However, in all the cases, the pseudopotential formalism also introduces an important approximation, requiring a pseudization of the valence  wave functions near the nuclei that makes them smoother and node free.

In the recent past, much work has been devoted to assess pseudopotential calculations for ground state \cite{Willand2013,Lejaeghere2014,Prandini2018,Lejaeghere2016} and excited-state properties with respect to all-electron methods, notably for self-energy calculations of quasiparticle band structure energies  \cite{Ku2002,Delaney2004,Tiago_2004,vanSchilfgaarde2006,Friedrich2006,Gomez_2008,Luppi_2008,Klimes2014,Friedrich2011,Friedrich2011_Erratum,Jiang2016,Jiang2018}. 
In the present work, we directly address the question of the validity of the pseudopotential approximation for XANES spectra of shallow-core edges (i.e., for electron binding energies smaller than $\sim$180 eV), investigating the limits of use of pseudo wave functions for shallow-core states in many-body BSE calculations.
It is clear that the description of deep core levels will be always out of reach for plane-wave basis methods. However, the high plane-wave cutoff required by semicore states can be now alleviated 
by the new generation of ultrasoft norm-conserving pseudopotentials\cite{Hamann_2013}.
Besides the promised lower computational cost for shallower core levels, an
advantage of pseudopotential plane-wave calculations with respect to all-electron methods is that they do not make any hypothesis concerning the localisation of the core hole inside the muffin tin of the absorbing atom \footnote{The same hypothesis is made when the core orbitals are obtained from a calculation of the isolated atom\cite{Shirley2004,Bloechl1994,Unzog2022}.}.

In particular, here we investigate the effects of the electron-hole interactions on the optical absorption and shallow-core XANES spectra of alumina. {\alo} is a  wide-gap insulator,  with many possible applications as a structural ceramic (e.g. as a replacement to SiO$_2$ gate oxide technology) and optical material (also thanks to the high-damage threshold for UV laser applications), and a prototypical system to investigate core-hole effects in XANES spectroscopy\cite{French_1990,French_1994,Tanaka_1996,Cabaret_1996,Ildefonse_1998,Mo_2000,vanBokhoven_2001}.

The article is organised as follows.
After a short description of the employed methodology in Sec.~\ref{sec:method}, comprising a review of the theoretical background (Sec.~\ref{ssec:theory}) and a summary of the computational details (Sec.~\ref{ssec:computational_details}), Sec.~\ref{sec:results} presents the results of the calculations together with their analysis. In Sec.~\ref{ssec:benchmark} pseudopotential calculations are assessed with respect to all-electron benchmarks for both optical and Al L$_{2,3}$ XANES spectra, while Sec.~\ref{ssec:interference} contains a discussion on the issue of the core-hole localisation in the muffin tin for the Al L$_{1}$ XANES spectrum.
Sec.~\ref{ssec:absorption} compares the calculated spectra with available experiments and analyses the effects of the electron-hole interactions on the spectra. 
Finally, Sec.~\ref{sec:conclusions} draws the conclusions summarizing the results of the work.

\section{\label{sec:method} Methodology} 
	
\subsection{\label{ssec:theory} Theoretical background}

In the framework of Green's function theory\cite{Martin2016}, the Bethe-Salpeter equation (BSE) yields the density response function from the solution of a Dyson-like equation for the two-particle correlation function\cite{Strinati1988}.
In the GW approximation (GWA) to the self-energy\cite{Hedin1965}, with a statically screened Coulomb interaction $W$, the BSE takes the form of an excitonic Hamiltonian\cite{Onida2002} in the basis  $\ket{vc{\bf k}}$ of transitions between occupied  $v{\bf k}$ and  unoccupied  bands $c{\bf k}$ (i.e., uncorrelated electron-hole pairs):
\beq 
\bra{vc{\bf k}} H_{\rm exc} \ket{v'c'{\bf k}'}=E_{vc{\bf k}} \delta_{vv'}\delta_{cc'}\delta_{{\bf k}{\bf k}'} + \bra{vc{\bf k}} \bar{v}_c-W \ket{v'c'{\bf k}'}.\label{eq:BSE} 
\eeq
Here  $E_{vc{\bf k}}=E_{c\bfk}-E_{v\bfk}$ are the interband transition energies calculated in the GWA, while $\bar v_c$  is the  Coulomb interaction without its macroscopic component (i.e., the component ${\bf G}=0$ in reciprocal space). The statically screened Coulomb interaction $W=\epsilon^{-1}v_c$ is usually calculated adopting the random-phase approximation (RPA) for the inverse  dielectric function $\epsilon^{-1}$.

The GWA-BSE is nowadays the state-of-the-art approach for the simulation, interpretation and prediction of optical spectra in solids\cite{Albrecht1998,Benedict1998,Rohlfing2000,Bechstedt2014,Martin2016}, and is  more and more used also for the simulation of core-level excitation spectra\cite{Vinson2011,Vinson2012,Gilmore2015,Gilmore2021,Geondzhian2015,Dashwood2021,Vinson2022,Olovsson2009,Olovsson2009b,Olovsson2011,Vorwerk2017,Vorwerk2019,Vorwerk2020,Laskowski2010,Yao2022,deGroot2021,Vorwerk_2022,Unzog2022}.
A great advantage of theory with respect to experiments is the possibility to 
separately suppress (or activate)  the various interactions at play in the materials, which allows one to single out their specific effect on the spectra and the materials properties.
By setting to zero the two electron-hole interactions, $\bar v_c$ and $-W$, the excitonic Hamiltonian~\eqref{eq:BSE} reduces to a diagonal matrix and corresponds to the independent-particle approximation (IPA). 
By switching on the electron-hole exchange interaction $\bar v_c$ in Eq.~\eqref{eq:BSE}, one retrieves the RPA. With respect to the IPA, the RPA includes the so-called crystal local field effects\cite{Adler1962,Wiser1963}. They are  related to the inhomogeneous charge response of materials through the induced microscopic Hartree potentials counteracting the external perturbations.
Finally, by also switching on the electron-hole direct interaction $-W$, the full BSE~\eqref{eq:BSE} describes excitonic effects, which are due to the electron-hole attraction\footnote{There is also the possibility to include $-W$ and exclude $\bar v_c$, which corresponds to the description of 
triplet excitations.}.
The electron-hole interactions contributing to the off-diagonal matrix elements of the BSE~\eqref{eq:BSE} give rise to a mixing of the independent-particle transitions, which is formally obtained from the solution of the eigenvalue equation for the excitonic hamiltonian: $H_{\rm exc} A_\lambda = E_\lambda A_\lambda$.

The absorption spectra, expressed both in the optical and XANES regimes by the imaginary part of the macroscopic dielectric function $\textrm{Im}\epsilon_M(\w)$ in the long wavelength limit $\bfq\to0$, in the so-called Tamm-Dancoff approximation can be directly written in terms of eigenvectors $A_\lambda$ and eigenvalues $E_\lambda$ of the BSE Hamiltonian~\eqref{eq:BSE} as:
\begin{equation}
	\textrm{Im}\epsilon_M(\w) =  \lim_{\qv\to 0}\frac{8\pi^2}{\Omega q^2} \sum_\lambda \left|\sum_{vc{\bf k}}  A_\lambda^{vc{\bf k}} \tilde{\rho}_{vc{\bf k}}(\qv) \right|^2 \delta(\w- E_\lambda ),
	\label{spectrumBSE}
\end{equation}
where  $\Omega$ is the crystal volume, and 
\begin{equation}
	\rhot_{vc{\bf k}}(\qv)= \int \varphi^*_{v\kv-\qv}(\rv) e^{-i\qv\cdot\rv}\varphi_{c\kv}(\rv) d\rv
	\label{eq:rhotw}
\end{equation}
are the independent-particle oscillator strengths.
Here the single-particle orbitals $\varphi_i$ are usually Kohn-Sham orbitals.
If the exciton energy $E_\lambda$ is smaller than the smallest independent-particle transition energy $E_{vc{\bf k}}$, the exciton $\lambda$ is said to be bound: the difference between $E_{vc{\bf k}}$ and $E_\lambda$ is its binding energy.

The contribution of each exciton $\lambda$ to the spectrum can be analysed by introducing the cumulative function:
\begin{equation}
    S_\lambda(\omega) =\lim_{\bfq\rightarrow 0} \frac{8\pi}{\Omega q^2} \;  \left| \sum_{vc{\bf k}}^{{E_{vc{\bf k}}<\omega}} A_\lambda^{vc{\bf k}} \tilde{\rho}_{vc{\bf k}}(\qv)    \right|^2.
     \label{eq:cumu}
\end{equation}
Since the eigenvectors $A_\lambda$ and the oscillator strengths $\tilde{\rho}(\bfq)$
are both complex quantities, the cumulative function~\eqref{eq:cumu} is not a monotonic function of $\w$.
The limit $S_\lambda(\w\to\infty)$ 
is the oscillator strength of the exciton $\lambda$ in the absorption spectrum.
If it is negligibly small, the exciton is said to be dark, otherwise it is called a bright exciton, for it contributes to the spectrum. 
Even in the $\bfq\to 0$ limit, the oscillator strengths $\tilde{\rho}(\bfq)$ depends on the direction of $\bfq$, so each 
exciton $\lambda$ can at the same time be a bright exciton in one polarization direction and dark in another. 

Finally, the investigation of the electron-hole correlation function for each exciton $\lambda$,
\begin{equation}
	\Psi_\lambda(\bfr_h,\bfr_e) = \sum_{vc{\bf k}}  A_\lambda^{vc{\bf k}}\phi^*_{v\bfk}(\bfr_h)\phi_{c\bfk}(\bfr_e),
	\label{excwf}
\end{equation}
gives information about the localisation in real space of the electron-hole pair, which results from the electron-hole attraction.
Assuming that the hole is in a specific position $\bfr_h=\bfr_h^0$, one can visualize the corresponding density distribution of the electron  $|\Psi_\lambda(\bfr_h^0,\bfr_e)|^2$.

\subsection{\label{ssec:computational_details} Computational details}

\begin{figure*}[!th]
	\includegraphics[width=1.0\textwidth]{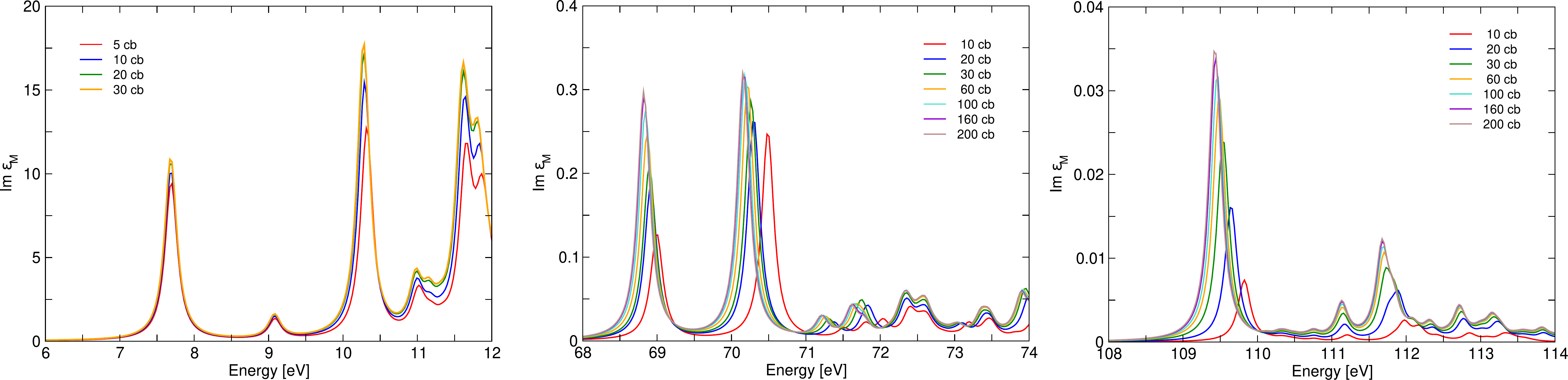}
	\caption[width=1.0\textwidth]{\label{fig:cb-convergence} Convergence of BSE absorption spectra with the number of unoccupied conduction bands (cb). (Left) Optical spectrum. (Middle) XANES at L$_{2,3}$ edge. (Right) XANES at L$_1$ edge.}
\end{figure*}

We have performed calculations using both a  pseudopotential (PP) plane-wave method and a full-potential all-electron (AE) linearized augmented plane-wave method.
AE calculations have been done in particular to assess the validity of PP calculations for the  core-level excitations (see Sec.~\ref{ssec:benchmark}). The converged BSE absorption spectra and their analysis (see Sec.~\ref{ssec:absorption}) have been then obtained in the PP framework.
In the pseudopotential case, we have used the {\tt Abinit} code\cite{Gonze2016} for the ground-state and screening calculations, and the \exc\ code\cite{EXCcode} for the BSE calculations.
In the all-electron case, we have used the \exciting\ code\cite{Gulans_2014} for  obtaining all the benchmark results.

The Kohn-Sham ground-state calculations have been performed within the local density approximation\cite{Kohn1965} (LDA).

We have employed norm-conserving Troullier-Martins \cite{Troullier_1991} (TM) and optimized norm-conserving Vanderbilt \cite{Hamann_2013,vanSetten_2018} (ONCVPSP) pseudopotentials.
In particular, for the absorption spectra a special TM pseudopotential\cite{Zhou_2020} treating also Al $2s$ and $2p$ states as valence electrons has been used.
Calculation with the ONCVPSP pseudopotential converged with 42 Hartree cutoff of the plane-wave expansion, while the hard TM pseudopotential required 320 Hartree.

The statically screened Coulomb interaction $W$ has been obtained (within the RPA) with the ONCVPSP pseudopotential  (without  Al $2s$ and $2p$ core levels), including 100 bands, and with a cutoff of 8 and 14.7~Hartree for the Kohn-Sham wave functions for the optical and shallow-core excitations, respectively. The size of the screening matrix in the plane-wave basis was 6~Hartree for the optical and 8~Hartree for the core spectrum.
We have verified that, contrary to calculations of the screened interaction for other materials like silicon\cite{Luppi_2008} or simple metals\cite{Sturm_1990,Quong_1993,Zhou_2018}, the effect of core polarization is negligible in {\alo}.

For the all-electron results, the ground-state calculations were performed using a plane wave cutoff, $R_{\rm MT}|\bfG+\bfk|_{\rm max}$, of 18~Hartree and muffin-tin (MT) spheres $R_{\rm MT}$ of 2 bohr and 1.45 bohr for aluminum and oxygen, respectively. The RPA screening was obtained with 100 conduction bands and a cutoff in the matrix size of 5~Hartree (maintaining the same cutoff of the ground state for the plane waves).

The GW band structure has been approximated within a scissor correction model. The LDA conduction bands have been rigidly shifted upwards by 2.64 eV, which corresponds to the band gap correction obtained within the perturbative G$_0$W$_0$ scheme by Marinopoulos and Gr\"uning\cite{Myrta_2011}.

The BSE calculations for the absorption spectra have been performed with shifted
$\bfk$-point grids (i.e., not containing high-symmetry $\bfk$ points), which allowed for a quicker convergence of the spectra\cite{Benedict1998}. 
The optical absorption spectrum converged with a 10$\times$10$\times$10 $\bfk$-point grid, 
while the XANES spectra at the Al L$_{2,3}$ and L$_1$ edges converged with  a 8$\times$8$\times$8 $\bfk$-point grid. 
The exciton analysis and plot have been instead done with a smaller $\Gamma$-centered 4$\times$4$\times$4 $ \bfk$-point grid.

The BSE spectra for the optical spectrum or the XANES spectra at the Al L$_{2,3}$ and L$_1$ edges had a different convergence rate with respect to the number of empty bands considered in the BSE hamiltonian.
Fig.~\ref{fig:cb-convergence} shows their convergence study (carried out here with a reduced number of $\bfk$ points in a $\Gamma$-centered 2$\times$2$\times$2 $\bfk$-point grid).
While the optical spectrum (left panel) quickly converges with the number of empty bands, the XANES spectra (middle and right panels) require many more empty bands, also to converge the lowest energy peak.
In the converged spectra, obtained with many more $\bfk$ points, this slow convergence is partially attenuated by the fact that the spectra become smoother. The optical absorption spectra have been thus obtained with 12 valence bands and 12 unoccupied bands. The XANES spectra at the L$_{2,3}$ and L$_1$ edges included all the corresponding core levels together with 30 unoccupied bands.
A 0.1~eV Lorentzian broadening has been applied to the spectra.

In the all-electron BSE calculations, we considered the same parameters used in the calculation of the screening: 9~Hartree for the wavefunction cutoff and 5~Hartree to describe the electron-hole interactions.
In the pseudopotential BSE calculations, we have used a cutoff of 30~Hartree (110~Hartree) for the Kohn-Sham wavefunctions expansion in the optical (semi-core) spectra, and 7.3~Hartree for the plane-wave representation of the electron-hole interactions.
We note that, as usual (see e.g.\cite{Lorin2021}), the plane-wave cutoffs for the BSE matrixelements can be significantly reduced with respect to the high cutoff needed for the ground-state calculation. Therefore, even for pseudopotential BSE calculations of shallow-core excitations, the limiting factor remains the large size of the BSE hamiltonian~\eqref{eq:BSE} in extended systems, which is given by the number of electron-hole transitions (i.e., the number of occupied bands $\times$ the number of unoccupied bands $\times$ the number of $\bfk$ points in the full Brillouin zone).

\section{\label{sec:results} Results}

\subsection{\label{sec:bands} Crystal and electronic structure of $\alpha$-Al$_2$O$_3$}

\begin{figure}[ht]
	\includegraphics[width=0.7\columnwidth]{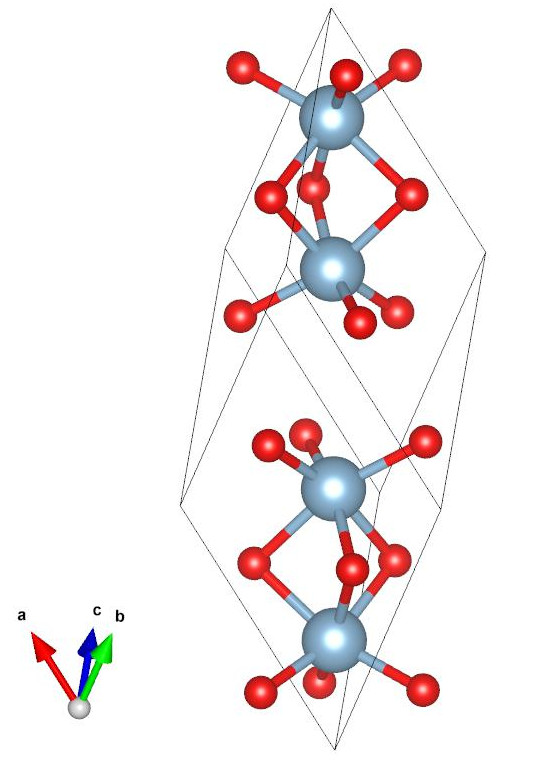}
	\caption{\label{fig:crystal} Primitive rhombohedral unit cell of the crystal structure of {\alo}. Red and grey balls represent O and Al atoms, respectively. The Al atoms are aligned along the cartesian $z$ axis, which is the vertical direction in the figure, while the O atoms belong to the $xy$ planes perpendicular to it.}
\end{figure}

The crystal structure of corundum {\alo} is trigonal (see Fig.~\ref{fig:crystal}). In the primitive rhombohedral unit cell (space group R$\bar{3}$c, number 167) there are two formula units. The corundum structure can also be viewed as a hexagonal cell containing six formula units with alternate layers of Al and O atoms in planes perpendicular to the hexagonal $c_H$ axis.
In the {\alo} structure all Al atoms occupy octahedral sites coordinated with 6 O atoms, which form two equilateral triangles located respectively slightly above and below each Al atom along the $c_H$ direction.

We adopted the experimental lattice parameters from Ref.~\cite{Newnham_1962}: $a_H = b_H$ = 4.7589~{\AA} and $c_H$ = 12.991~{\AA} in the hexagonal unit cell, which corresponds to $a_R$ = 5.128~{\AA} and $\alpha=55.287 \degree$ in the rhombohedral primitive cell. 
In the reference frame used in the simulations, the hexagonal $c_H$ axis is aligned along the cartesian $z$ axis, which is the vertical direction in Fig.~\ref{fig:crystal}.

\begin{figure}[th]
	\includegraphics[width=1.0\columnwidth]{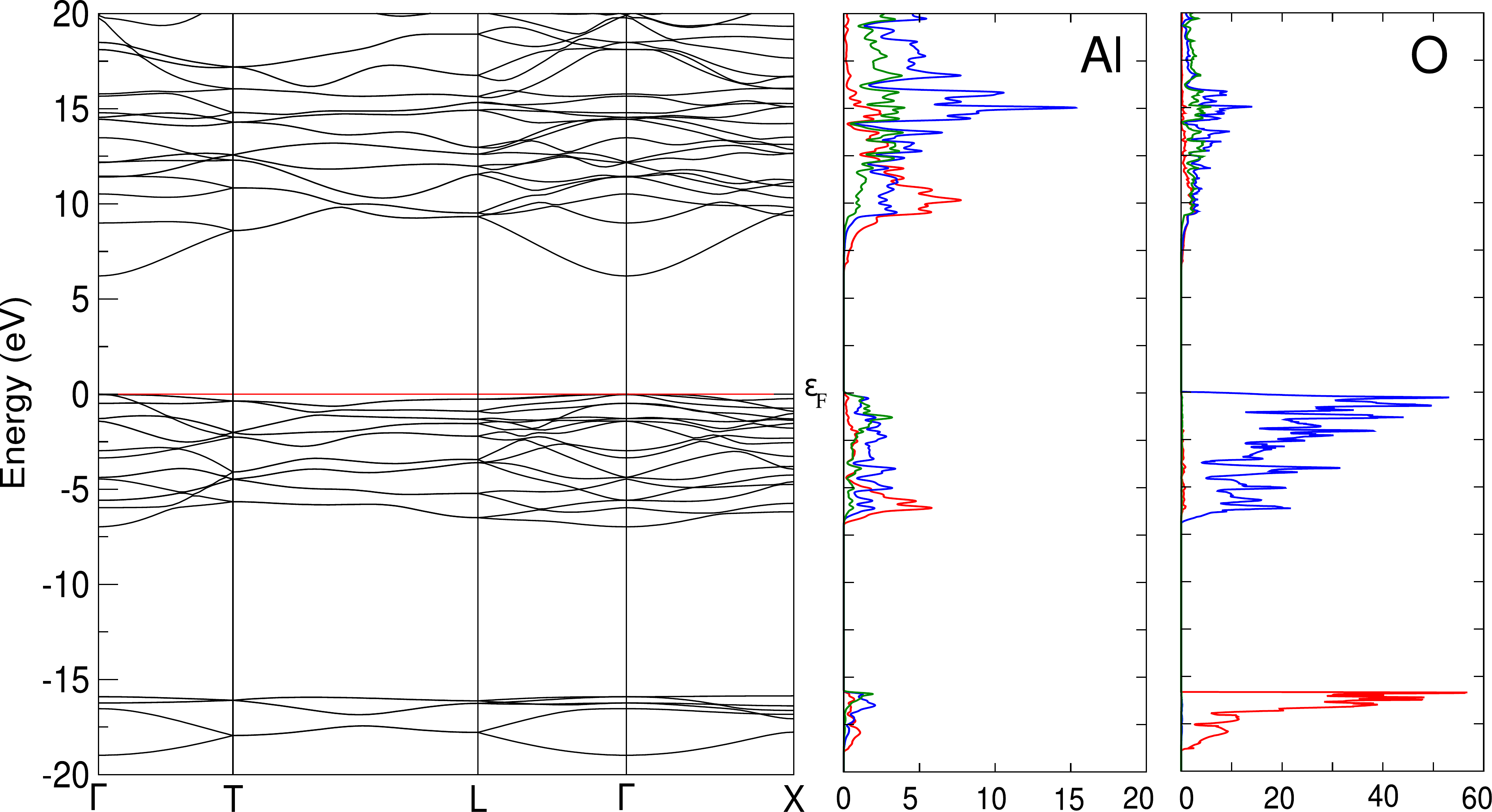}
	\caption{\label{fig:bands} (Left) LDA Kohn-Sham band structure of {\alo}. The top of the valence band has been set to zero. Density of states projected on (middle) Al  and (right) O atoms, resolved in the angular components: $s$ (red), $p$ (blue) and $d$ (green).}
\end{figure}

The left panel of Fig.~\ref{fig:bands} shows the Kohn-Sham LDA band structure along a high symmetry path in the first Brillouin zone, together with the projected density of states on the Al (middle panel) and O (right panel) atoms. 
{\alo} has a direct bandgap at the $\Gamma$ point, which amounts to 6.21~eV in the LDA. This value is in very good agreement with the result of Ref. \onlinecite{Mackrodt_2020} obtained with the same experimental lattice parameters. 
Calculations\cite{Ahuja_2004,Myrta_2011,Santos_2015,Mackrodt_2020} that adopt a crystal structure optimised within the LDA, rather than the experimental one, instead obtain larger band gaps. In particular, the difference with respect to Ref. \onlinecite{Myrta_2011} is 0.51 eV. We refer to Ref.  \onlinecite{Santos_2015} for a detailed analysis of the dependence on the band gap on the lattice parameters.
As usual, the Kohn-Sham band gap underestimates the experimental fundamental gap, estimated to be  9.57 eV
from temperature-dependent vacuum ultraviolet (VUV) spectroscopy\cite{French_1994} and 9.6 eV from conductivity measurements\cite{Will_1992}.

The 6  bands located between -19~eV and  -15.9~eV are the O $2s$ states, while the upper 18 valence bands, starting at  $\sim$ -7~eV, are mostly due to O $2p$ states, partially hybridised with Al states. The valence bands are quite flat along the entire path.
The bottom conduction band consists of Al $3s$ hybridised with O $3s$ at the $\Gamma$ point and also with O $2p$ elsewhere, showing a strong dispersion around the  $\Gamma$ point. The higher conduction bands have mainly Al $3p$ and $3d$ character, also hybridised with O states.  
This overview of the electronic properties  confirms the intermediate covalent-ionic nature of the chemical bond in {\alo}.

Finally, the Al $2p$ and $2s$ core levels (not shown in Fig.~\ref{fig:bands}) in LDA are located 61.7~eV and 99.4~eV below the top valence, which, as usual, largely underestimates the experimental values\cite{Crist_2004} of 70.7~eV and 115.6~eV, respectively. 
The calculations do not include the spin-orbit coupling, so the $2p_{1/2}$ and $2p_{3/2}$ levels are not split.
In all cases, we have verified that pseudopotential and all-electron calculations give the same band structures and core-level energies. 

\subsection{\label{ssec:benchmark} All-electron benchmark at the Al L$_{2,3}$ edge}

\begin{figure*}[ht]
	\includegraphics[width=2.0\columnwidth]{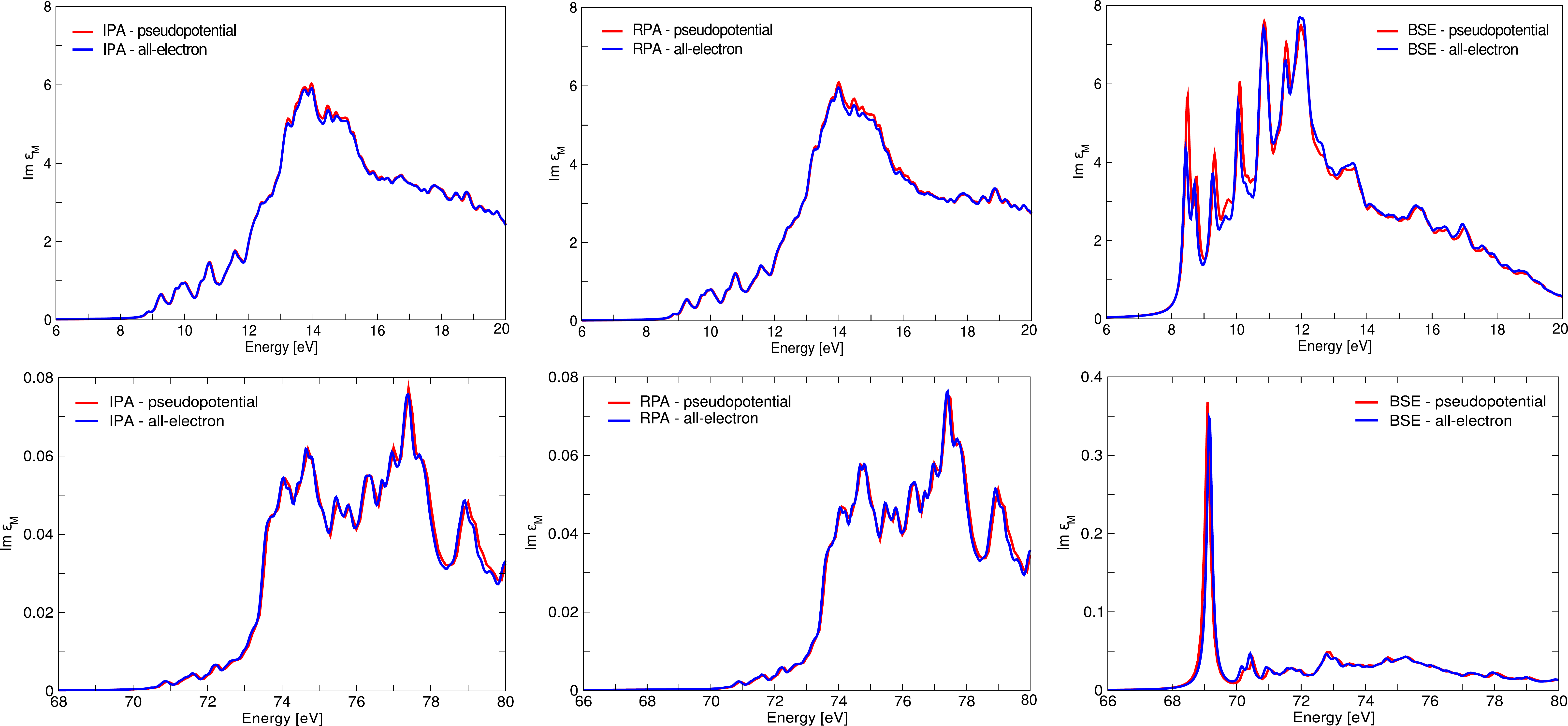}
	\caption{\label{fig:dp-vs-exc} Comparison of absorption spectra calculated with pseudopotential (red lines) and all-electron (blue lines) methods, using a $\Gamma$-centered  $8\times8\times8$ $\bfk$-point grid, (left panels) in the independent particle approximation (IPA), (middle panels) in the random-phase approximation (RPA), and (right panels) from the full Bethe-Salpeter equation (BSE). (Upper panels) Optical spectra (with 12 valence bands and 20 conduction bands). (Bottom panels) XANES spectra at Al L$_{2,3}$ edge (with 12 core levels and 12 conduction bands).} 
\end{figure*}

One of the main goals of this work is to demonstrate that shallow-core spectra can be calculated with high accuracy also using the pseudopotential  approximation. 
Even when reliable pseudopotentials provide accurate results in ground-state DFT calculations for the charge density or the total energy, or in GW calculations for quasiparticle excitations, the question whether they also give accurate XANES spectra should be examined carefully.
In DFT or GW calculations, the quantities of interest result from many integrals where the details of the single matrix elements are not crucial.  Instead, optical absorption and XANES spectra additionally reflect directly the quality of the individual dipole matrix elements in Eq.~\eqref{eq:rhotw}, which is a much more stringent requirement.
In the XANES regime, indeed, excitation energies are substantially higher than those found in optical spectra, which implies that the matrix elements involve regions of space closer to the ionic cores. In a norm-conserving pseudopotential approach, the true shallow core wavefunction is replaced by a pseudo-wavefunction. While their norms are identical, the true wavefunction and the pseudo-wavefunction are different for distances to the ionic cores that are smaller than the cutoff radii. Therefore, as a consequence of the pseudization of the wavefunctions, the dipole matrix elements \eqref{eq:rhotw} for the XANES spectra could be inaccurate in a pseudopotential scheme, even for accurate 
pseudopotentials that are reliable for DFT and GW calculations (such as the ones used in the present work). The limits of the accuracy of pseudopotentials for XANES spectra of shallow core levels, to the
best of our knowledge, have not been explored so far.

We used full-potential all-electron calculations, considered as the gold-standard computational method~\cite{Draxl-Puschnig_2002, Gulans_2014}, to benchmark the optical and core spectra calculated with PPs. In order to perform this comparison properly, for both optical absorption and Al L$_{2,3}$  edge XANES spectra, the same choice has been made for the occupied states included explicitly in the two calculations, and the resulting spectra were converged consistently in the two cases (see Sec. \ref{ssec:computational_details}). 

\begin{figure}[ht]
	\includegraphics[width=0.9\columnwidth]{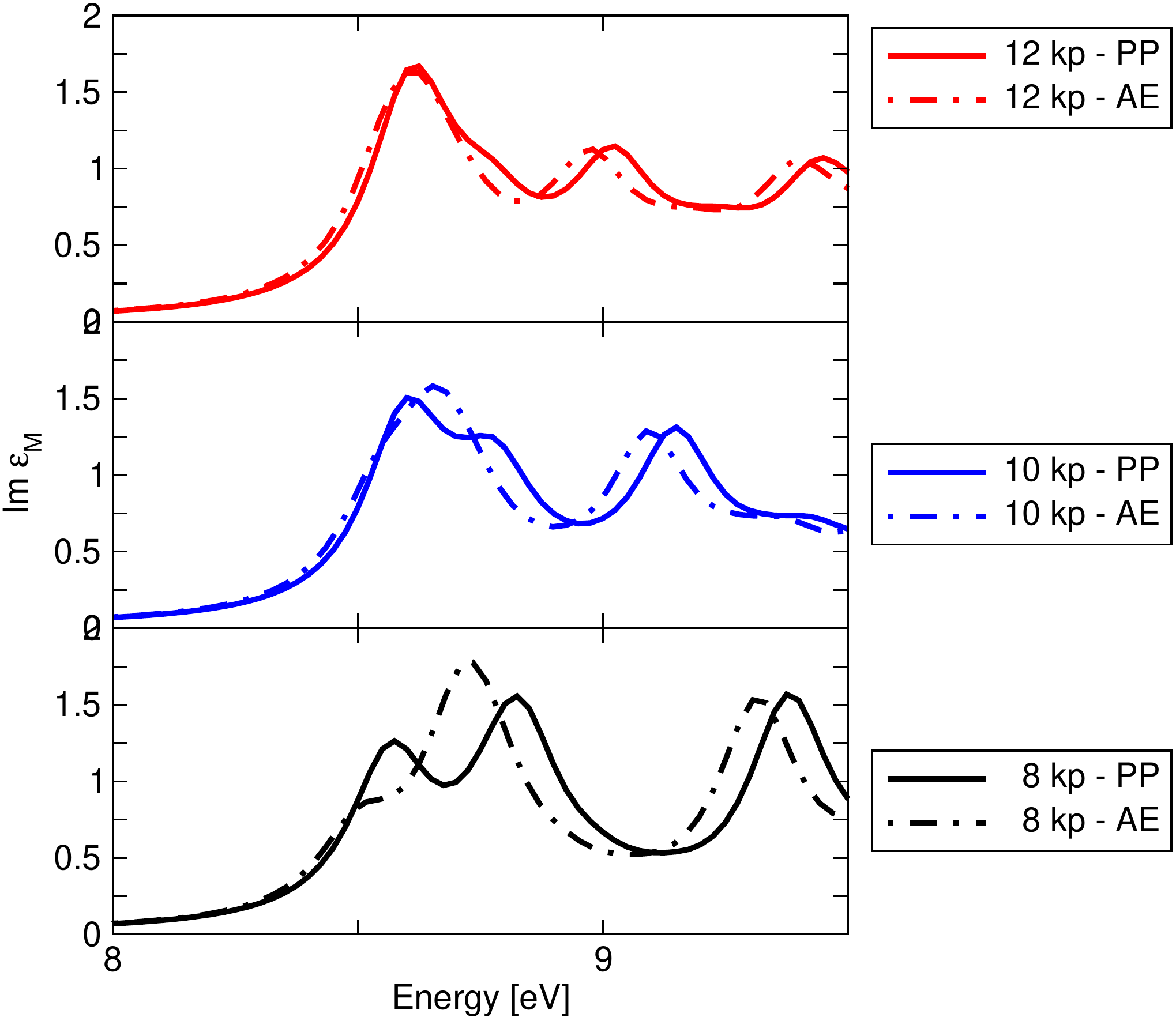}
	\caption{\label{fig:2cb-kp-convergence} Convergence of BSE optical spectra calculated with pseudopotential (PP) (solid lines) and all-electron (AE) (dot-dashed lines) methods (with 2 conduction and 2 valence bands), for increasing number of $\bfk$ points ($\Gamma$-centered grids with 8, 10 and 12 divisions  for bottom, central and top  panel, respectively).}
\end{figure}

The valence and L$_{2,3}$ spectra obtained at different levels of approximations, IPA, RPA and BSE, are shown in the top and bottom panels of Fig.~\ref{fig:dp-vs-exc}, respectively.
We can make several observations for both optical absorption and XANES spectra:
i) The results of the left panels of Fig.~\ref{fig:dp-vs-exc} show that the pseudopotential approximation reproduces the all-electron spectra with excellent accuracy within the IPA.   
ii) For the RPA spectra (central panels) we find a similar result. This is in part related to the fact that local fields effects are not important in the energy ranges considered here. 
iii) Finally, also the full BSE calculations  (right panels) with the two approaches are in very good agreement. To have an idea of what can be expected, we note that a recent comparison  between different BSE calculations of XAS spectra  presented larger discrepancies\cite{Unzog2022}.
The origin of the residual difference between the BSE spectra in  Fig.~\ref{fig:dp-vs-exc}
lies in the different treatment between the two codes of the integrable singularity of the diagonal matrix elements of electron-hole attraction $-W$ in Eq.~\eqref{eq:BSE}, calculated in reciprocal space, when $\bfk-\bfk'=\bfq=0$ and the reciprocal-lattice vectors are $\bfG=\bfG'=0$. We note that the different treatment of this singularity has been  recently identified as the origin of numerically different results  
also  in a comparison among different codes for GW  quasiparticle calculations \cite{Rangel_2020}.
This singularity is, in fact, eliminated in the BSE by evaluating the integral
$$
-\frac{4\pi}{\Omega} \epsilon^{-1}_{\bfG=0,\bfG'=0}(\bfq\to0) \frac{1}{(2\pi)^3}\int_{\Omega_{\bfq=0}}d\bfq \frac{1}{q^2}, 
$$
where $\Omega_{\bfq=0} = \Omega_{BZ}/N_k$. In order to carry out, numerically or analytically, this integral, one has to define the shape for the little volume $\Omega_{\bfq=0}$ around the origin of the Brillouin zone and, in anisotropic materials, choose the $\bfq\to0$ direction  in order to evaluate the inverse dielectric function $\epsilon^{-1}(\bfq\to0)$.
The different details about how this integral is performed can be found in Ref.\cite{Albrecht1999} and Refs. \cite{Puschnig2002}, for \exc\ and \exciting, respectively.
If we exclude this singular contribution, the two BSE results become superimposed, as in the IPA case. In addition, this contribution  vanishes (more or less rapidly according to the kind of exciton\cite{Fuchs2008,Gorelov2023}) in the convergency with $\bfk$ points. Fig.~\ref{fig:2cb-kp-convergence} indeed shows that the differences in the spectra obtained with the two codes tend to vanish with increasing number of $\bfk$ points. 

Most importantly for the scope of the present work, for this detailed comparison we find that the differences between the PP and AE codes remain always of the same order of magnitude for both valence and shallow-core spectra.
Therefore, in summary, we can safely conclude that the benchmarks with respect to the all-electron approach show that pseudopotential calculations for optical and XANES spectroscopies for semicore states are reliable and accurate.
	
\subsection{\label{ssec:interference} Interference effects at the Al L$_{1}$ edge} 

\begin{figure*}[ht]
	\includegraphics[width=0.66\columnwidth]{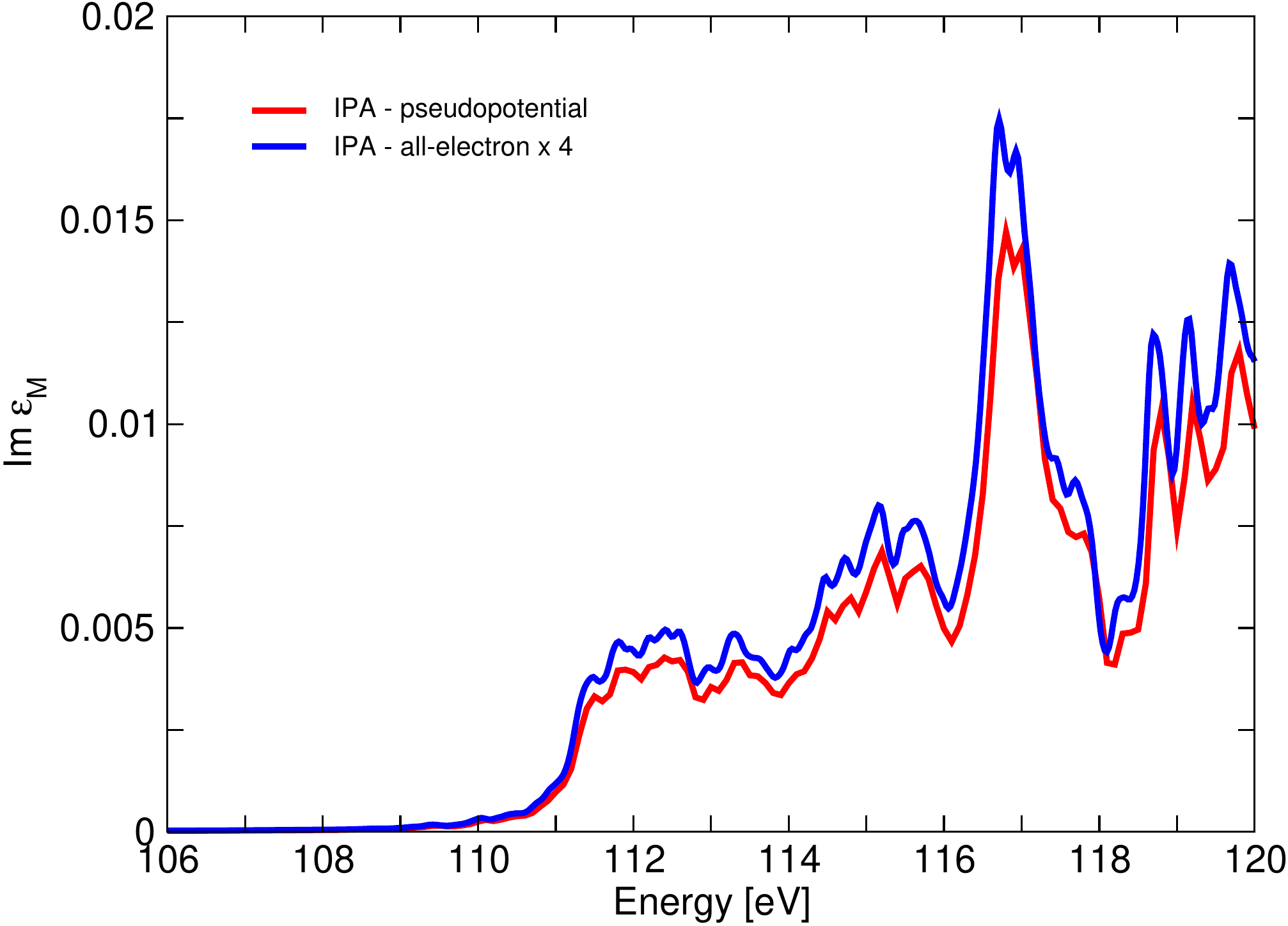}
	\includegraphics[width=0.66\columnwidth]{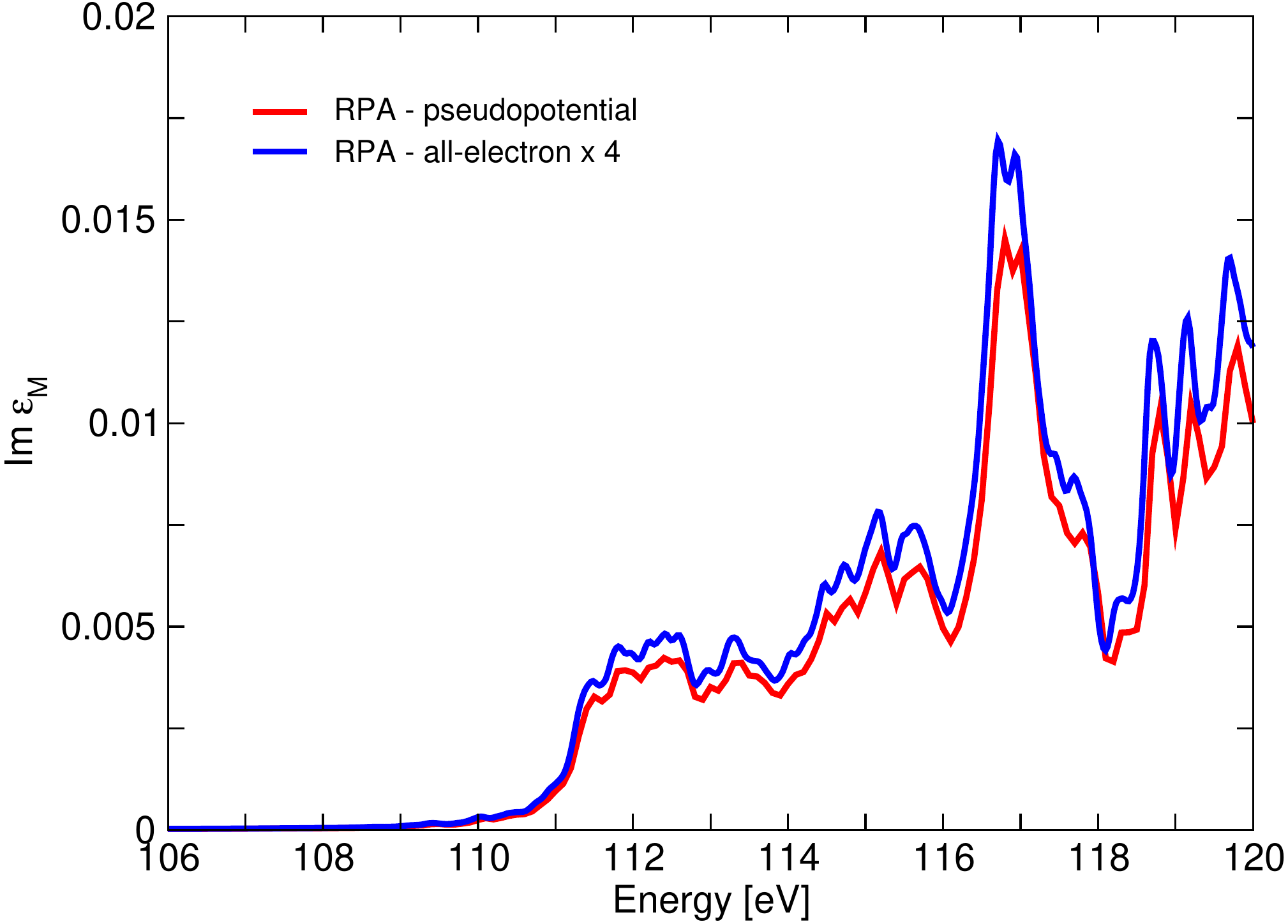}
	\includegraphics[width=0.66\columnwidth]{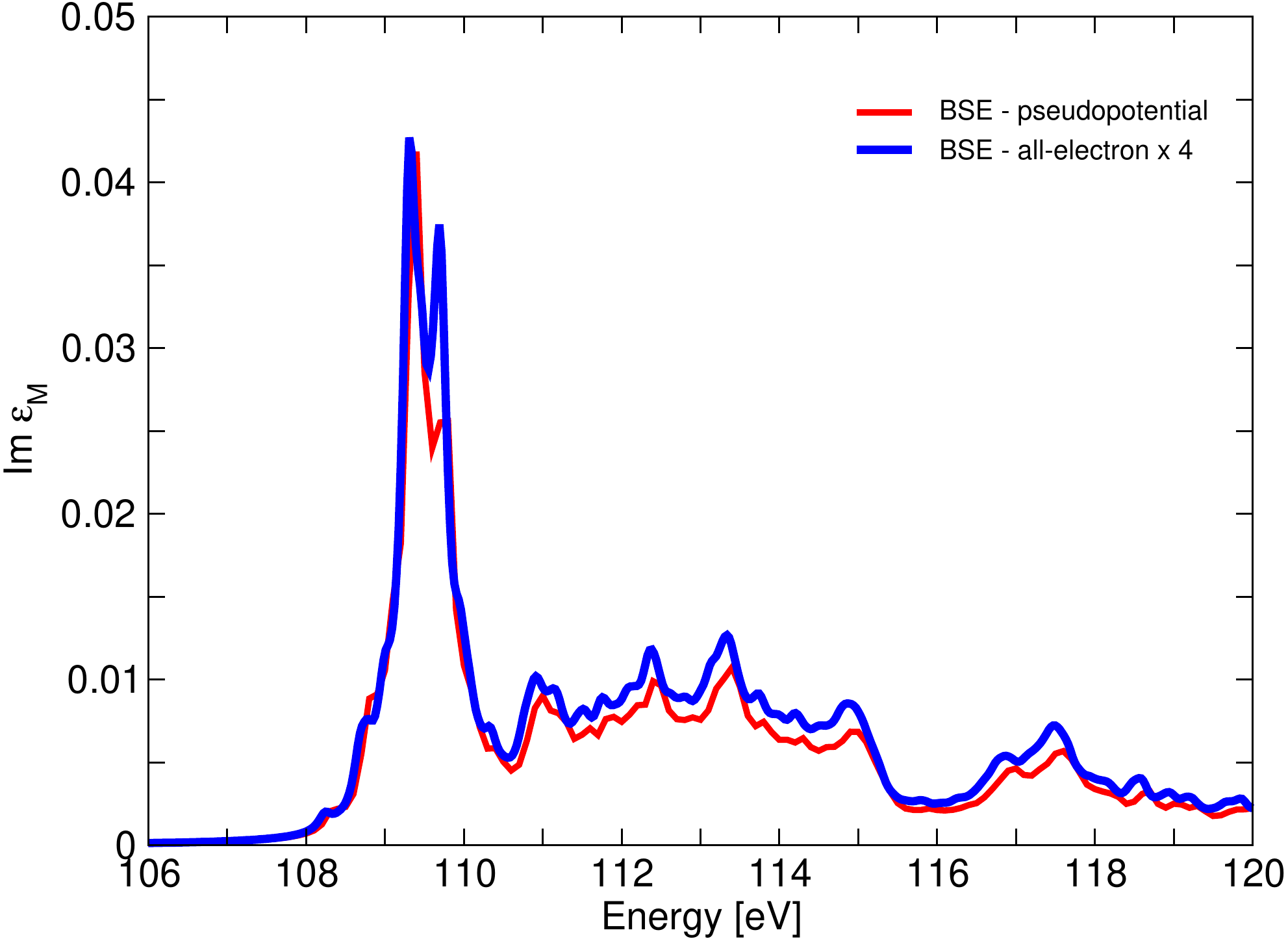}
	\caption{\label{fig:dp-vs-exc-l1} XANES spectra at the Al L$_{1}$ edge calculated with \exc\ (pseudopotential code) and \exciting\ (all-electron code). All the calculations are performed using a $\Gamma$-centered $8\times8\times8$ grid of $\bfk$ points and 30 unoccupied bands. In \exc\ we include the 4 $2s$ bands corresponding to the 4 Al atoms, while in \exciting\ we include only one $2s$ level (i.e., the $2s$ state on the Al atom where the core hole is created). For this reason, the spectra of \exciting\ are multiplied by 4.} 
\end{figure*}

The comparison between all-electron and pseudopotential calculations is more delicate for the Al L$_1$ edge, since in this case the semicore electrons are treated differently by the two codes. While \exciting\ considers the Al $2s$ states as core states, entirely contained inside the muffin tin, in \exc\ they are treated in the same manner as the valence electrons.  Moreover, \exciting\ fixes the core hole at a single specific atomic site, whereas \exc\ treats all the atomic sites on equal footing.

One of the limitations of the linearized augmented-plane-wave (LAPW) method is that it could give a wrong description of semicore states when they are considered to be contained only inside the muffin-tin (MT) sphere, but they actually overlap significantly with valence electrons or are too extended to be entirely contained inside the MT~\cite{muffin-tin_1993, Gulans_2014}. In order to overcome this problem, local orbitals are included to complement the basis. However, the quality of this basis set depends on the choice of energy parameters~\cite{local-orbital_1991, Gulans_2014}. 
In addition, interference effects between different atomic sites could play an important role, but they would not be straightforwardly taken into account when considering the core states only inside a single muffin-tin~\cite{Vorwerk_2022}. 
For all these reasons, since we have already  validated the pseudopotential approach for the Al L$_{2,3}$ edge spectra, we can now conversely use \exc\ to benchmark the  Al L$_1$ edge spectra obtained with \exciting.

The  L$_1$ edge absorption spectra calculated  using different levels of approximations are shown in Fig.~\ref{fig:dp-vs-exc-l1}. 
Notice that in \exc\ all the 4 bands corresponding to the 2s state of the 4 Al atoms need to be included in the calculations in order to properly represent the electronic transitions, while in \exciting\ only one occupied level is considered: the $2s$ state of the Al atom where the core hole is sitting. For a quantitative comparison between the two approaches,
since there are 4 equivalent Al atoms in the unit cell, the spectrum obtained with \exciting\ has been multiplied by 4. 

In all level of approximations, the pseudopotential and all-electron spectra differ slightly (and more than in the optical or L$_{2,3}$ edge cases),
showing that interference effects among the Al atoms come to play. These interferences are quite small in {\alo}, also because the Al atoms lie at equivalent positions in the unit cell, but they are in any case clearly detectable in the comparison in Fig.~\ref{fig:dp-vs-exc-l1}. It has been shown that these effects can be even  more significant  in other materials \cite{Vorwerk_2022}. 
While including these effects is still possible with \exciting\ (and with all approaches that fix the core hole in a specific position) by doing multiple calculations and generalizing Eq.~\eqref{spectrumBSE}, as described in detail in Ref.~\onlinecite{Vorwerk_2022},
interferences are naturally taken into account by 
pseudopotential approaches, where all atoms are treated on the same footing and only a single calculation is  needed.

\subsection{\label{ssec:absorption} Optical and XANES spectra: valence and shallow-core excitations}

\subsubsection{Comparison with experiments}

\begin{figure*}[ht]
	\includegraphics[width=\columnwidth]{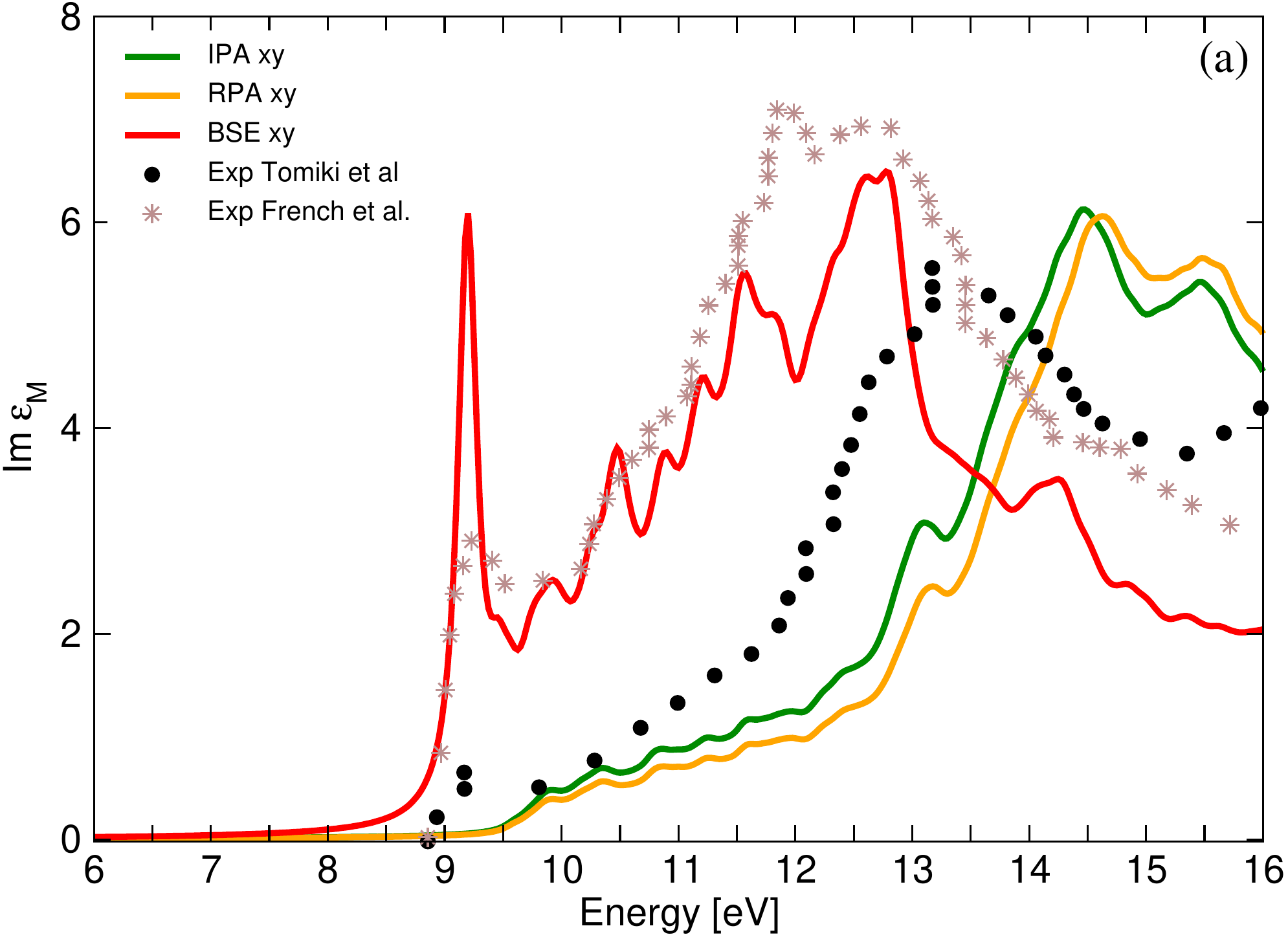}
    \includegraphics[width=\columnwidth]{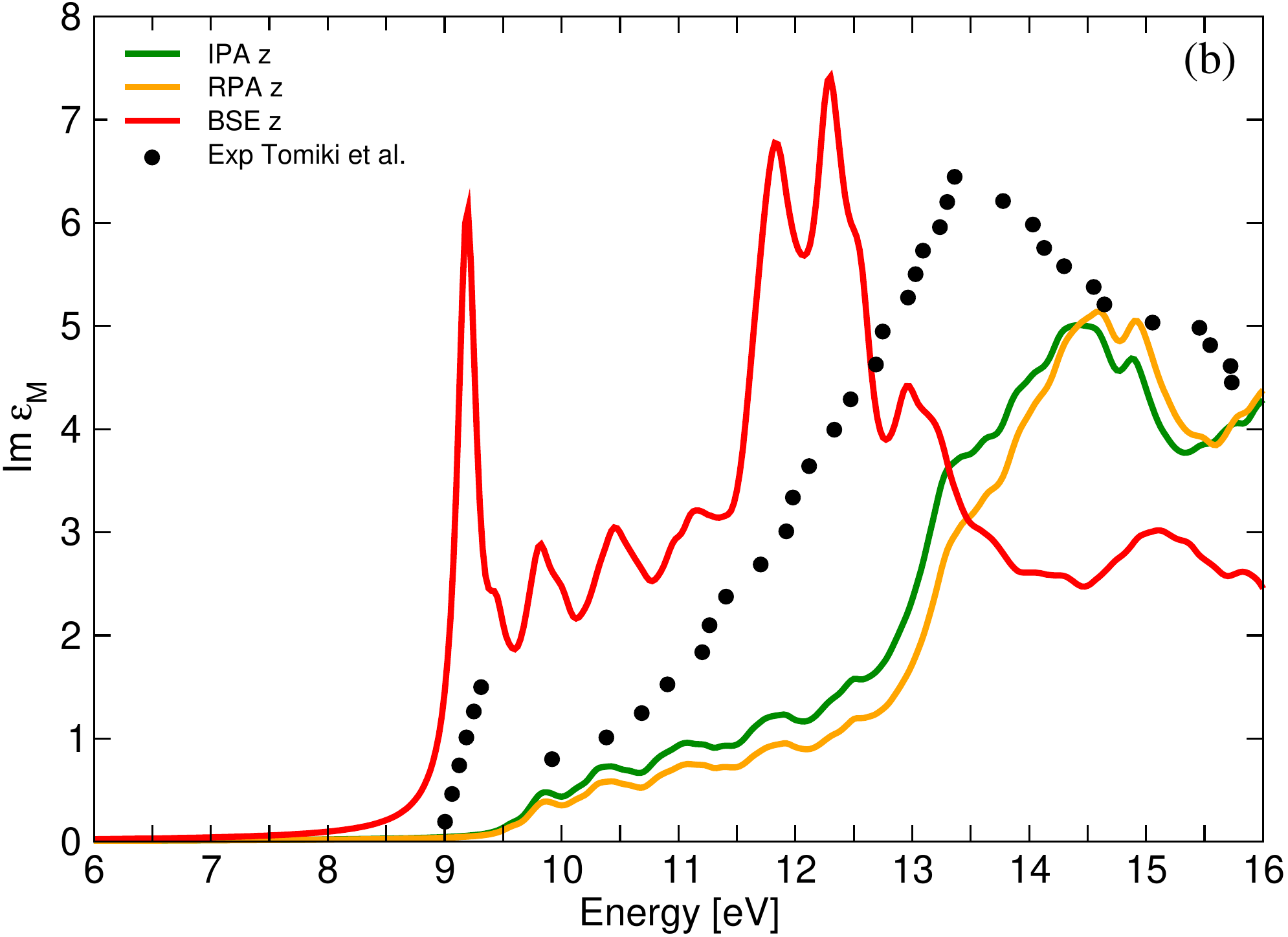}
	\includegraphics[width=\columnwidth]{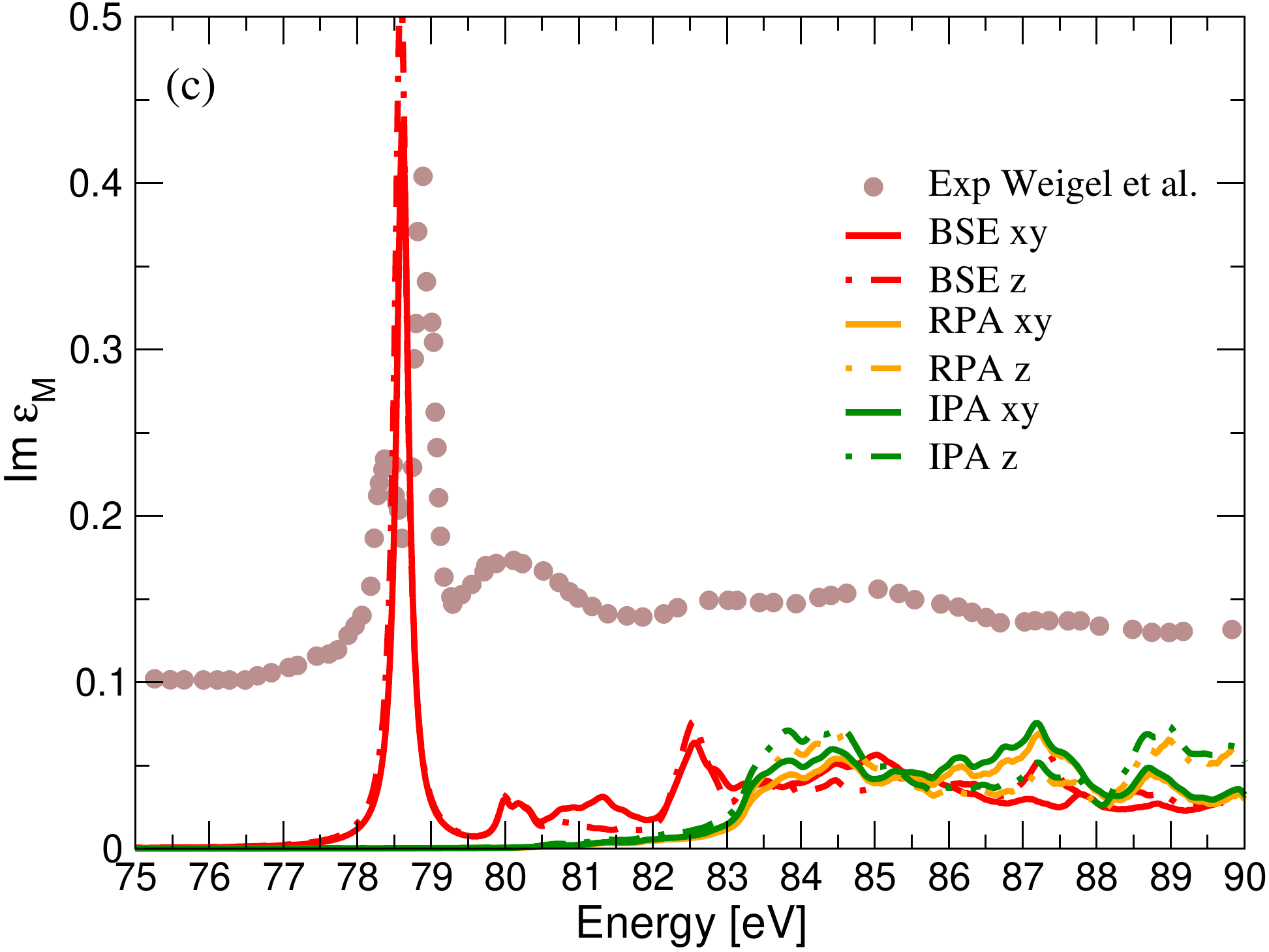}
    \includegraphics[width=\columnwidth]{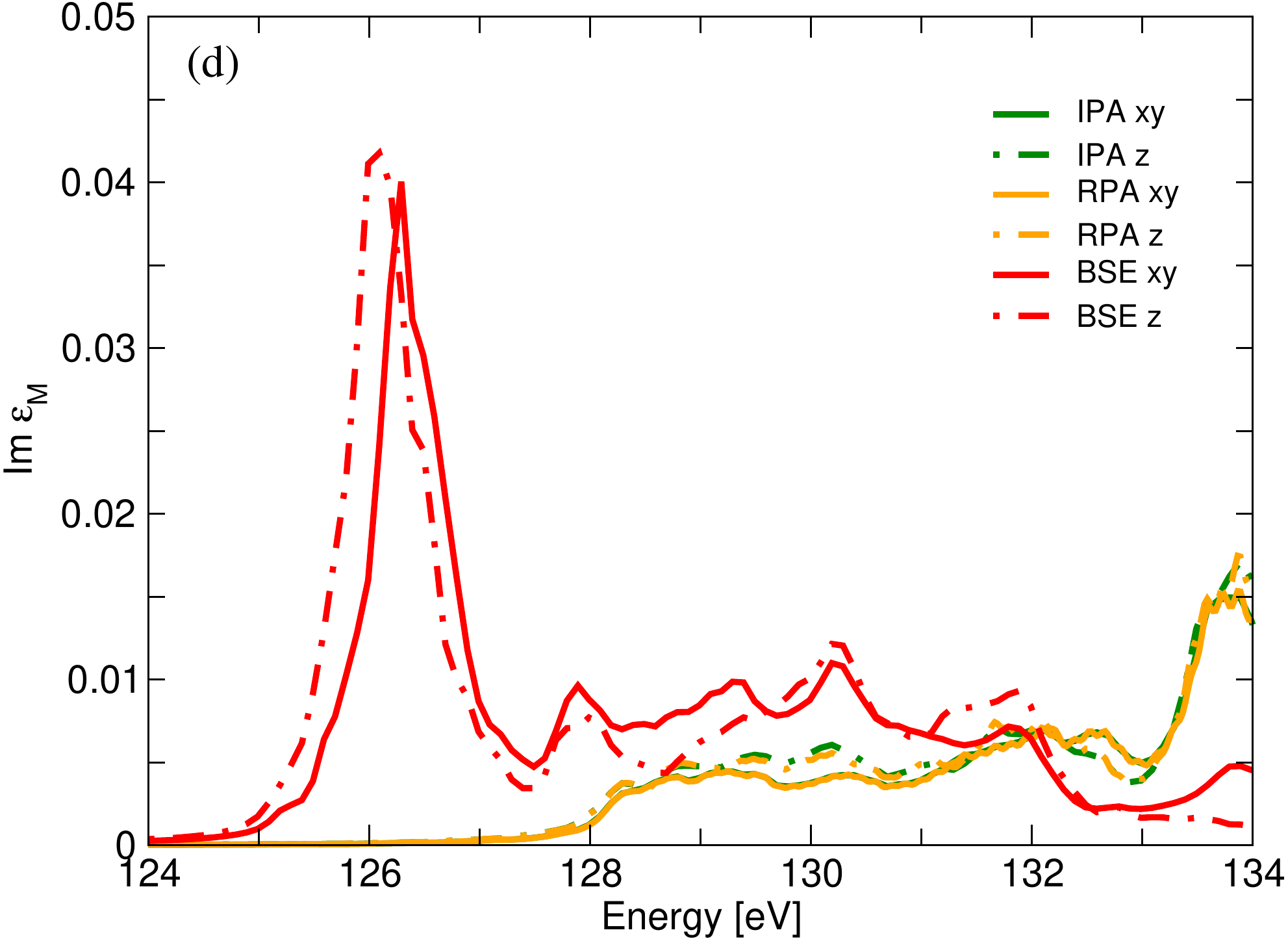}
	\caption{\label{fig:absorption} Comparison of theoretical results with experimental data from Tomiki {\it et al.}  \protect\cite{Tomiki_1993} and French {\it et al.}\protect\cite{French_1998} for the optical absorption, and Weigel  {\it et al.} \protect\cite{Weigel_2008} for the XANES at the L$_{2,3}$ edge.
	The calculated spectra are obtained in the independent particle approximation (IPA),  green lines, in the random-phase approximation (RPA),  orange lines, and from the solution of the Bethe-Salpeter equation (BSE),  red lines. 
	Optical absorption spectra for polarization in the  (a)  $xy$ plane and (b) in the  $z$ direction: the calculated spectra have been blueshifted by 0.7 eV.
    (c) Absorption spectra at the L$_{2,3}$ edge in the $xy$ (solid lines) and $z$ (dot-dashed lines) polarizations compared to the isotropic XANES experimental spectrum\cite{Weigel_2008}, to which a vertical offset has been added for improved clarity.  (d) Absorption spectra at the L$_1$ edge in the $xy$ (solid lines) and $z$ (dot-dashed lines) polarizations.}
\end{figure*}

Fig.~\ref{fig:absorption} compares the calculated absorption spectra $\text{Im} \epsilon_M(\w)$ with experiment, for both the optical absorption corresponding to valence excitations and the XANES spectrum of the shallow-core excitations at the Al L$_{2,3}$ edge. 
The same figure also displays the results of the calculations at the Al L$_{1}$ edge, where, to best of our knowledge, no experimental XANES spectra are available for {\alo}, since this core level excitation is less commonly studied than the Al K edge\cite{Cabaret_1996,Ildefonse_1998,vanBokhoven_1999,Mizoguchi_2009}.
In all cases, the presence of sharp and pronounced peaks at the onset of the BSE spectra (red lines), which are absent in the RPA and IPA spectra (orange and green lines), is an evidence of strong excitonic effects.    
Taking into account the electron-hole attraction in the BSE is the key to bring the calculations in agreement with experiment.

As already discussed in Ref. \onlinecite{Myrta_2011}, for the optical absorption in the polarization direction perpendicular to the $z$ axis (i.e. in the $xy$ plane), where two VUV spectroscopy experiments \cite{Tomiki_1993,French_1998} are available,
there are large discrepancies between  the experimental spectra themselves [see Fig.~\ref{fig:absorption}(a)].
They agree on the position of the absorption onset and the presence of a sharp peak at $\sim$~9.2 eV, while they largely differ in the intensities of the various spectral features.
Those differences can be attributed to the fact that both absorption spectra have been obtained from measured reflectivity data using the Kramers-Kroning relations, which introduces uncertainties in the  $\text{Im}\epsilon_M(\w)$ spectra.
The calculated optical spectra in Fig.~\ref{fig:absorption}(a)-(b) have been blueshifted by 0.7 eV. 
This underestimation of the onset of the absorption spectrum is a manifestation 
 of the underestimation of the band gap 
 by the perturbative G$_0$W$_0$ approach, which is a systematic error for large gap materials\cite{vanSchilfgaarde_2006}. 
As a matter of fact, the 2.64 eV scissor correction that we have employed here, which is taken from the G$_0$W$_0$ calculation in Ref. \onlinecite{Myrta_2011}, underestimates the band gap correction to the LDA.
The BSE calculation  in Ref. \onlinecite{Myrta_2011} is also in very good agreement with the present result: the 
difference in the peak positions 
is actually due to the LDA band gap difference (see Sec.~\ref{sec:bands}).
The BSE spectrum  in the $xy$ polarization reproduces well the spectral shape measured by French {\it et al.}\cite{French_1998}, while there are larger differences with the experimental spectra in both polarizations measured by Tomiki {\it et al.}\cite{Tomiki_1993}.

At the Al L$_{2,3}$ edge, see Fig.~\ref{fig:absorption}(c), the calculated spectra have been blueshifted by 9.75 eV, which matches well the needed correction to the LDA Al $2p$ core level energy (see Sec.~\ref{sec:bands}).
The calculations neglect the spin-orbit coupling and therefore miss the splitting of the main peak into a doublet separated by 0.47 eV in the high-resolution experimental XANES spectrum from Ref. \onlinecite{Weigel_2008} (which also agrees well with previous experiments \cite{OBrien-Al2O3-L23_1991,OBrien_1993,Tomiki_1993}).
In the spectra, the first, most prominent, excitonic peak is followed by a series of lower intensity peaks.
While the absolute intensity of the experimental spectrum is arbitrary, the relative intensity of the first and second peaks gives information about the coordination number of Al and the nature of the chemical bond: a lower symmetry enhances the intensity of the second peak. Moreover, a lower coordination shifts the edge to lower energies, while higher bond ionicity shifts the edge to higher energies \cite{vanBokhoven_2001,Weigel_2008}.

At the Al L$_1$ edge there is no available experiment. Therefore, the curves in Fig.~\ref{fig:absorption}(d) have been shifted by 19.5 eV, in order for the smallest independent-particle transition energy, from the Al $2s$ band to the bottom-conduction band, to match the experimental value of 125.2 eV, which corresponds to the sum of the fundamental band gap plus the binding energy of the Al $2s$ states \cite{French_1994,Crist_2004} (see Sec.~\ref{sec:bands}).   
We find that the main prominent excitonic peak in the BSE spectra is preceded by a pre-edge structure, more evident in the $xy$ direction (solid lines).
At the Al K edge, which mainly probes the analogous $1s\to3p$ transition, there has been much work to explain the origin of a similar prepeak structure\cite{Cabaret_1996,Mo_2000,Cabaret_2005,Cabaret_2009,Brouder2010,Manuel_2012,Nemausat_2016,Delhommaye_2021}, which has been finally interpreted in terms of atomic vibrations enabling monopole transitions to unoccupied Al $3s$ states.
In the present case, the calculations do not take into account the coupling with atomic vibrations and nevertheless the BSE spectra show a prepeak structure. This finding therefore calls for a detailed comparison with other calculations including atomic vibrations and, possibly, experiments at the Al L$_1$ edge.

\subsubsection{Anisotropy and local field effects}

The {\alo} crystal is optically uniaxial. As shown by 
Fig.~\ref{fig:absorption}(a)-(b), at the onset of the optical spectrum the anisotropy is rather small, while it becomes larger for higher energy features.
The lowest energy exciton is visible along the $z$ polarization, while it is dark in the perpendicular $xy$ polarization. It is separated by $\sim$ 25 meV from a pair of degenerate excitons that are visible in the perpendicular $xy$ direction and, conversely, dark in the $z$ direction.  Tomiki {\it et al.}\cite{Tomiki_1993} experimentally determined a similar splitting of the exciton peaks in the two polarization directions (35 meV at room temperature and 86 meV at 10 K).
We find that the binding energy of these excitons is of order of 0.3 eV, which is 
more than twice the 0.13 eV value estimated from temperature-dependent VUV spectroscopy\cite{French_1994}.
A similar splitting of the lowest energy exciton occurs also at the L$_{2,3}$ edge\cite{Tomiki_1993}, where 
its binding energy largely increases up to 1.6 eV.
For the optical and the L$_{2,3}$ cases, both the lowest energy exciton in the BSE spectrum and the excitation at the smallest independent-particle transition energy in the IPA spectrum have a significant oscillator strength.
Instead, at the L$_1$ edge the lowest energy transitions have a $2s\to3s$ character and are dipole forbidden. 
We find that the binding energy of the lowest dark exciton at the L$_1$ edge is 1.2 eV.
The lowest bright excitons in the $z$ and $xy$ polarization directions are located 1.6 eV and 1.8 eV above it, respectively. They belong to the prepeak in the spectrum.
In this case, we define their binding energy as the difference with respect to the corresponding first allowed transition in the IPA spectrum: it amounts to 0.6 eV. The splitting of the main exciton peak in the two polarizations is also the largest one at the L$_1$ edge, being more than 0.2 eV.

By comparing the RPA and IPA optical spectra, orange and green lines in Fig.~\ref{fig:absorption}(a)-(b), respectively, we note that the effect of crystal local fields is quite small for both polarizations, in contrast to typical layered van  der Waals materials like  graphite, where local field effects are strong for the polarization along the hexagonal axis\cite{Marinopoulos_2002}.
Marinopoulos and Gr\"uning\cite{Myrta_2011} also found that local field effects are not essential to describe satisfactorily the low energy part of the experimental spectra, whereas they become crucial for higher energies (above 16 eV, not shown in Fig.~\ref{fig:absorption}), in correspondence to the excitation of the more localised O $2s$ electrons.
Indeed, the degree of electron localisation directly correlates with the degree of charge inhomogeneity, which is a key factor for the induced microscopic local fields.
One may therefore expect that the excitation spectra of shallow-core levels, which are even more localised, should be more affected by local field effects. This phenomenon has been in fact observed for many shallow-core levels\cite{Vast_2002,Dash_2007,Huotari_2010,Cudazzo_2014,Ruotsalainen_2021}.
However, in {\alo} for both the L$_{2,3}$ and L$_1$ edges
the comparison of the absorption spectra calculated within the RPA and in the IPA shows that local field effects are actually negligible\footnote{It is well known that local field effects, expressed as electron-hole exchange interaction in the BSE framework, are essential to get the correct branching ratios between L$_{2}$ and L$_{3}$ components, see e.g.  \cite{Gilmore2015,Vinson2012,Ankudinov_2005}.
However, in the present case the neglect of spin-orbit coupling does not allow us to resolve the two components.  For {\alo} an electron–hole exchange energy of 0.3 eV  has been estimated \cite{Weigel_2008,OBrien-Al2O3-L23_1991}.} (even weaker than in the optical regime). 
We can understand this result by noticing that the intensity of the L$_{2,3}$ and L$_1$  absorption spectra  is one or two orders of magnitude smaller than for the optical absorption. This large intensity difference reflects the fact that Al $2p$ and $2s$ states are much less polarizable than valence states. 
Therefore, even though their electronic charge is much more localized and inhomogeneous, local fields associated to the excitations of Al $2p$ and $2s$ are small because they are weakly polarizable, which also leads to weak induced potentials.

\subsubsection{Analysis of excitonic effects}
\begin{figure*}	
	\includegraphics[height=4.0cm,keepaspectratio,trim= 0.5cm 1.2cm 0.75cm 0.5cm]{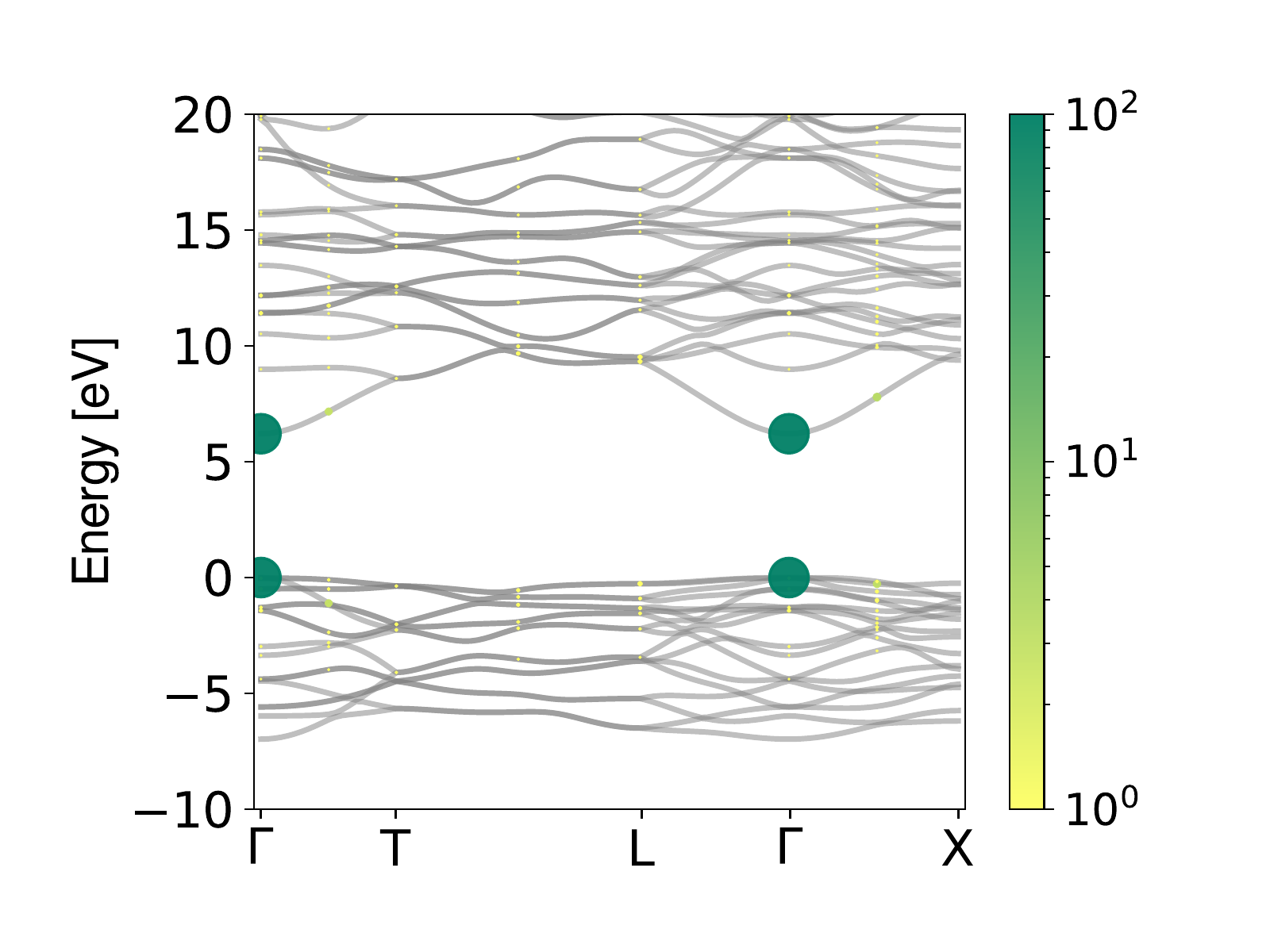}
	\includegraphics[height=4.0cm,keepaspectratio,trim= 0.5cm 1.2cm 0.75cm 0.5cm]{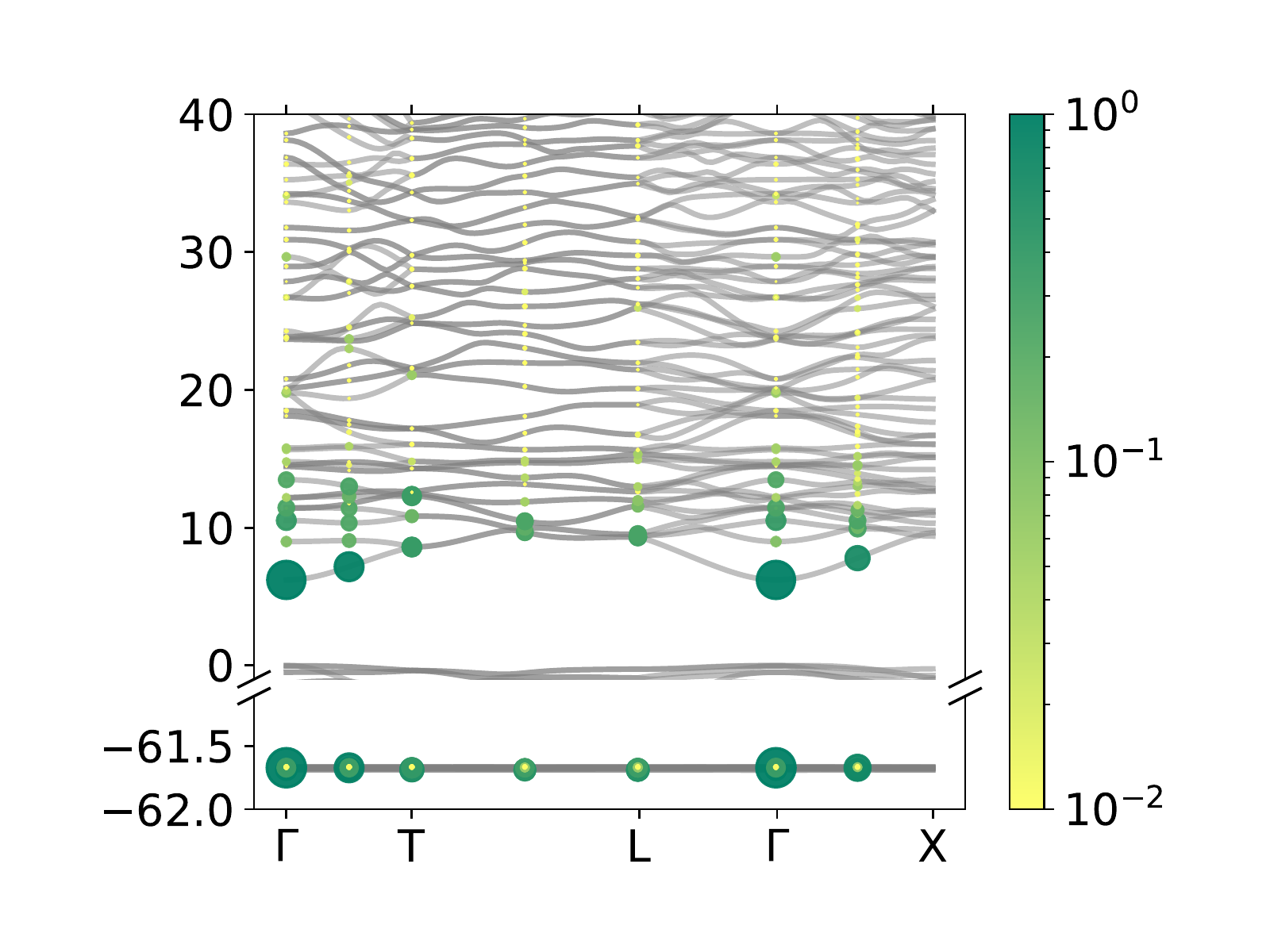}
	\includegraphics[height=4.0cm,keepaspectratio,trim= 0.5cm 1.2cm 0.75cm 0.5cm]{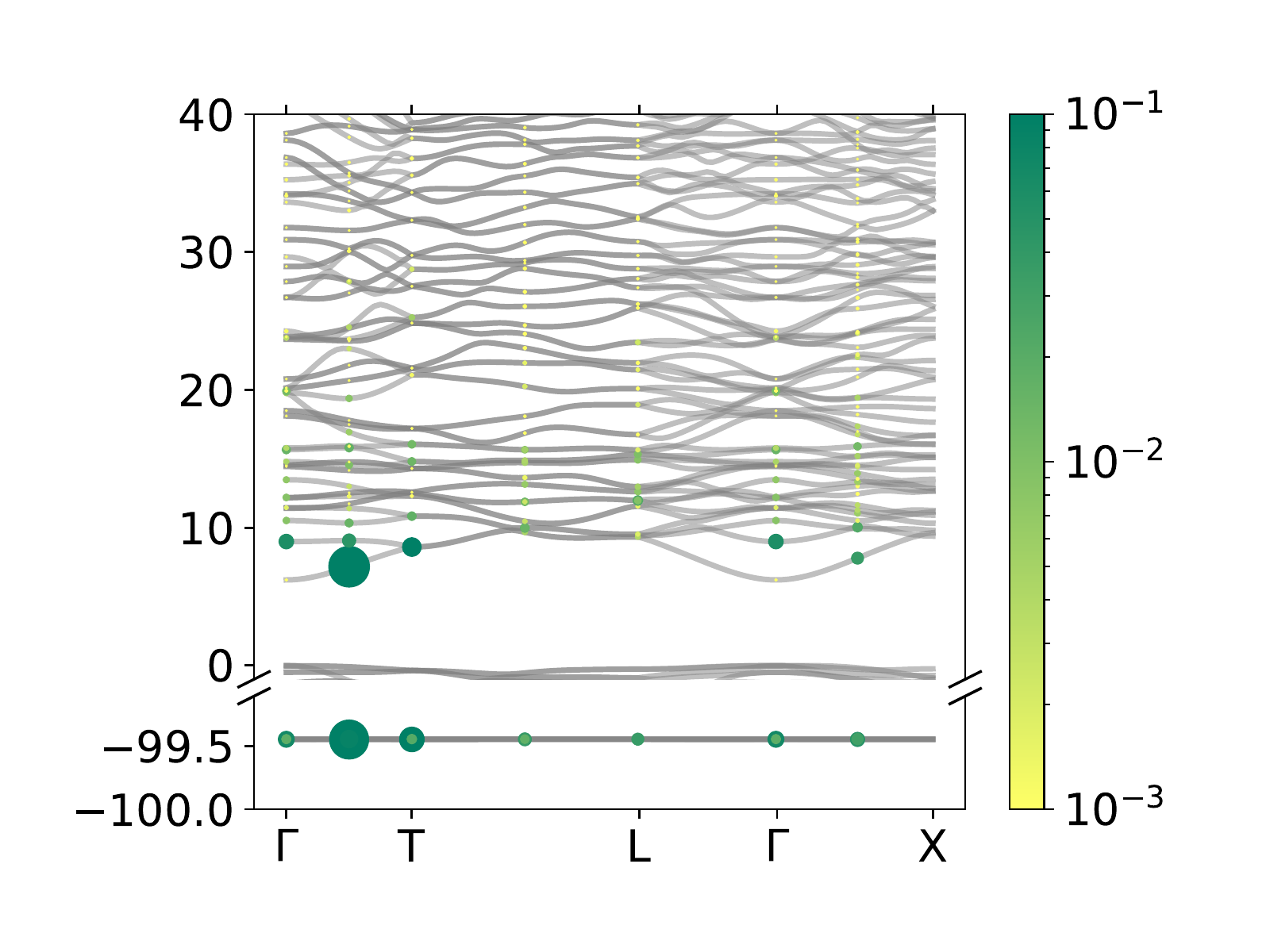}
	\caption{\label{fig:Arho} Contributions of independent transitions to the lowest energy bright  exciton  intensity in the absorption spectra: (left) for the optical spectrum; (middle) for the XANES at theL$_{2,3}$ edge; (right) for the XANES at the L$_{1}$ edge. The three plots are performed for polarization in the $z$ direction. The size of the circles is proportional to $\vert \tilde\rho_{vc{\bfk}} A_{\lambda}^{vc{\bfk}}\vert$.} 
\end{figure*}

Excitonic effects in solids can be understood as the result of the mixing of the independent-particle, vertical interband transitions at various $\bfk$ points in the Brillouin zone, which are weighted by the excitonic coefficients $A_{vc\bfk}^\lambda$, i.e., the eigenvectors of the excitonic Hamiltonian~\eqref{eq:BSE}.
The analysis of the excitonic coefficients therefore directly informs on the character of the exciton.

Fig.~\ref{fig:Arho} represents, projected on the LDA band structure, the partial contributions $\left| A_{vc\bfk}^\lambda \tilde\rho_{vc\bfk} \right|$ 
to the oscillator strength of the lowest energy bright excitons in the absorption spectra of Fig.~\ref{fig:absorption}.

Each independent-particle transition ${v\bfk\to c\bfk}$ is represented by a pair of circles, one in the occupied band $v$ and one in the unoccupied band $c$, whose size is proportional to the value of the contribution.
For the optical spectrum (left panel of Fig.~\ref{fig:Arho}), we consider the exciton giving rise to the first peak in the absorption spectrum in the $z$ polarization. Our analysis shows that the largest contribution stems from the top-valence bottom-conduction transition at the $\Gamma$ point, in correspondence to the direct band gap. The next $\bfk$ points along the L$\Gamma$X line in the conduction band give a contribution that is already 10 times smaller. The others are even smaller. 
This is due to the fact that for this exciton the top-valence bottom-conduction transition at the $\Gamma$ point has the predominant coefficient $A^{vc{\bfk}}_\lambda$, together with a large
single-particle oscillator strength $\tilde\rho_{vc\bfk}$ in the $z$ direction.
Instead, the same $\tilde\rho_{vc\bfk}$ is negligibly small in the $x$ or $y$ direction, explaining why the same exciton is dark in the $xy$ plane.

For the L$_{2,3}$ and L$_1$ excitation spectra, all the $\bfk$ points for the corresponding core levels are involved in the spectra, as one may expect from the fact that the core levels are not dispersive.
Also for first exciton peak in the L$_{2,3}$ XANES spectrum (middle panel of Fig.~\ref{fig:Arho}), the lowest conduction band at the $\Gamma$ point gives the largest contribution, having a large Al $3s$ character (see Sec.~\ref{fig:bands}).
However, in this case the other $\bfk$ points of the bottom conduction band and the higher conduction bands significantly contribute to the spectrum as well.
This illustrate the deviation from a simple independent-particle picture of a Al $2p\to3s$ atomic transition, 
since many transitions are mixed together to 
produce the excitonic peak at the onset of the L$_{2,3}$ XANES spectrum.

For the L$_1$ XANES spectrum (right panel of Fig.~\ref{fig:Arho}), we consider the first bright exciton in the $z$ polarization direction, which belongs to the prepeak in the spectrum in Fig.~\ref{fig:absorption}(d).
Contrary to the other two cases, the bottom-conduction band at the $\Gamma$ point gives no contribution, consistently with the $2s\to3s$ character of the transition, which is dipole forbidden.
The largest contributions are instead given by the $\bfk$ points along the $\Gamma$T line of the bottom conduction band, which  have $3p$ character as well.  Even in this case higher conduction bands contribute significantly to the intensity of the excitonic prepeak.

\begin{figure}[ht]
	\includegraphics[width=0.8\columnwidth]{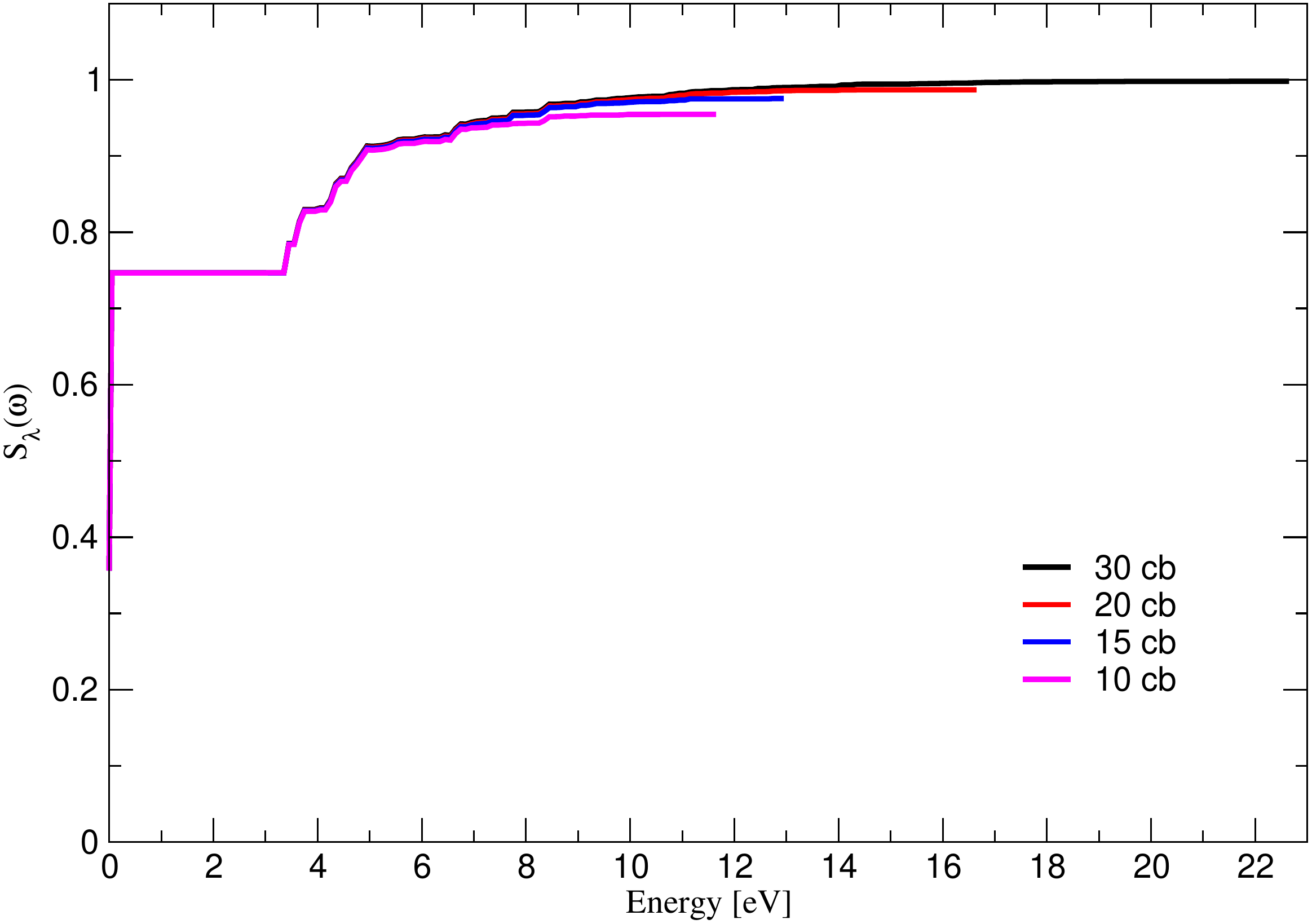}
 	\includegraphics[width=0.8\columnwidth]{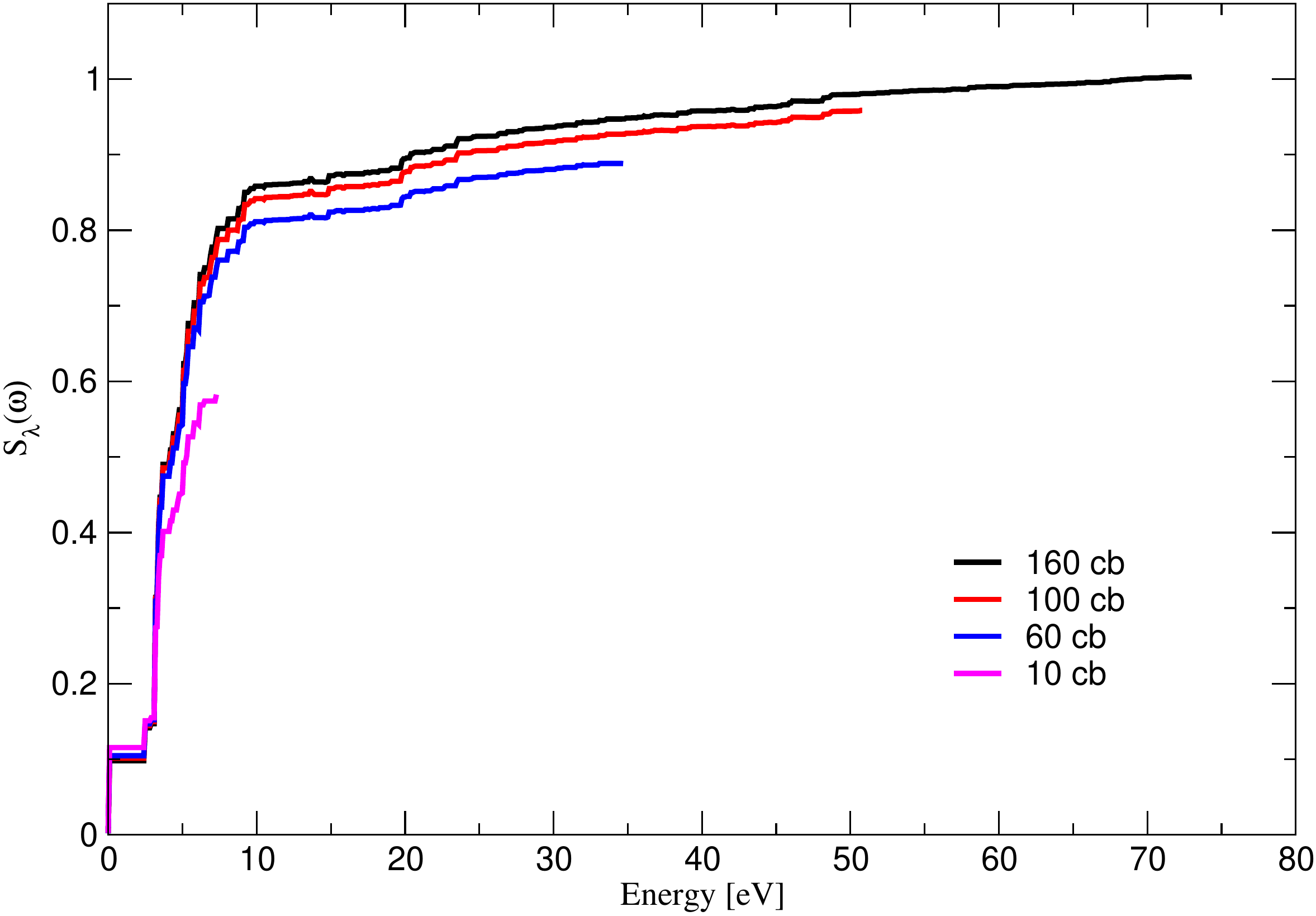}
	\caption{\label{fig:sum-Arho} Cumulative sums $S_\lambda(\w)$ as a function of number of conduction bands (cb) in the BSE hamiltonian for the lowest energy bright exciton in the $z$ direction for (top panel) the optical spectrum (bottom panel) and the XANES spectrum at the L$_{2,3}$ edge. In each case, the zero of the energy axis has been set  to the smallest independent-particle transition energy and  $S_\lambda(\w)$ has been normalised to its largest value.}
\end{figure}

The plot in Fig.~\ref{fig:sum-Arho} of the cumulative sums $S_\lambda(\w)$, see Eq.~\eqref{eq:cumu}, as a function of the number of conduction bands explains the different convergence behavior between the optical and   XANES spectra shown in Fig. \ref{fig:cb-convergence}.
By increasing the number of conduction bands in the BSE Hamiltonian~\eqref{eq:BSE},  the largest possible independent-particle transition energy progressively increases.
Therefore, the curves for larger numbers of conduction bands extend to higher energies.
However, in the case of the optical spectrum (top panel), the cumulative sum $S_\lambda(\w)$ rapidly converges to the final result. Already considering transition energies within 12 eV from the smallest one and including 15 conduction bands in the BSE hamiltonian give a result of the oscillator strength very close to 100\%.
Instead, in the case of the  L$_{2,3}$ edge (bottom panel), the range of transition energies needed to get close to 100\% has to be much larger, of the order of 50 eV above the smallest transition energy. 
Moreover, the various curves in the bottom panel of Fig.~\ref{fig:sum-Arho} do not overlap, as it is the case for the optical spectrum in the upper panel. This behavior indicates  that at the  L$_{2,3}$ edge interband transitions to higher conduction bands in the BSE hamiltonian mix together with transitions to lower conductions bands, which  affects the behavior of the cumulative sum $S_\lambda(\w)$ also at lower energies.
The reason of this strong mixing is the fact that  at the  L$_{2,3}$ edge there are many interband transitions with similar  intensity. 
This, in turns, explains why the convergence of the XANES spectrum with the number of conduction bands is slow (see Fig.~\ref{fig:cb-convergence}), and requires extra care.

\begin{figure}[th]
	\includegraphics[width=1.0\columnwidth]{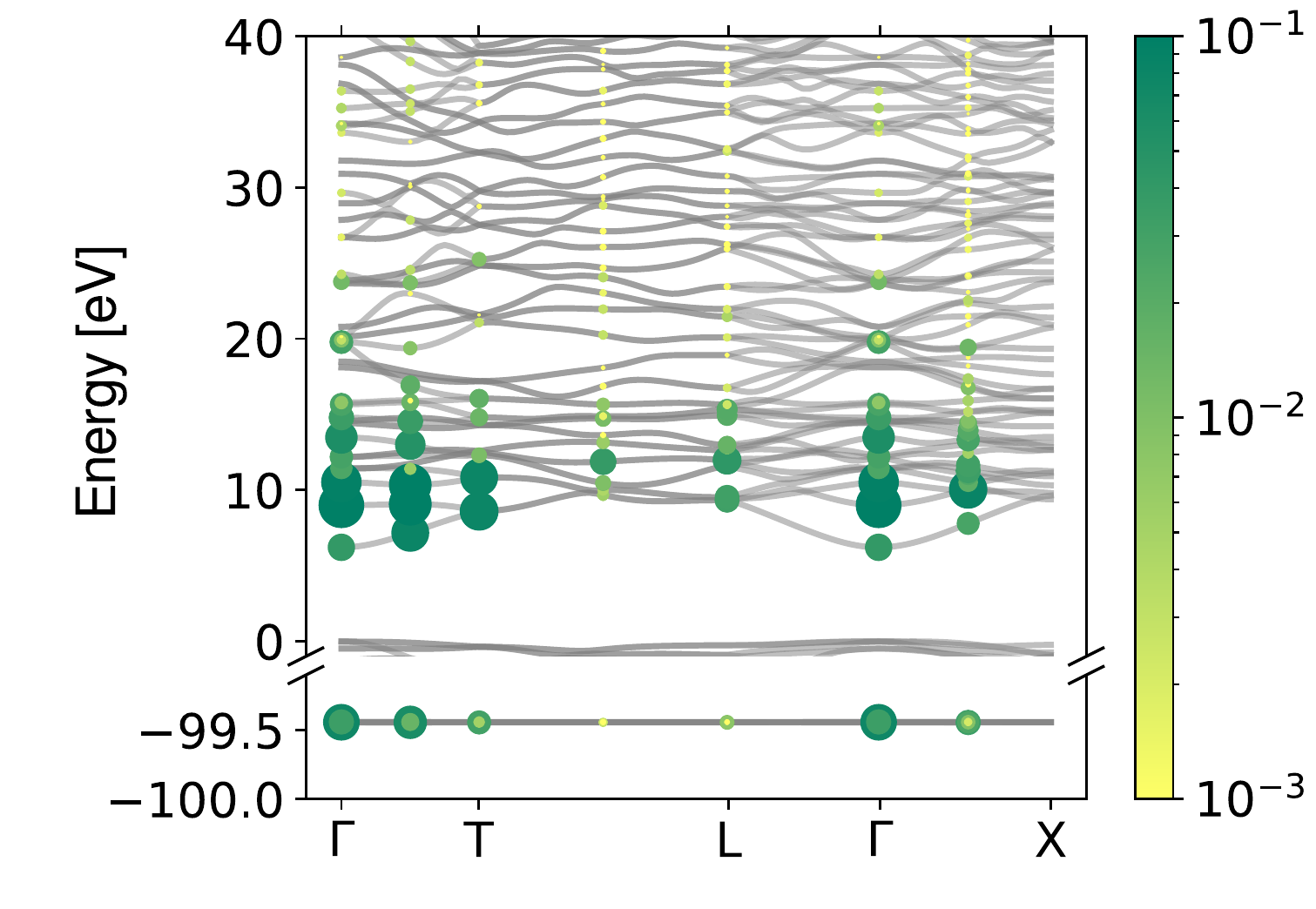}
	\includegraphics[width=0.68\columnwidth]{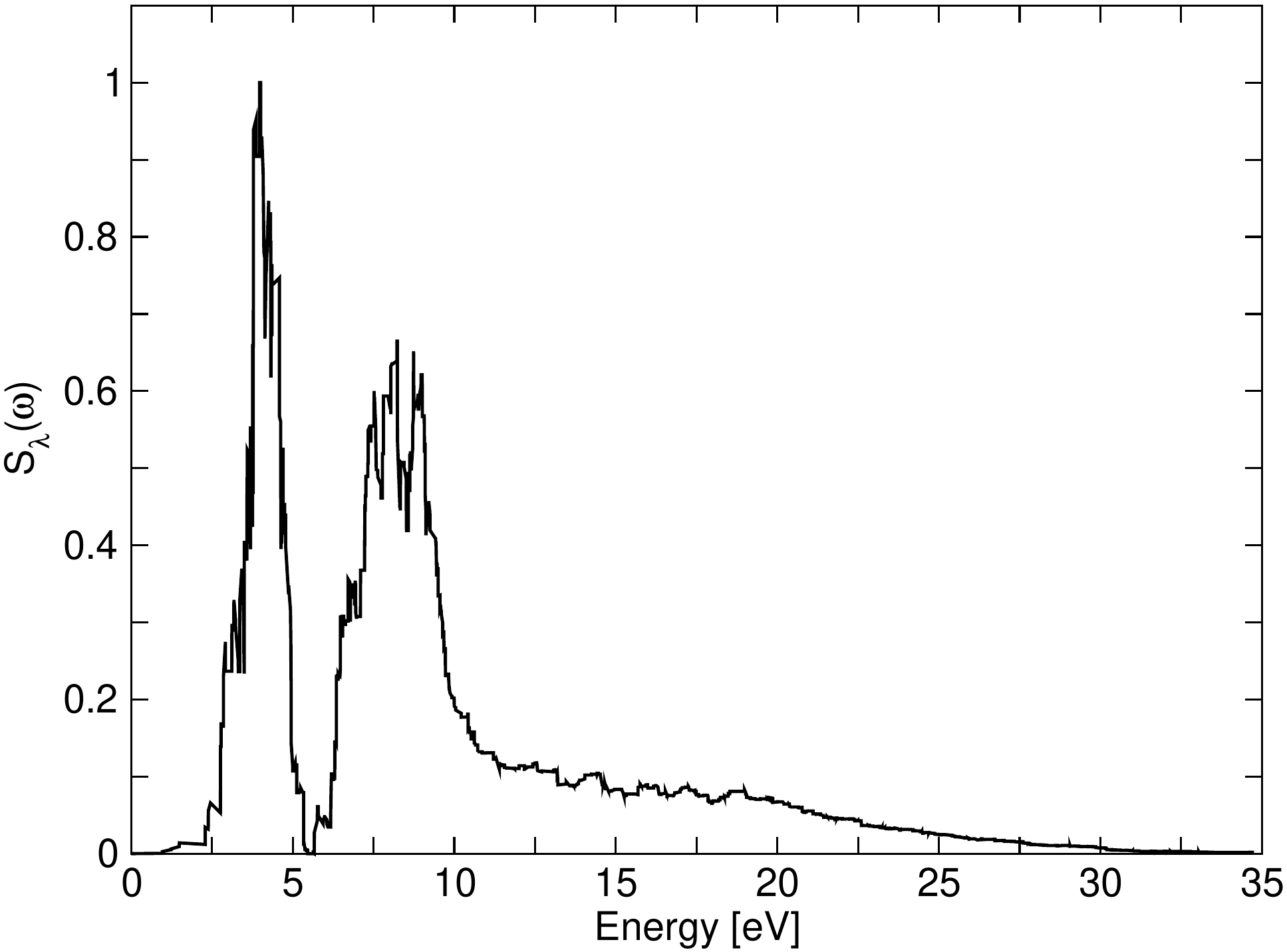}
	\caption{\label{fig:dark-l1} Contributions of independent transitions to the dipole strength of the lowest energy dark  exciton  in the XANES spectrum at the L$_1$ edge. (Top panel) The size of the circle is proportional to $\vert \tilde\rho_{vck} A_{\lambda}\vert$. (Bottom panel) Corresponding cumulative sum $S_\lambda(\w)$. The zero of the energy axis has been set to the smallest independent-particle transition energy and the intensity normalised to the largest value.}
\end{figure}

The lowest-energy dark excitons, both in the optical spectrum and the L$_{2,3}$ edge, have a cumulative sum $S_\lambda(\w)$ that is always close to zero. It means that all the independent-particle oscillator strengths $\tilde\rho_{vc\bfk}$ are always small, indicating dipole forbidden transitions.
The situation is instead different for the lowest dark exciton at the  L$_{1}$ edge.
In this case, some transitions to the lowest conduction bands have a weak but not zero contribution $\vert \tilde\rho_{vck} A_{\lambda}\vert$ to the spectrum, as shown by their representation on the LDA  band structure in the top panel of Fig.~\ref{fig:dark-l1}.
The corresponding cumulative sum $S_\lambda(\w)$, bottom panel of  Fig.~\ref{fig:dark-l1}, is indeed not always zero: it has even two distinct peaks, before progressively decreasing to zero, giving rise to a dark exciton. This suggests the occurrence of destructive interference of contributions $\tilde\rho_{vck} A_{\lambda}$  of different sign, involving transitions over a large range of energy. Moreover, it also shows that including not enough conduction bands in the BSE hamiltonian~\eqref{eq:BSE} would produce a weak excitonic peak in the spectrum. It is another indication that an independent-particle picture is here inadequate, whereas the strong electron-hole interaction manifest itself as the (positive or negative) interference of many electron-hole pairs.

\begin{figure*}
	\includegraphics[width=0.62\columnwidth]{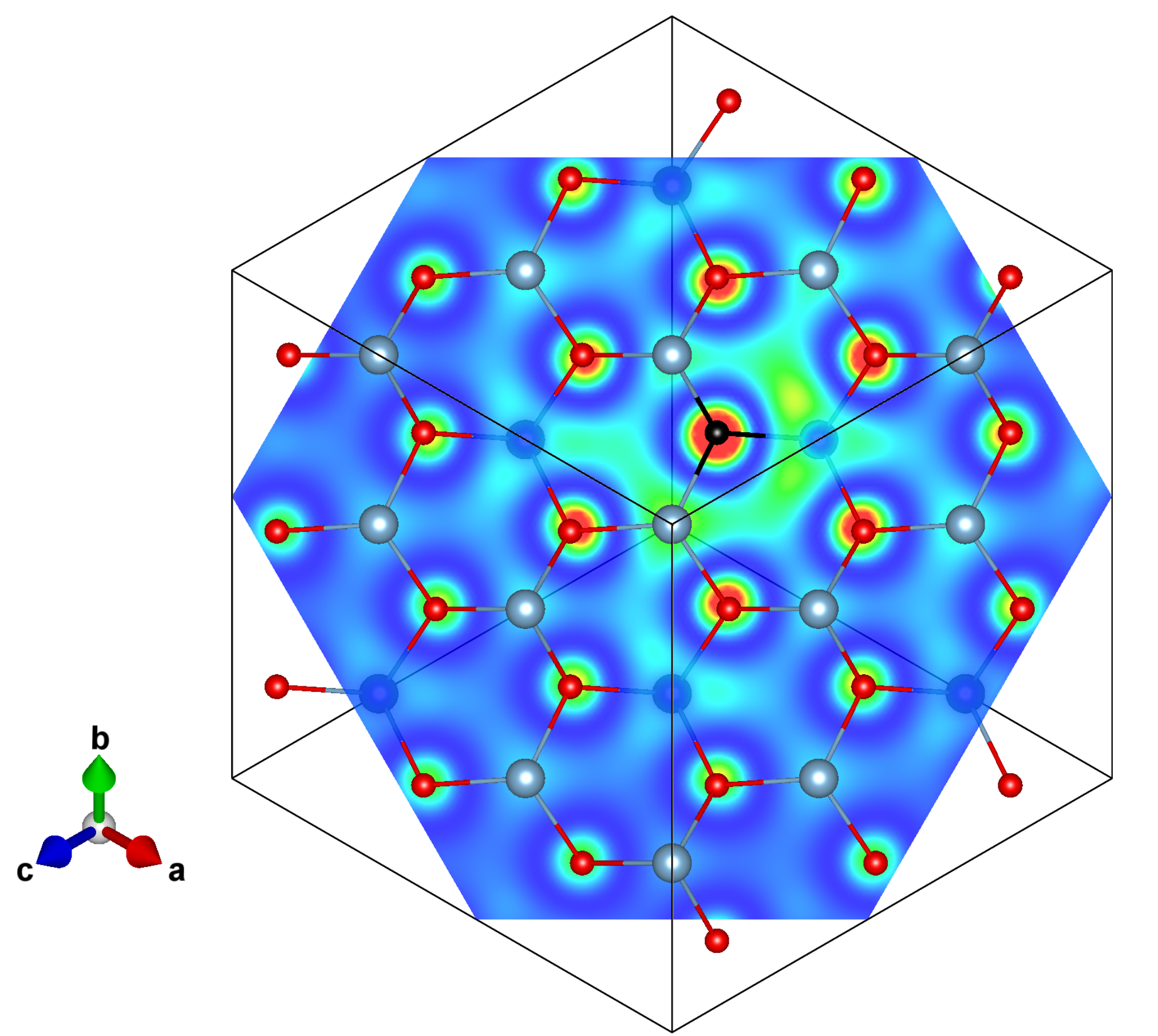}
	\includegraphics[width=0.62\columnwidth]{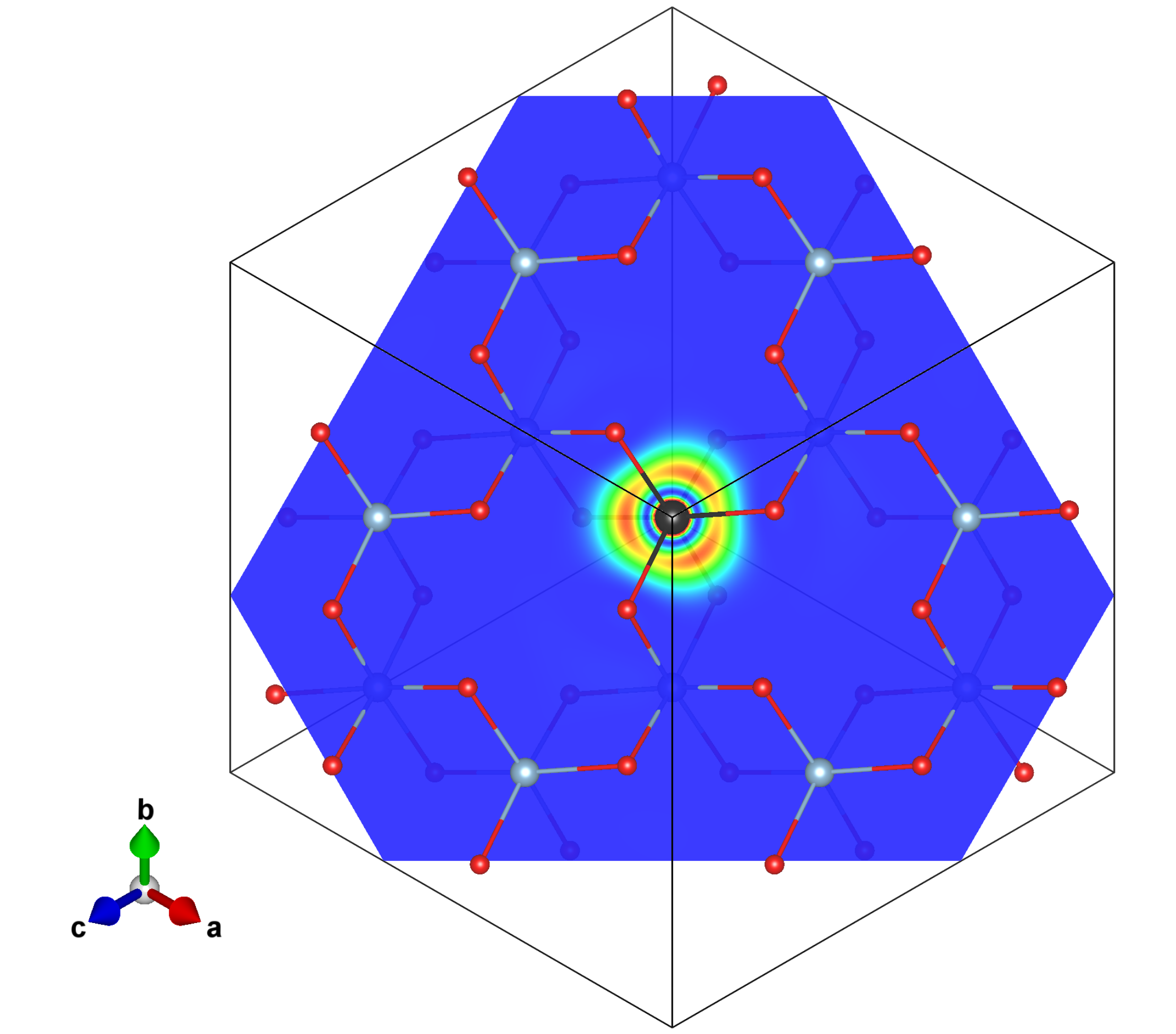}
	\includegraphics[width=0.62\columnwidth]{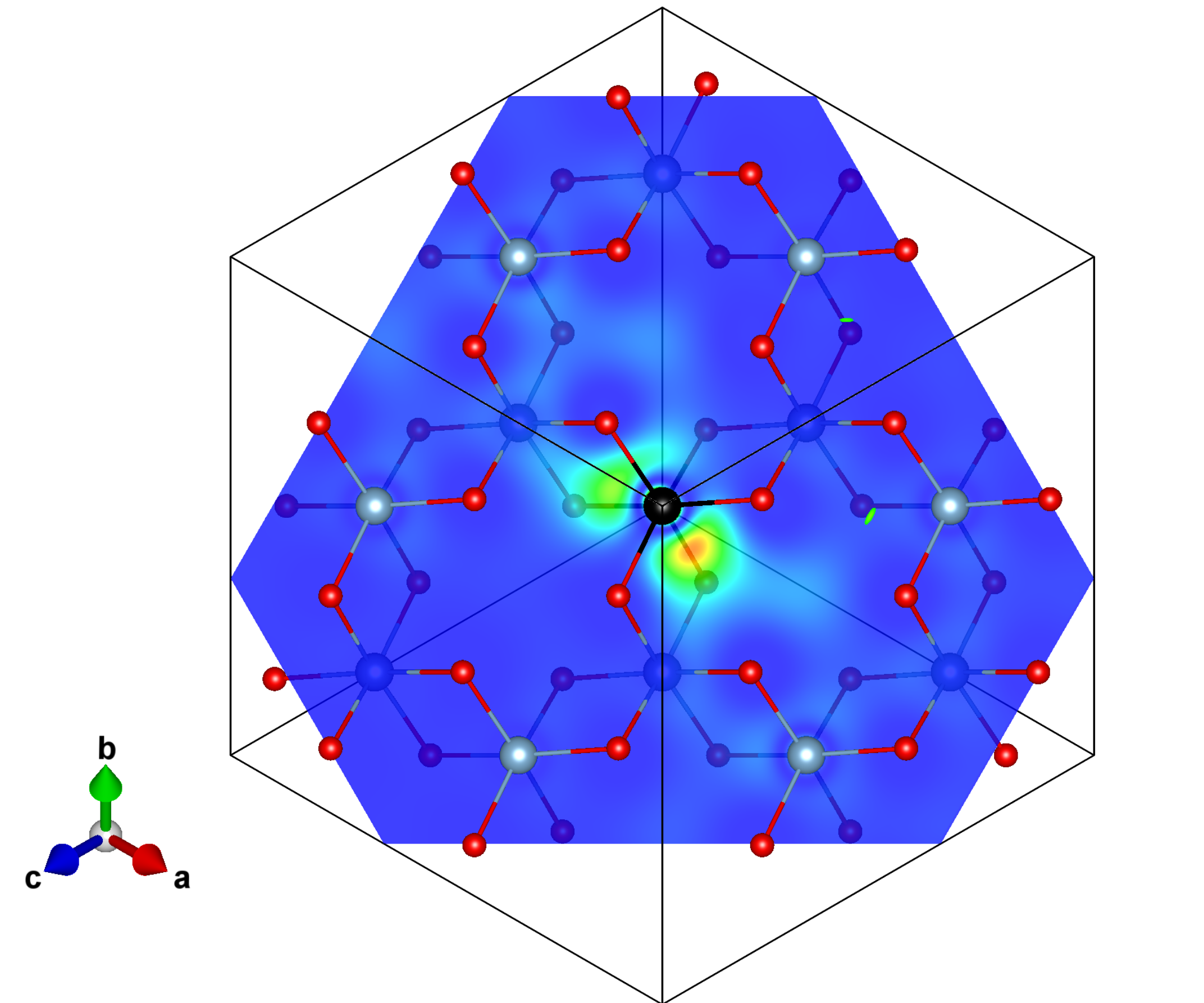}
	\includegraphics[width=0.1\columnwidth]{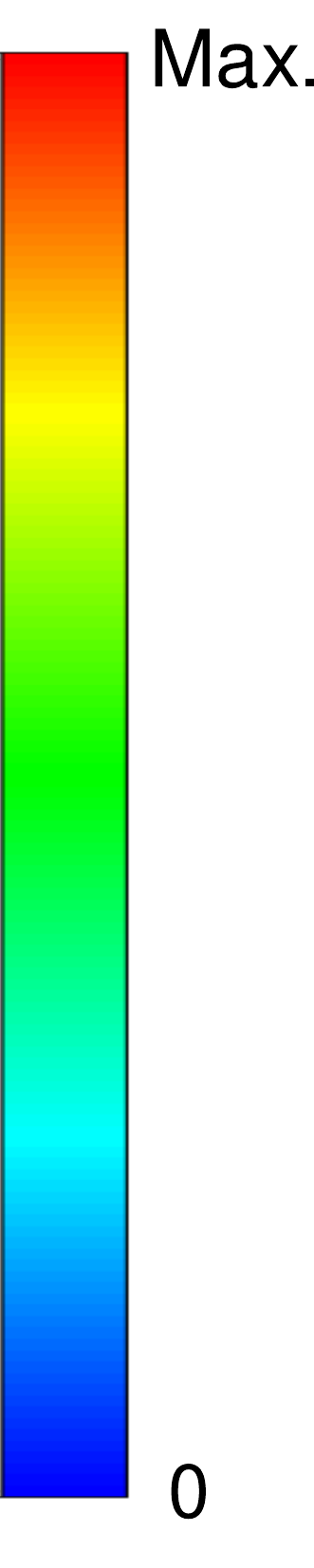}
	\caption{\label{fig:wf} Exciton correlation function $\Psi_\lambda(\bfr_h,\bfr_e)$ for the lowest bright exciton in the (left) optical and (middle) L$_{2,3}$ edge spectra and (right) at the prepeak at the L$_1$ edge. The position of the hole  $\bfr_h$  is fixed at $\bfr_h^0$ (see black ball). The color plots show the corresponding electron density distribution  $|\Psi_\lambda(\bfr_h^0,\bfr_e)|^2$ in the $xy$ plane perpendicular to the $z$ axis contaning the hole. 
    In order to avoid nodes of the orbitals, the hole position has been slightly displaced from an oxygen atom for the optical exciton, and from an aluminum atom for the L$_{23}$ and $L_1$ edge by 0.05~$a_R$ along the $c$ direction. The intensity goes from 0 to the maximum value of the square of the excitonic wavefunctions. }
\end{figure*}

Fig.~\ref{fig:wf} displays the electron density distribution $|\Psi_\lambda(\bfr_h^0,\bfr_e)|^2$ for a fixed position of the hole $\bfr_h^0$ for the wavefunction of the lowest bright excitons in the spectra. In the color plots, we consider a cut of the three-dimensional distribution in the $xy$ plane, perpendicular to the $z$ axis, containing the hole. In all cases, the hole position (represented by the black ball in Fig.~\ref{fig:wf}) has been chosen slightly away from the atoms along the rhombohedral $c$ direction, in order to avoid the nodes of the orbitals. This is the reason why the electron distribution is not symmetrical around the hole.
For an uncorrelated electron-hole pair, the electron density would be delocalised all over the crystal, corresponding to a Bloch wavefunction.
The effect of the electron-hole correlation is instead to localise the electron density around the hole.

For the optical spectrum (left panel), the hole has been placed near an O atom, consistently with the main character of the valence band (see Sec.~\ref{sec:bands}).
Here we discover that the electron charge is also surprisingly located at the O atoms, and quite delocalised in the $xy$ plane. 
This picture is indeed in contrast with the naive expectation of a charge transfer O $\to$ Al nature of the exciton, which is based on the largely ionic character of the electronic properties of {\alo}. 
However, the strong Al-O hybridisation of the bottom conduction bands makes it possible for the exciton to localise entirely on the O atoms.
The nature of the exciton in {\alo} therefore turns out to be  similar to what found\cite{Rohlfing1998,Gatti2013} in other ionic materials like LiF, where, analogously, for a hole fixed at a F atom,  the electron charge is located mainly on F atoms as well.

Finally, the other two panels of Fig.~\ref{fig:wf} show the wavefunctions of the first bright exciton at the L$_{2,3}$ edge and 
in the prepeak of the L$_1$ edge.
In both cases, the hole is localised close to an Al atom.
For the L$_{2,3}$ edge, the resulting electron charge has the shape of an Al $3s$ orbital. For the L$_1$ edge, the cut shows  a distorted Al $3p$ orbital pointing to the next neighbor O atom. In these cases, the electron charge is entirely localised around the same Al site, displaying the atomic nature and the Frenkel character of the core excitons.

\section{\label{sec:conclusions} Conclusions}

In summary, we have presented a  pseudopotential computational scheme that permits one to  evaluate optical and XANES spectra on the same footing, using the same basis set for valence and
shallow-core electrons. We have validated the approach by comparison with full potential all-electron 
calculations, at three different levels of theory, independent-particle approximation, RPA and full
excitonic calculation, within the BSE formalism. 
We have applied this approach to study the optical and semicore excitations of corundum \alo, a
promising material for its optical and structural properties. 
Both  optical and XANES spectra present strong many-body effects that require the highest 
level of theory for an accurate description. In particular, the intense exciton feature in the XANES spectrum at the L$_1$ edge could be measurable. To the best of our knowledge, this edge has never been measured. Our prediction therefore opens up new avenues for the measurement of unoccupied Al 3p states also by means of soft x-ray spectroscopy.

The  anisotropy in the optical regime reveals a different order of excitons in the $z$ and perpendicular $xy$ directions: the first exciton is bright along $z$, followed by dark excitons, while it is the opposite in the perpendicular $xy$ direction. This indicates a fully linear dichroic character of the spectra. The same splitting occurs also for the L$_{2,3}$ edge.
Moreover, we find a pre-edge feature at the Al L$_1$ edge that has a purely electronic nature. This seems to contrast with the origin of the pre-edge observed in XANES spectra at the Al K edge attributed to the coupling with phonons. Our work therefore calls for a careful examination of the role of phonons at the Al L$_1$ edge as well.
Finally, the main exciton peak in the optical absorption spectrum of {\alo} does not have a O$\rightarrow$Al charge transfer character, as one would expect from the largely ionic nature of the electronic structure of \alo. Instead, as a consequence of the electron-hole correlation, the exciton is localised on O atoms only.  

This work opens the way for an accurate description of other shallow-core spectroscopies within a pseudopotential scheme, such as electron energy loss near-edge structures (ELNES) or  x-ray Raman scattering (XRS). Finally, treating shallow-core, valence, and conduction states on the same footing can be particularly useful to describe resonant inelastic x-ray scattering (RIXS).   

\begin{acknowledgments}
We acknowledge valuable discussions with Christian Vorwerk.
We thank the French Agence Nationale de la Recherche (ANR) for financial support (Grant Agreements No.  ANR-19-CE30-0011). Computational time was granted by GENCI (Project  No.  544).   
\end{acknowledgments}

\bibliography{bib}

\begin{thebibliography}{145}%
\makeatletter
\providecommand \@ifxundefined [1]{%
 \@ifx{#1\undefined}
}%
\providecommand \@ifnum [1]{%
 \ifnum #1\expandafter \@firstoftwo
 \else \expandafter \@secondoftwo
 \fi
}%
\providecommand \@ifx [1]{%
 \ifx #1\expandafter \@firstoftwo
 \else \expandafter \@secondoftwo
 \fi
}%
\providecommand \natexlab [1]{#1}%
\providecommand \enquote  [1]{``#1''}%
\providecommand \bibnamefont  [1]{#1}%
\providecommand \bibfnamefont [1]{#1}%
\providecommand \citenamefont [1]{#1}%
\providecommand \href@noop [0]{\@secondoftwo}%
\providecommand \href [0]{\begingroup \@sanitize@url \@href}%
\providecommand \@href[1]{\@@startlink{#1}\@@href}%
\providecommand \@@href[1]{\endgroup#1\@@endlink}%
\providecommand \@sanitize@url [0]{\catcode `\\12\catcode `\$12\catcode
  `\&12\catcode `\#12\catcode `\^12\catcode `\_12\catcode `\%12\relax}%
\providecommand \@@startlink[1]{}%
\providecommand \@@endlink[0]{}%
\providecommand \url  [0]{\begingroup\@sanitize@url \@url }%
\providecommand \@url [1]{\endgroup\@href {#1}{\urlprefix }}%
\providecommand \urlprefix  [0]{URL }%
\providecommand \Eprint [0]{\href }%
\providecommand \doibase [0]{http://dx.doi.org/}%
\providecommand \selectlanguage [0]{\@gobble}%
\providecommand \bibinfo  [0]{\@secondoftwo}%
\providecommand \bibfield  [0]{\@secondoftwo}%
\providecommand \translation [1]{[#1]}%
\providecommand \BibitemOpen [0]{}%
\providecommand \bibitemStop [0]{}%
\providecommand \bibitemNoStop [0]{.\EOS\space}%
\providecommand \EOS [0]{\spacefactor3000\relax}%
\providecommand \BibitemShut  [1]{\csname bibitem#1\endcsname}%
\let\auto@bib@innerbib\@empty
\bibitem [{\citenamefont {van Bokhoven}\ and\ \citenamefont
  {Lamberti}(2016)}]{vanBokhoven2016}%
  \BibitemOpen
  \bibinfo {editor} {\bibfnamefont {J.}~\bibnamefont {van Bokhoven}}\ and\
  \bibinfo {editor} {\bibfnamefont {C.}~\bibnamefont {Lamberti}},\ eds.,\
  \href@noop {} {\emph {\bibinfo {title} {X-Ray Absorption and X-Ray Emission
  Spectroscopy: Theory and Applications}}}\ (\bibinfo  {publisher} {Wiley},\
  \bibinfo {year} {2016})\BibitemShut {NoStop}%
\bibitem [{\citenamefont {{de Groot}}\ \emph {et~al.}(2021)\citenamefont {{de
  Groot}}, \citenamefont {Elnaggar}, \citenamefont {Frati}, \citenamefont {pan
  Wang}, \citenamefont {Delgado-Jaime}, \citenamefont {{van Veenendaal}},
  \citenamefont {Fernandez-Rodriguez}, \citenamefont {Haverkort}, \citenamefont
  {Green}, \citenamefont {{van der Laan}}, \citenamefont {Kvashnin},
  \citenamefont {Hariki}, \citenamefont {Ikeno}, \citenamefont
  {Ramanantoanina}, \citenamefont {Daul}, \citenamefont {Delley}, \citenamefont
  {Odelius}, \citenamefont {Lundberg}, \citenamefont {Kuhn}, \citenamefont
  {Bokarev}, \citenamefont {Shirley}, \citenamefont {Vinson}, \citenamefont
  {Gilmore}, \citenamefont {Stener}, \citenamefont {Fronzoni}, \citenamefont
  {Decleva}, \citenamefont {Kruger}, \citenamefont {Retegan}, \citenamefont
  {Joly}, \citenamefont {Vorwerk}, \citenamefont {Draxl}, \citenamefont
  {Rehr},\ and\ \citenamefont {Tanaka}}]{deGroot2021}%
  \BibitemOpen
  \bibfield  {author} {\bibinfo {author} {\bibfnamefont {F.~M.}\ \bibnamefont
  {{de Groot}}}, \bibinfo {author} {\bibfnamefont {H.}~\bibnamefont
  {Elnaggar}}, \bibinfo {author} {\bibfnamefont {F.}~\bibnamefont {Frati}},
  \bibinfo {author} {\bibfnamefont {R.}~\bibnamefont {pan Wang}}, \bibinfo
  {author} {\bibfnamefont {M.~U.}\ \bibnamefont {Delgado-Jaime}}, \bibinfo
  {author} {\bibfnamefont {M.}~\bibnamefont {{van Veenendaal}}}, \bibinfo
  {author} {\bibfnamefont {J.}~\bibnamefont {Fernandez-Rodriguez}}, \bibinfo
  {author} {\bibfnamefont {M.~W.}\ \bibnamefont {Haverkort}}, \bibinfo {author}
  {\bibfnamefont {R.~J.}\ \bibnamefont {Green}}, \bibinfo {author}
  {\bibfnamefont {G.}~\bibnamefont {{van der Laan}}}, \bibinfo {author}
  {\bibfnamefont {Y.}~\bibnamefont {Kvashnin}}, \bibinfo {author}
  {\bibfnamefont {A.}~\bibnamefont {Hariki}}, \bibinfo {author} {\bibfnamefont
  {H.}~\bibnamefont {Ikeno}}, \bibinfo {author} {\bibfnamefont
  {H.}~\bibnamefont {Ramanantoanina}}, \bibinfo {author} {\bibfnamefont
  {C.}~\bibnamefont {Daul}}, \bibinfo {author} {\bibfnamefont {B.}~\bibnamefont
  {Delley}}, \bibinfo {author} {\bibfnamefont {M.}~\bibnamefont {Odelius}},
  \bibinfo {author} {\bibfnamefont {M.}~\bibnamefont {Lundberg}}, \bibinfo
  {author} {\bibfnamefont {O.}~\bibnamefont {Kuhn}}, \bibinfo {author}
  {\bibfnamefont {S.~I.}\ \bibnamefont {Bokarev}}, \bibinfo {author}
  {\bibfnamefont {E.}~\bibnamefont {Shirley}}, \bibinfo {author} {\bibfnamefont
  {J.}~\bibnamefont {Vinson}}, \bibinfo {author} {\bibfnamefont
  {K.}~\bibnamefont {Gilmore}}, \bibinfo {author} {\bibfnamefont
  {M.}~\bibnamefont {Stener}}, \bibinfo {author} {\bibfnamefont
  {G.}~\bibnamefont {Fronzoni}}, \bibinfo {author} {\bibfnamefont
  {P.}~\bibnamefont {Decleva}}, \bibinfo {author} {\bibfnamefont
  {P.}~\bibnamefont {Kruger}}, \bibinfo {author} {\bibfnamefont
  {M.}~\bibnamefont {Retegan}}, \bibinfo {author} {\bibfnamefont
  {Y.}~\bibnamefont {Joly}}, \bibinfo {author} {\bibfnamefont {C.}~\bibnamefont
  {Vorwerk}}, \bibinfo {author} {\bibfnamefont {C.}~\bibnamefont {Draxl}},
  \bibinfo {author} {\bibfnamefont {J.}~\bibnamefont {Rehr}}, \ and\ \bibinfo
  {author} {\bibfnamefont {A.}~\bibnamefont {Tanaka}},\ }\href {\doibase
  https://doi.org/10.1016/j.elspec.2021.147061} {\bibfield  {journal} {\bibinfo
   {journal} {Journal of Electron Spectroscopy and Related Phenomena}\ }\textbf
  {\bibinfo {volume} {249}},\ \bibinfo {pages} {147061} (\bibinfo {year}
  {2021})}\BibitemShut {NoStop}%
\bibitem [{\citenamefont {Fujikawa}(1983)}]{Fujikawa1983}%
  \BibitemOpen
  \bibfield  {author} {\bibinfo {author} {\bibfnamefont {T.}~\bibnamefont
  {Fujikawa}},\ }\href {\doibase 10.1143/JPSJ.52.4001} {\bibfield  {journal}
  {\bibinfo  {journal} {Journal of the Physical Society of Japan}\ }\textbf
  {\bibinfo {volume} {52}},\ \bibinfo {pages} {4001} (\bibinfo {year}
  {1983})}\BibitemShut {NoStop}%
\bibitem [{\citenamefont {Tyson}\ \emph {et~al.}(1992)\citenamefont {Tyson},
  \citenamefont {Hodgson}, \citenamefont {Natoli},\ and\ \citenamefont
  {Benfatto}}]{Tyson1992}%
  \BibitemOpen
  \bibfield  {author} {\bibinfo {author} {\bibfnamefont {T.~A.}\ \bibnamefont
  {Tyson}}, \bibinfo {author} {\bibfnamefont {K.~O.}\ \bibnamefont {Hodgson}},
  \bibinfo {author} {\bibfnamefont {C.~R.}\ \bibnamefont {Natoli}}, \ and\
  \bibinfo {author} {\bibfnamefont {M.}~\bibnamefont {Benfatto}},\ }\href
  {\doibase 10.1103/PhysRevB.46.5997} {\bibfield  {journal} {\bibinfo
  {journal} {Phys. Rev. B}\ }\textbf {\bibinfo {volume} {46}},\ \bibinfo
  {pages} {5997} (\bibinfo {year} {1992})}\BibitemShut {NoStop}%
\bibitem [{\citenamefont {Ahlers}\ \emph {et~al.}(1998)\citenamefont {Ahlers},
  \citenamefont {Schütz}, \citenamefont {Popescu},\ and\ \citenamefont
  {Ebert}}]{Ahlers1998}%
  \BibitemOpen
  \bibfield  {author} {\bibinfo {author} {\bibfnamefont {D.}~\bibnamefont
  {Ahlers}}, \bibinfo {author} {\bibfnamefont {G.}~\bibnamefont {Schütz}},
  \bibinfo {author} {\bibfnamefont {V.}~\bibnamefont {Popescu}}, \ and\
  \bibinfo {author} {\bibfnamefont {H.}~\bibnamefont {Ebert}},\ }\href
  {\doibase 10.1063/1.367853} {\bibfield  {journal} {\bibinfo  {journal}
  {Journal of Applied Physics}\ }\textbf {\bibinfo {volume} {83}},\ \bibinfo
  {pages} {7082} (\bibinfo {year} {1998})}\BibitemShut {NoStop}%
\bibitem [{\citenamefont {Rehr}\ and\ \citenamefont {Albers}(2000)}]{Rehr2000}%
  \BibitemOpen
  \bibfield  {author} {\bibinfo {author} {\bibfnamefont {J.~J.}\ \bibnamefont
  {Rehr}}\ and\ \bibinfo {author} {\bibfnamefont {R.~C.}\ \bibnamefont
  {Albers}},\ }\href {\doibase 10.1103/RevModPhys.72.621} {\bibfield  {journal}
  {\bibinfo  {journal} {Rev. Mod. Phys.}\ }\textbf {\bibinfo {volume} {72}},\
  \bibinfo {pages} {621} (\bibinfo {year} {2000})}\BibitemShut {NoStop}%
\bibitem [{\citenamefont {Rehr}\ \emph {et~al.}(2009)\citenamefont {Rehr},
  \citenamefont {Kas}, \citenamefont {Prange}, \citenamefont {Sorini},
  \citenamefont {Takimoto},\ and\ \citenamefont {Vila}}]{Rehr2009}%
  \BibitemOpen
  \bibfield  {author} {\bibinfo {author} {\bibfnamefont {J.~J.}\ \bibnamefont
  {Rehr}}, \bibinfo {author} {\bibfnamefont {J.~J.}\ \bibnamefont {Kas}},
  \bibinfo {author} {\bibfnamefont {M.~P.}\ \bibnamefont {Prange}}, \bibinfo
  {author} {\bibfnamefont {A.~P.}\ \bibnamefont {Sorini}}, \bibinfo {author}
  {\bibfnamefont {Y.}~\bibnamefont {Takimoto}}, \ and\ \bibinfo {author}
  {\bibfnamefont {F.}~\bibnamefont {Vila}},\ }\href {\doibase
  https://doi.org/10.1016/j.crhy.2008.08.004} {\bibfield  {journal} {\bibinfo
  {journal} {Comptes Rendus Physique}\ }\textbf {\bibinfo {volume} {10}},\
  \bibinfo {pages} {548} (\bibinfo {year} {2009})}\BibitemShut {NoStop}%
\bibitem [{\citenamefont {Rehr}\ \emph {et~al.}(2010)\citenamefont {Rehr},
  \citenamefont {Kas}, \citenamefont {Vila}, \citenamefont {Prange},\ and\
  \citenamefont {Jorissen}}]{Rehr2010}%
  \BibitemOpen
  \bibfield  {author} {\bibinfo {author} {\bibfnamefont {J.~J.}\ \bibnamefont
  {Rehr}}, \bibinfo {author} {\bibfnamefont {J.~J.}\ \bibnamefont {Kas}},
  \bibinfo {author} {\bibfnamefont {F.~D.}\ \bibnamefont {Vila}}, \bibinfo
  {author} {\bibfnamefont {M.~P.}\ \bibnamefont {Prange}}, \ and\ \bibinfo
  {author} {\bibfnamefont {K.}~\bibnamefont {Jorissen}},\ }\href {\doibase
  10.1039/B926434E} {\bibfield  {journal} {\bibinfo  {journal} {Phys. Chem.
  Chem. Phys.}\ }\textbf {\bibinfo {volume} {12}},\ \bibinfo {pages} {5503}
  (\bibinfo {year} {2010})}\BibitemShut {NoStop}%
\bibitem [{\citenamefont {de~Groot}(2005)}]{deGroot2005}%
  \BibitemOpen
  \bibfield  {author} {\bibinfo {author} {\bibfnamefont {F.}~\bibnamefont
  {de~Groot}},\ }\href {\doibase https://doi.org/10.1016/j.ccr.2004.03.018}
  {\bibfield  {journal} {\bibinfo  {journal} {Coordination Chemistry Reviews}\
  }\textbf {\bibinfo {volume} {249}},\ \bibinfo {pages} {31} (\bibinfo {year}
  {2005})},\ \bibinfo {note} {synchrotron Radiation in Inorganic and
  Bioinorganic Chemistry}\BibitemShut {NoStop}%
\bibitem [{\citenamefont {De~Groot}\ and\ \citenamefont
  {Kotani}(2008)}]{DeGroot2008}%
  \BibitemOpen
  \bibfield  {author} {\bibinfo {author} {\bibfnamefont {F.}~\bibnamefont
  {De~Groot}}\ and\ \bibinfo {author} {\bibfnamefont {A.}~\bibnamefont
  {Kotani}},\ }\href@noop {} {\emph {\bibinfo {title} {Core level spectroscopy
  of solids}}}\ (\bibinfo  {publisher} {CRC press},\ \bibinfo {year}
  {2008})\BibitemShut {NoStop}%
\bibitem [{\citenamefont {Haverkort}\ \emph {et~al.}(2012)\citenamefont
  {Haverkort}, \citenamefont {Zwierzycki},\ and\ \citenamefont
  {Andersen}}]{Haverkort2012}%
  \BibitemOpen
  \bibfield  {author} {\bibinfo {author} {\bibfnamefont {M.~W.}\ \bibnamefont
  {Haverkort}}, \bibinfo {author} {\bibfnamefont {M.}~\bibnamefont
  {Zwierzycki}}, \ and\ \bibinfo {author} {\bibfnamefont {O.~K.}\ \bibnamefont
  {Andersen}},\ }\href {\doibase 10.1103/PhysRevB.85.165113} {\bibfield
  {journal} {\bibinfo  {journal} {Phys. Rev. B}\ }\textbf {\bibinfo {volume}
  {85}},\ \bibinfo {pages} {165113} (\bibinfo {year} {2012})}\BibitemShut
  {NoStop}%
\bibitem [{\citenamefont {Mo}\ and\ \citenamefont {Ching}(2000)}]{Mo_2000}%
  \BibitemOpen
  \bibfield  {author} {\bibinfo {author} {\bibfnamefont {S.-D.}\ \bibnamefont
  {Mo}}\ and\ \bibinfo {author} {\bibfnamefont {W.~Y.}\ \bibnamefont {Ching}},\
  }\href {\doibase 10.1103/PhysRevB.62.7901} {\bibfield  {journal} {\bibinfo
  {journal} {Phys. Rev. B}\ }\textbf {\bibinfo {volume} {62}},\ \bibinfo
  {pages} {7901} (\bibinfo {year} {2000})}\BibitemShut {NoStop}%
\bibitem [{\citenamefont {Gougoussis}\ \emph {et~al.}(2009)\citenamefont
  {Gougoussis}, \citenamefont {Calandra}, \citenamefont {Seitsonen},\ and\
  \citenamefont {Mauri}}]{Gougoussis2009}%
  \BibitemOpen
  \bibfield  {author} {\bibinfo {author} {\bibfnamefont {C.}~\bibnamefont
  {Gougoussis}}, \bibinfo {author} {\bibfnamefont {M.}~\bibnamefont
  {Calandra}}, \bibinfo {author} {\bibfnamefont {A.~P.}\ \bibnamefont
  {Seitsonen}}, \ and\ \bibinfo {author} {\bibfnamefont {F.}~\bibnamefont
  {Mauri}},\ }\href {\doibase 10.1103/PhysRevB.80.075102} {\bibfield  {journal}
  {\bibinfo  {journal} {Phys. Rev. B}\ }\textbf {\bibinfo {volume} {80}},\
  \bibinfo {pages} {075102} (\bibinfo {year} {2009})}\BibitemShut {NoStop}%
\bibitem [{\citenamefont {Taillefumier}\ \emph {et~al.}(2002)\citenamefont
  {Taillefumier}, \citenamefont {Cabaret}, \citenamefont {Flank},\ and\
  \citenamefont {Mauri}}]{Taillefumier2002}%
  \BibitemOpen
  \bibfield  {author} {\bibinfo {author} {\bibfnamefont {M.}~\bibnamefont
  {Taillefumier}}, \bibinfo {author} {\bibfnamefont {D.}~\bibnamefont
  {Cabaret}}, \bibinfo {author} {\bibfnamefont {A.-M.}\ \bibnamefont {Flank}},
  \ and\ \bibinfo {author} {\bibfnamefont {F.}~\bibnamefont {Mauri}},\ }\href
  {\doibase 10.1103/PhysRevB.66.195107} {\bibfield  {journal} {\bibinfo
  {journal} {Phys. Rev. B}\ }\textbf {\bibinfo {volume} {66}},\ \bibinfo
  {pages} {195107} (\bibinfo {year} {2002})}\BibitemShut {NoStop}%
\bibitem [{\citenamefont {Bun\ifmmode~\u{a}\else \u{a}\fi{}u}\ and\
  \citenamefont {Calandra}(2013)}]{Bunau2013}%
  \BibitemOpen
  \bibfield  {author} {\bibinfo {author} {\bibfnamefont {O.}~\bibnamefont
  {Bun\ifmmode~\u{a}\else \u{a}\fi{}u}}\ and\ \bibinfo {author} {\bibfnamefont
  {M.}~\bibnamefont {Calandra}},\ }\href {\doibase 10.1103/PhysRevB.87.205105}
  {\bibfield  {journal} {\bibinfo  {journal} {Phys. Rev. B}\ }\textbf {\bibinfo
  {volume} {87}},\ \bibinfo {pages} {205105} (\bibinfo {year}
  {2013})}\BibitemShut {NoStop}%
\bibitem [{\citenamefont {Mazevet}\ \emph {et~al.}(2010)\citenamefont
  {Mazevet}, \citenamefont {Torrent}, \citenamefont {Recoules},\ and\
  \citenamefont {Jollet}}]{Mazevet2010}%
  \BibitemOpen
  \bibfield  {author} {\bibinfo {author} {\bibfnamefont {S.}~\bibnamefont
  {Mazevet}}, \bibinfo {author} {\bibfnamefont {M.}~\bibnamefont {Torrent}},
  \bibinfo {author} {\bibfnamefont {V.}~\bibnamefont {Recoules}}, \ and\
  \bibinfo {author} {\bibfnamefont {F.}~\bibnamefont {Jollet}},\ }\href
  {\doibase https://doi.org/10.1016/j.hedp.2009.06.004} {\bibfield  {journal}
  {\bibinfo  {journal} {High Energy Density Physics}\ }\textbf {\bibinfo
  {volume} {6}},\ \bibinfo {pages} {84} (\bibinfo {year} {2010})}\BibitemShut
  {NoStop}%
\bibitem [{\citenamefont {Hetényi}\ \emph {et~al.}(2004)\citenamefont
  {Hetényi}, \citenamefont {De~Angelis}, \citenamefont {Giannozzi},\ and\
  \citenamefont {Car}}]{Hetenyi2004}%
  \BibitemOpen
  \bibfield  {author} {\bibinfo {author} {\bibfnamefont {B.}~\bibnamefont
  {Hetényi}}, \bibinfo {author} {\bibfnamefont {F.}~\bibnamefont
  {De~Angelis}}, \bibinfo {author} {\bibfnamefont {P.}~\bibnamefont
  {Giannozzi}}, \ and\ \bibinfo {author} {\bibfnamefont {R.}~\bibnamefont
  {Car}},\ }\href {\doibase 10.1063/1.1703526} {\bibfield  {journal} {\bibinfo
  {journal} {The Journal of Chemical Physics}\ }\textbf {\bibinfo {volume}
  {120}},\ \bibinfo {pages} {8632} (\bibinfo {year} {2004})}\BibitemShut
  {NoStop}%
\bibitem [{\citenamefont {Prendergast}\ and\ \citenamefont
  {Galli}(2006)}]{Prendergast2006}%
  \BibitemOpen
  \bibfield  {author} {\bibinfo {author} {\bibfnamefont {D.}~\bibnamefont
  {Prendergast}}\ and\ \bibinfo {author} {\bibfnamefont {G.}~\bibnamefont
  {Galli}},\ }\href {\doibase 10.1103/PhysRevLett.96.215502} {\bibfield
  {journal} {\bibinfo  {journal} {Phys. Rev. Lett.}\ }\textbf {\bibinfo
  {volume} {96}},\ \bibinfo {pages} {215502} (\bibinfo {year}
  {2006})}\BibitemShut {NoStop}%
\bibitem [{\citenamefont {Gao}\ \emph {et~al.}(2009)\citenamefont {Gao},
  \citenamefont {Pickard}, \citenamefont {Perlov},\ and\ \citenamefont
  {Milman}}]{Gao_2009}%
  \BibitemOpen
  \bibfield  {author} {\bibinfo {author} {\bibfnamefont {S.-P.}\ \bibnamefont
  {Gao}}, \bibinfo {author} {\bibfnamefont {C.~J.}\ \bibnamefont {Pickard}},
  \bibinfo {author} {\bibfnamefont {A.}~\bibnamefont {Perlov}}, \ and\ \bibinfo
  {author} {\bibfnamefont {V.}~\bibnamefont {Milman}},\ }\href {\doibase
  10.1088/0953-8984/21/10/104203} {\bibfield  {journal} {\bibinfo  {journal}
  {Journal of Physics: Condensed Matter}\ }\textbf {\bibinfo {volume} {21}},\
  \bibinfo {pages} {104203} (\bibinfo {year} {2009})}\BibitemShut {NoStop}%
\bibitem [{\citenamefont {Prentice}\ \emph {et~al.}(2020)\citenamefont
  {Prentice}, \citenamefont {Aarons}, \citenamefont {Womack}, \citenamefont
  {Allen}, \citenamefont {Andrinopoulos}, \citenamefont {Anton}, \citenamefont
  {Bell}, \citenamefont {Bhandari}, \citenamefont {Bramley}, \citenamefont
  {Charlton}, \citenamefont {Clements}, \citenamefont {Cole}, \citenamefont
  {Constantinescu}, \citenamefont {Corsetti}, \citenamefont {Dubois},
  \citenamefont {Duff}, \citenamefont {Escartín}, \citenamefont {Greco},
  \citenamefont {Hill}, \citenamefont {Lee}, \citenamefont {Linscott},
  \citenamefont {O'Regan}, \citenamefont {Phipps}, \citenamefont {Ratcliff},
  \citenamefont {Serrano}, \citenamefont {Tait}, \citenamefont {Teobaldi},
  \citenamefont {Vitale}, \citenamefont {Yeung}, \citenamefont {Zuehlsdorff},
  \citenamefont {Dziedzic}, \citenamefont {Haynes}, \citenamefont {Hine},
  \citenamefont {Mostofi}, \citenamefont {Payne},\ and\ \citenamefont
  {Skylaris}}]{Prentice2020}%
  \BibitemOpen
  \bibfield  {author} {\bibinfo {author} {\bibfnamefont {J.~C.~A.}\
  \bibnamefont {Prentice}}, \bibinfo {author} {\bibfnamefont {J.}~\bibnamefont
  {Aarons}}, \bibinfo {author} {\bibfnamefont {J.~C.}\ \bibnamefont {Womack}},
  \bibinfo {author} {\bibfnamefont {A.~E.~A.}\ \bibnamefont {Allen}}, \bibinfo
  {author} {\bibfnamefont {L.}~\bibnamefont {Andrinopoulos}}, \bibinfo {author}
  {\bibfnamefont {L.}~\bibnamefont {Anton}}, \bibinfo {author} {\bibfnamefont
  {R.~A.}\ \bibnamefont {Bell}}, \bibinfo {author} {\bibfnamefont
  {A.}~\bibnamefont {Bhandari}}, \bibinfo {author} {\bibfnamefont {G.~A.}\
  \bibnamefont {Bramley}}, \bibinfo {author} {\bibfnamefont {R.~J.}\
  \bibnamefont {Charlton}}, \bibinfo {author} {\bibfnamefont {R.~J.}\
  \bibnamefont {Clements}}, \bibinfo {author} {\bibfnamefont {D.~J.}\
  \bibnamefont {Cole}}, \bibinfo {author} {\bibfnamefont {G.}~\bibnamefont
  {Constantinescu}}, \bibinfo {author} {\bibfnamefont {F.}~\bibnamefont
  {Corsetti}}, \bibinfo {author} {\bibfnamefont {S.~M.-M.}\ \bibnamefont
  {Dubois}}, \bibinfo {author} {\bibfnamefont {K.~K.~B.}\ \bibnamefont {Duff}},
  \bibinfo {author} {\bibfnamefont {J.~M.}\ \bibnamefont {Escartín}}, \bibinfo
  {author} {\bibfnamefont {A.}~\bibnamefont {Greco}}, \bibinfo {author}
  {\bibfnamefont {Q.}~\bibnamefont {Hill}}, \bibinfo {author} {\bibfnamefont
  {L.~P.}\ \bibnamefont {Lee}}, \bibinfo {author} {\bibfnamefont
  {E.}~\bibnamefont {Linscott}}, \bibinfo {author} {\bibfnamefont {D.~D.}\
  \bibnamefont {O'Regan}}, \bibinfo {author} {\bibfnamefont {M.~J.~S.}\
  \bibnamefont {Phipps}}, \bibinfo {author} {\bibfnamefont {L.~E.}\
  \bibnamefont {Ratcliff}}, \bibinfo {author} {\bibfnamefont {A.~R.}\
  \bibnamefont {Serrano}}, \bibinfo {author} {\bibfnamefont {E.~W.}\
  \bibnamefont {Tait}}, \bibinfo {author} {\bibfnamefont {G.}~\bibnamefont
  {Teobaldi}}, \bibinfo {author} {\bibfnamefont {V.}~\bibnamefont {Vitale}},
  \bibinfo {author} {\bibfnamefont {N.}~\bibnamefont {Yeung}}, \bibinfo
  {author} {\bibfnamefont {T.~J.}\ \bibnamefont {Zuehlsdorff}}, \bibinfo
  {author} {\bibfnamefont {J.}~\bibnamefont {Dziedzic}}, \bibinfo {author}
  {\bibfnamefont {P.~D.}\ \bibnamefont {Haynes}}, \bibinfo {author}
  {\bibfnamefont {N.~D.~M.}\ \bibnamefont {Hine}}, \bibinfo {author}
  {\bibfnamefont {A.~A.}\ \bibnamefont {Mostofi}}, \bibinfo {author}
  {\bibfnamefont {M.~C.}\ \bibnamefont {Payne}}, \ and\ \bibinfo {author}
  {\bibfnamefont {C.-K.}\ \bibnamefont {Skylaris}},\ }\href {\doibase
  10.1063/5.0004445} {\bibfield  {journal} {\bibinfo  {journal} {The Journal of
  Chemical Physics}\ }\textbf {\bibinfo {volume} {152}},\ \bibinfo {pages}
  {174111} (\bibinfo {year} {2020})},\ \Eprint
  {http://arxiv.org/abs/https://doi.org/10.1063/5.0004445}
  {https://doi.org/10.1063/5.0004445} \BibitemShut {NoStop}%
\bibitem [{\citenamefont {Hjalmarson}\ \emph {et~al.}(1981)\citenamefont
  {Hjalmarson}, \citenamefont {B\"uttner},\ and\ \citenamefont
  {Dow}}]{Hjalmarson1981}%
  \BibitemOpen
  \bibfield  {author} {\bibinfo {author} {\bibfnamefont {H.~P.}\ \bibnamefont
  {Hjalmarson}}, \bibinfo {author} {\bibfnamefont {H.}~\bibnamefont
  {B\"uttner}}, \ and\ \bibinfo {author} {\bibfnamefont {J.~D.}\ \bibnamefont
  {Dow}},\ }\href {\doibase 10.1103/PhysRevB.24.6010} {\bibfield  {journal}
  {\bibinfo  {journal} {Phys. Rev. B}\ }\textbf {\bibinfo {volume} {24}},\
  \bibinfo {pages} {6010} (\bibinfo {year} {1981})}\BibitemShut {NoStop}%
\bibitem [{\citenamefont {Lie}\ \emph {et~al.}(1999)\citenamefont {Lie},
  \citenamefont {Brydson},\ and\ \citenamefont {Davock}}]{Lie1999}%
  \BibitemOpen
  \bibfield  {author} {\bibinfo {author} {\bibfnamefont {K.}~\bibnamefont
  {Lie}}, \bibinfo {author} {\bibfnamefont {R.}~\bibnamefont {Brydson}}, \ and\
  \bibinfo {author} {\bibfnamefont {H.}~\bibnamefont {Davock}},\ }\href
  {\doibase 10.1103/PhysRevB.59.5361} {\bibfield  {journal} {\bibinfo
  {journal} {Phys. Rev. B}\ }\textbf {\bibinfo {volume} {59}},\ \bibinfo
  {pages} {5361} (\bibinfo {year} {1999})}\BibitemShut {NoStop}%
\bibitem [{\citenamefont {Triguero}\ \emph {et~al.}(1998)\citenamefont
  {Triguero}, \citenamefont {Pettersson},\ and\ \citenamefont
  {\AA{}gren}}]{Triguero1998}%
  \BibitemOpen
  \bibfield  {author} {\bibinfo {author} {\bibfnamefont {L.}~\bibnamefont
  {Triguero}}, \bibinfo {author} {\bibfnamefont {L.~G.~M.}\ \bibnamefont
  {Pettersson}}, \ and\ \bibinfo {author} {\bibfnamefont {H.}~\bibnamefont
  {\AA{}gren}},\ }\href {\doibase 10.1103/PhysRevB.58.8097} {\bibfield
  {journal} {\bibinfo  {journal} {Phys. Rev. B}\ }\textbf {\bibinfo {volume}
  {58}},\ \bibinfo {pages} {8097} (\bibinfo {year} {1998})}\BibitemShut
  {NoStop}%
\bibitem [{\citenamefont {Klein}\ \emph {et~al.}(2021)\citenamefont {Klein},
  \citenamefont {Hall},\ and\ \citenamefont {Maurer}}]{Klein_2021}%
  \BibitemOpen
  \bibfield  {author} {\bibinfo {author} {\bibfnamefont {B.~P.}\ \bibnamefont
  {Klein}}, \bibinfo {author} {\bibfnamefont {S.~J.}\ \bibnamefont {Hall}}, \
  and\ \bibinfo {author} {\bibfnamefont {R.~J.}\ \bibnamefont {Maurer}},\
  }\href {\doibase 10.1088/1361-648X/abdf00} {\bibfield  {journal} {\bibinfo
  {journal} {Journal of Physics: Condensed Matter}\ }\textbf {\bibinfo {volume}
  {33}},\ \bibinfo {pages} {154005} (\bibinfo {year} {2021})}\BibitemShut
  {NoStop}%
\bibitem [{\citenamefont {Rehr}\ \emph {et~al.}(2005)\citenamefont {Rehr},
  \citenamefont {Soininen},\ and\ \citenamefont {Shirley}}]{Rehr2005}%
  \BibitemOpen
  \bibfield  {author} {\bibinfo {author} {\bibfnamefont {J.~J.}\ \bibnamefont
  {Rehr}}, \bibinfo {author} {\bibfnamefont {J.~A.}\ \bibnamefont {Soininen}},
  \ and\ \bibinfo {author} {\bibfnamefont {E.~L.}\ \bibnamefont {Shirley}},\
  }\href {\doibase 10.1238/Physica.Topical.115a00207} {\bibfield  {journal}
  {\bibinfo  {journal} {Physica Scripta}\ }\textbf {\bibinfo {volume} {2005}},\
  \bibinfo {pages} {207} (\bibinfo {year} {2005})}\BibitemShut {NoStop}%
\bibitem [{\citenamefont {Liang}\ \emph {et~al.}(2017)\citenamefont {Liang},
  \citenamefont {Vinson}, \citenamefont {Pemmaraju}, \citenamefont {Drisdell},
  \citenamefont {Shirley},\ and\ \citenamefont {Prendergast}}]{Liang2017}%
  \BibitemOpen
  \bibfield  {author} {\bibinfo {author} {\bibfnamefont {Y.}~\bibnamefont
  {Liang}}, \bibinfo {author} {\bibfnamefont {J.}~\bibnamefont {Vinson}},
  \bibinfo {author} {\bibfnamefont {S.}~\bibnamefont {Pemmaraju}}, \bibinfo
  {author} {\bibfnamefont {W.~S.}\ \bibnamefont {Drisdell}}, \bibinfo {author}
  {\bibfnamefont {E.~L.}\ \bibnamefont {Shirley}}, \ and\ \bibinfo {author}
  {\bibfnamefont {D.}~\bibnamefont {Prendergast}},\ }\href {\doibase
  10.1103/PhysRevLett.118.096402} {\bibfield  {journal} {\bibinfo  {journal}
  {Phys. Rev. Lett.}\ }\textbf {\bibinfo {volume} {118}},\ \bibinfo {pages}
  {096402} (\bibinfo {year} {2017})}\BibitemShut {NoStop}%
\bibitem [{\citenamefont {Onida}\ \emph {et~al.}(2002)\citenamefont {Onida},
  \citenamefont {Reining},\ and\ \citenamefont {Rubio}}]{Onida2002}%
  \BibitemOpen
  \bibfield  {author} {\bibinfo {author} {\bibfnamefont {G.}~\bibnamefont
  {Onida}}, \bibinfo {author} {\bibfnamefont {L.}~\bibnamefont {Reining}}, \
  and\ \bibinfo {author} {\bibfnamefont {A.}~\bibnamefont {Rubio}},\ }\href
  {\doibase 10.1103/RevModPhys.74.601} {\bibfield  {journal} {\bibinfo
  {journal} {Rev. Mod. Phys.}\ }\textbf {\bibinfo {volume} {74}},\ \bibinfo
  {pages} {601} (\bibinfo {year} {2002})}\BibitemShut {NoStop}%
\bibitem [{\citenamefont {Besley}\ \emph {et~al.}(2009)\citenamefont {Besley},
  \citenamefont {Peach},\ and\ \citenamefont {Tozer}}]{Besley_2009}%
  \BibitemOpen
  \bibfield  {author} {\bibinfo {author} {\bibfnamefont {N.~A.}\ \bibnamefont
  {Besley}}, \bibinfo {author} {\bibfnamefont {M.~J.~G.}\ \bibnamefont
  {Peach}}, \ and\ \bibinfo {author} {\bibfnamefont {D.~J.}\ \bibnamefont
  {Tozer}},\ }\href {\doibase 10.1039/B912718F} {\bibfield  {journal} {\bibinfo
   {journal} {Phys. Chem. Chem. Phys.}\ }\textbf {\bibinfo {volume} {11}},\
  \bibinfo {pages} {10350} (\bibinfo {year} {2009})}\BibitemShut {NoStop}%
\bibitem [{\citenamefont {Bun\ifmmode~\u{a}\else \u{a}\fi{}u}\ and\
  \citenamefont {Joly}(2012)}]{Bunau2012}%
  \BibitemOpen
  \bibfield  {author} {\bibinfo {author} {\bibfnamefont {O.}~\bibnamefont
  {Bun\ifmmode~\u{a}\else \u{a}\fi{}u}}\ and\ \bibinfo {author} {\bibfnamefont
  {Y.}~\bibnamefont {Joly}},\ }\href {\doibase 10.1103/PhysRevB.85.155121}
  {\bibfield  {journal} {\bibinfo  {journal} {Phys. Rev. B}\ }\textbf {\bibinfo
  {volume} {85}},\ \bibinfo {pages} {155121} (\bibinfo {year}
  {2012})}\BibitemShut {NoStop}%
\bibitem [{\citenamefont {Bunău}\ and\ \citenamefont
  {Joly}(2012)}]{Bunau2012b}%
  \BibitemOpen
  \bibfield  {author} {\bibinfo {author} {\bibfnamefont {O.}~\bibnamefont
  {Bunău}}\ and\ \bibinfo {author} {\bibfnamefont {Y.}~\bibnamefont {Joly}},\
  }\href {\doibase 10.1088/0953-8984/24/21/215502} {\bibfield  {journal}
  {\bibinfo  {journal} {Journal of Physics: Condensed Matter}\ }\textbf
  {\bibinfo {volume} {24}},\ \bibinfo {pages} {215502} (\bibinfo {year}
  {2012})}\BibitemShut {NoStop}%
\bibitem [{\citenamefont {Strinati}(1982)}]{Strinati1982}%
  \BibitemOpen
  \bibfield  {author} {\bibinfo {author} {\bibfnamefont {G.}~\bibnamefont
  {Strinati}},\ }\href {\doibase 10.1103/PhysRevLett.49.1519} {\bibfield
  {journal} {\bibinfo  {journal} {Phys. Rev. Lett.}\ }\textbf {\bibinfo
  {volume} {49}},\ \bibinfo {pages} {1519} (\bibinfo {year}
  {1982})}\BibitemShut {NoStop}%
\bibitem [{\citenamefont {Strinati}(1984)}]{Strinati1984}%
  \BibitemOpen
  \bibfield  {author} {\bibinfo {author} {\bibfnamefont {G.}~\bibnamefont
  {Strinati}},\ }\href {\doibase 10.1103/PhysRevB.29.5718} {\bibfield
  {journal} {\bibinfo  {journal} {Phys. Rev. B}\ }\textbf {\bibinfo {volume}
  {29}},\ \bibinfo {pages} {5718} (\bibinfo {year} {1984})}\BibitemShut
  {NoStop}%
\bibitem [{\citenamefont {Shirley}(1998)}]{Shirley1998}%
  \BibitemOpen
  \bibfield  {author} {\bibinfo {author} {\bibfnamefont {E.~L.}\ \bibnamefont
  {Shirley}},\ }\href {\doibase 10.1103/PhysRevLett.80.794} {\bibfield
  {journal} {\bibinfo  {journal} {Phys. Rev. Lett.}\ }\textbf {\bibinfo
  {volume} {80}},\ \bibinfo {pages} {794} (\bibinfo {year} {1998})}\BibitemShut
  {NoStop}%
\bibitem [{\citenamefont {Carlisle}\ \emph {et~al.}(1999)\citenamefont
  {Carlisle}, \citenamefont {Shirley}, \citenamefont {Terminello},
  \citenamefont {Jia}, \citenamefont {Callcott}, \citenamefont {Ederer},
  \citenamefont {Perera},\ and\ \citenamefont {Himpsel}}]{Carlisle1999}%
  \BibitemOpen
  \bibfield  {author} {\bibinfo {author} {\bibfnamefont {J.~A.}\ \bibnamefont
  {Carlisle}}, \bibinfo {author} {\bibfnamefont {E.~L.}\ \bibnamefont
  {Shirley}}, \bibinfo {author} {\bibfnamefont {L.~J.}\ \bibnamefont
  {Terminello}}, \bibinfo {author} {\bibfnamefont {J.~J.}\ \bibnamefont {Jia}},
  \bibinfo {author} {\bibfnamefont {T.~A.}\ \bibnamefont {Callcott}}, \bibinfo
  {author} {\bibfnamefont {D.~L.}\ \bibnamefont {Ederer}}, \bibinfo {author}
  {\bibfnamefont {R.~C.~C.}\ \bibnamefont {Perera}}, \ and\ \bibinfo {author}
  {\bibfnamefont {F.~J.}\ \bibnamefont {Himpsel}},\ }\href {\doibase
  10.1103/PhysRevB.59.7433} {\bibfield  {journal} {\bibinfo  {journal} {Phys.
  Rev. B}\ }\textbf {\bibinfo {volume} {59}},\ \bibinfo {pages} {7433}
  (\bibinfo {year} {1999})}\BibitemShut {NoStop}%
\bibitem [{\citenamefont {Shirley}(2000)}]{Shirley2000}%
  \BibitemOpen
  \bibfield  {author} {\bibinfo {author} {\bibfnamefont {E.}~\bibnamefont
  {Shirley}},\ }\href {\doibase https://doi.org/10.1016/S0022-3697(99)00333-9}
  {\bibfield  {journal} {\bibinfo  {journal} {Journal of Physics and Chemistry
  of Solids}\ }\textbf {\bibinfo {volume} {61}},\ \bibinfo {pages} {445}
  (\bibinfo {year} {2000})}\BibitemShut {NoStop}%
\bibitem [{\citenamefont {Martin}\ \emph {et~al.}(2016)\citenamefont {Martin},
  \citenamefont {Reining},\ and\ \citenamefont {Ceperley}}]{Martin2016}%
  \BibitemOpen
  \bibfield  {author} {\bibinfo {author} {\bibfnamefont {R.~M.}\ \bibnamefont
  {Martin}}, \bibinfo {author} {\bibfnamefont {L.}~\bibnamefont {Reining}}, \
  and\ \bibinfo {author} {\bibfnamefont {D.~M.}\ \bibnamefont {Ceperley}},\
  }\href@noop {} {\emph {\bibinfo {title} {Interacting Electrons: Theory and
  Computational Approaches}}}\ (\bibinfo  {publisher} {Cambridge University
  Press},\ \bibinfo {year} {2016})\BibitemShut {NoStop}%
\bibitem [{\citenamefont {Bechstedt}(2014)}]{Bechstedt2014}%
  \BibitemOpen
  \bibfield  {author} {\bibinfo {author} {\bibfnamefont {F.}~\bibnamefont
  {Bechstedt}},\ }\href@noop {} {\emph {\bibinfo {title} {Many-Body Approach to
  Electronic Excitations: Concepts and Applications}}},\ Springer Series in
  Solid-State Sciences\ (\bibinfo  {publisher} {Springer Berlin Heidelberg},\
  \bibinfo {year} {2014})\BibitemShut {NoStop}%
\bibitem [{\citenamefont {Botti}\ \emph {et~al.}(2007)\citenamefont {Botti},
  \citenamefont {Schindlmayr}, \citenamefont {Sole},\ and\ \citenamefont
  {Reining}}]{Botti_2007}%
  \BibitemOpen
  \bibfield  {author} {\bibinfo {author} {\bibfnamefont {S.}~\bibnamefont
  {Botti}}, \bibinfo {author} {\bibfnamefont {A.}~\bibnamefont {Schindlmayr}},
  \bibinfo {author} {\bibfnamefont {R.~D.}\ \bibnamefont {Sole}}, \ and\
  \bibinfo {author} {\bibfnamefont {L.}~\bibnamefont {Reining}},\ }\href
  {\doibase 10.1088/0034-4885/70/3/R02} {\bibfield  {journal} {\bibinfo
  {journal} {Reports on Progress in Physics}\ }\textbf {\bibinfo {volume}
  {70}},\ \bibinfo {pages} {357} (\bibinfo {year} {2007})}\BibitemShut
  {NoStop}%
\bibitem [{\citenamefont {Wills}\ \emph {et~al.}(2010)\citenamefont {Wills},
  \citenamefont {Alouani}, \citenamefont {Andersson}, \citenamefont {Delin},
  \citenamefont {Eriksson},\ and\ \citenamefont {Grechnyev}}]{Wills2010}%
  \BibitemOpen
  \bibfield  {author} {\bibinfo {author} {\bibfnamefont {J.}~\bibnamefont
  {Wills}}, \bibinfo {author} {\bibfnamefont {M.}~\bibnamefont {Alouani}},
  \bibinfo {author} {\bibfnamefont {P.}~\bibnamefont {Andersson}}, \bibinfo
  {author} {\bibfnamefont {A.}~\bibnamefont {Delin}}, \bibinfo {author}
  {\bibfnamefont {O.}~\bibnamefont {Eriksson}}, \ and\ \bibinfo {author}
  {\bibfnamefont {O.}~\bibnamefont {Grechnyev}},\ }\href@noop {} {\emph
  {\bibinfo {title} {Full-Potential Electronic Structure Method: Energy and
  Force Calculations with Density Functional and Dynamical Mean Field
  Theory}}},\ Springer Series in Solid-State Sciences\ (\bibinfo  {publisher}
  {Springer Berlin Heidelberg},\ \bibinfo {year} {2010})\BibitemShut {NoStop}%
\bibitem [{\citenamefont {Andersen}(1975)}]{Andersen_1975}%
  \BibitemOpen
  \bibfield  {author} {\bibinfo {author} {\bibfnamefont {O.~K.}\ \bibnamefont
  {Andersen}},\ }\href {\doibase 10.1103/PhysRevB.12.3060} {\bibfield
  {journal} {\bibinfo  {journal} {Phys. Rev. B}\ }\textbf {\bibinfo {volume}
  {12}},\ \bibinfo {pages} {3060} (\bibinfo {year} {1975})}\BibitemShut
  {NoStop}%
\bibitem [{\citenamefont {Sjöstedt}\ \emph {et~al.}(2000)\citenamefont
  {Sjöstedt}, \citenamefont {Nordström},\ and\ \citenamefont
  {Singh}}]{Sjostedt2000}%
  \BibitemOpen
  \bibfield  {author} {\bibinfo {author} {\bibfnamefont {E.}~\bibnamefont
  {Sjöstedt}}, \bibinfo {author} {\bibfnamefont {L.}~\bibnamefont
  {Nordström}}, \ and\ \bibinfo {author} {\bibfnamefont {D.}~\bibnamefont
  {Singh}},\ }\href {\doibase https://doi.org/10.1016/S0038-1098(99)00577-3}
  {\bibfield  {journal} {\bibinfo  {journal} {Solid State Communications}\
  }\textbf {\bibinfo {volume} {114}},\ \bibinfo {pages} {15} (\bibinfo {year}
  {2000})}\BibitemShut {NoStop}%
\bibitem [{\citenamefont {Madsen}\ \emph {et~al.}(2001)\citenamefont {Madsen},
  \citenamefont {Blaha}, \citenamefont {Schwarz}, \citenamefont {Sj\"ostedt},\
  and\ \citenamefont {Nordstr\"om}}]{Madsen2001}%
  \BibitemOpen
  \bibfield  {author} {\bibinfo {author} {\bibfnamefont {G.~K.~H.}\
  \bibnamefont {Madsen}}, \bibinfo {author} {\bibfnamefont {P.}~\bibnamefont
  {Blaha}}, \bibinfo {author} {\bibfnamefont {K.}~\bibnamefont {Schwarz}},
  \bibinfo {author} {\bibfnamefont {E.}~\bibnamefont {Sj\"ostedt}}, \ and\
  \bibinfo {author} {\bibfnamefont {L.}~\bibnamefont {Nordstr\"om}},\ }\href
  {\doibase 10.1103/PhysRevB.64.195134} {\bibfield  {journal} {\bibinfo
  {journal} {Phys. Rev. B}\ }\textbf {\bibinfo {volume} {64}},\ \bibinfo
  {pages} {195134} (\bibinfo {year} {2001})}\BibitemShut {NoStop}%
\bibitem [{\citenamefont {Payne}\ \emph {et~al.}(1992)\citenamefont {Payne},
  \citenamefont {Teter}, \citenamefont {Allan}, \citenamefont {Arias},\ and\
  \citenamefont {Joannopoulos}}]{Payne_1992}%
  \BibitemOpen
  \bibfield  {author} {\bibinfo {author} {\bibfnamefont {M.~C.}\ \bibnamefont
  {Payne}}, \bibinfo {author} {\bibfnamefont {M.~P.}\ \bibnamefont {Teter}},
  \bibinfo {author} {\bibfnamefont {D.~C.}\ \bibnamefont {Allan}}, \bibinfo
  {author} {\bibfnamefont {T.~A.}\ \bibnamefont {Arias}}, \ and\ \bibinfo
  {author} {\bibfnamefont {J.~D.}\ \bibnamefont {Joannopoulos}},\ }\href
  {\doibase 10.1103/RevModPhys.64.1045} {\bibfield  {journal} {\bibinfo
  {journal} {Rev. Mod. Phys.}\ }\textbf {\bibinfo {volume} {64}},\ \bibinfo
  {pages} {1045} (\bibinfo {year} {1992})}\BibitemShut {NoStop}%
\bibitem [{\citenamefont {Willand}\ \emph {et~al.}(2013)\citenamefont
  {Willand}, \citenamefont {Kvashnin}, \citenamefont {Genovese}, \citenamefont
  {Vázquez-Mayagoitia}, \citenamefont {Deb}, \citenamefont {Sadeghi},
  \citenamefont {Deutsch},\ and\ \citenamefont {Goedecker}}]{Willand2013}%
  \BibitemOpen
  \bibfield  {author} {\bibinfo {author} {\bibfnamefont {A.}~\bibnamefont
  {Willand}}, \bibinfo {author} {\bibfnamefont {Y.~O.}\ \bibnamefont
  {Kvashnin}}, \bibinfo {author} {\bibfnamefont {L.}~\bibnamefont {Genovese}},
  \bibinfo {author} {\bibfnamefont {A.}~\bibnamefont {Vázquez-Mayagoitia}},
  \bibinfo {author} {\bibfnamefont {A.~K.}\ \bibnamefont {Deb}}, \bibinfo
  {author} {\bibfnamefont {A.}~\bibnamefont {Sadeghi}}, \bibinfo {author}
  {\bibfnamefont {T.}~\bibnamefont {Deutsch}}, \ and\ \bibinfo {author}
  {\bibfnamefont {S.}~\bibnamefont {Goedecker}},\ }\href {\doibase
  10.1063/1.4793260} {\bibfield  {journal} {\bibinfo  {journal} {The Journal of
  Chemical Physics}\ }\textbf {\bibinfo {volume} {138}},\ \bibinfo {pages}
  {104109} (\bibinfo {year} {2013})}\BibitemShut {NoStop}%
\bibitem [{\citenamefont {Lejaeghere}\ \emph {et~al.}(2014)\citenamefont
  {Lejaeghere}, \citenamefont {Speybroeck}, \citenamefont {Oost},\ and\
  \citenamefont {Cottenier}}]{Lejaeghere2014}%
  \BibitemOpen
  \bibfield  {author} {\bibinfo {author} {\bibfnamefont {K.}~\bibnamefont
  {Lejaeghere}}, \bibinfo {author} {\bibfnamefont {V.~V.}\ \bibnamefont
  {Speybroeck}}, \bibinfo {author} {\bibfnamefont {G.~V.}\ \bibnamefont
  {Oost}}, \ and\ \bibinfo {author} {\bibfnamefont {S.}~\bibnamefont
  {Cottenier}},\ }\href {\doibase 10.1080/10408436.2013.772503} {\bibfield
  {journal} {\bibinfo  {journal} {Critical Reviews in Solid State and Materials
  Sciences}\ }\textbf {\bibinfo {volume} {39}},\ \bibinfo {pages} {1} (\bibinfo
  {year} {2014})}\BibitemShut {NoStop}%
\bibitem [{\citenamefont {Prandini}\ \emph {et~al.}(2018)\citenamefont
  {Prandini}, \citenamefont {Marrazzo}, \citenamefont {Castelli}, \citenamefont
  {Mounet},\ and\ \citenamefont {Marzari}}]{Prandini2018}%
  \BibitemOpen
  \bibfield  {author} {\bibinfo {author} {\bibfnamefont {G.}~\bibnamefont
  {Prandini}}, \bibinfo {author} {\bibfnamefont {A.}~\bibnamefont {Marrazzo}},
  \bibinfo {author} {\bibfnamefont {I.~E.}\ \bibnamefont {Castelli}}, \bibinfo
  {author} {\bibfnamefont {N.}~\bibnamefont {Mounet}}, \ and\ \bibinfo {author}
  {\bibfnamefont {N.}~\bibnamefont {Marzari}},\ }\href {\doibase
  10.1038/s41524-018-0127-2} {\bibfield  {journal} {\bibinfo  {journal} {npj
  Computational Materials}\ }\textbf {\bibinfo {volume} {4}},\ \bibinfo {pages}
  {2057} (\bibinfo {year} {2018})}\BibitemShut {NoStop}%
\bibitem [{\citenamefont {Lejaeghere}\ \emph {et~al.}(2016)\citenamefont
  {Lejaeghere}, \citenamefont {Bihlmayer}, \citenamefont {Björkman},
  \citenamefont {Blaha}, \citenamefont {Blügel}, \citenamefont {Blum},
  \citenamefont {Caliste}, \citenamefont {Castelli}, \citenamefont {Clark},
  \citenamefont {Corso}, \citenamefont {de~Gironcoli}, \citenamefont {Deutsch},
  \citenamefont {Dewhurst}, \citenamefont {Marco}, \citenamefont {Draxl},
  \citenamefont {Dułak}, \citenamefont {Eriksson}, \citenamefont
  {Flores-Livas}, \citenamefont {Garrity}, \citenamefont {Genovese},
  \citenamefont {Giannozzi}, \citenamefont {Giantomassi}, \citenamefont
  {Goedecker}, \citenamefont {Gonze}, \citenamefont {Grånäs}, \citenamefont
  {Gross}, \citenamefont {Gulans}, \citenamefont {Gygi}, \citenamefont
  {Hamann}, \citenamefont {Hasnip}, \citenamefont {Holzwarth}, \citenamefont
  {Iuşan}, \citenamefont {Jochym}, \citenamefont {Jollet}, \citenamefont
  {Jones}, \citenamefont {Kresse}, \citenamefont {Koepernik}, \citenamefont
  {Küçükbenli}, \citenamefont {Kvashnin}, \citenamefont {Locht},
  \citenamefont {Lubeck}, \citenamefont {Marsman}, \citenamefont {Marzari},
  \citenamefont {Nitzsche}, \citenamefont {Nordström}, \citenamefont {Ozaki},
  \citenamefont {Paulatto}, \citenamefont {Pickard}, \citenamefont {Poelmans},
  \citenamefont {Probert}, \citenamefont {Refson}, \citenamefont {Richter},
  \citenamefont {Rignanese}, \citenamefont {Saha}, \citenamefont {Scheffler},
  \citenamefont {Schlipf}, \citenamefont {Schwarz}, \citenamefont {Sharma},
  \citenamefont {Tavazza}, \citenamefont {Thunström}, \citenamefont
  {Tkatchenko}, \citenamefont {Torrent}, \citenamefont {Vanderbilt},
  \citenamefont {van Setten}, \citenamefont {Speybroeck}, \citenamefont
  {Wills}, \citenamefont {Yates}, \citenamefont {Zhang},\ and\ \citenamefont
  {Cottenier}}]{Lejaeghere2016}%
  \BibitemOpen
  \bibfield  {author} {\bibinfo {author} {\bibfnamefont {K.}~\bibnamefont
  {Lejaeghere}}, \bibinfo {author} {\bibfnamefont {G.}~\bibnamefont
  {Bihlmayer}}, \bibinfo {author} {\bibfnamefont {T.}~\bibnamefont
  {Björkman}}, \bibinfo {author} {\bibfnamefont {P.}~\bibnamefont {Blaha}},
  \bibinfo {author} {\bibfnamefont {S.}~\bibnamefont {Blügel}}, \bibinfo
  {author} {\bibfnamefont {V.}~\bibnamefont {Blum}}, \bibinfo {author}
  {\bibfnamefont {D.}~\bibnamefont {Caliste}}, \bibinfo {author} {\bibfnamefont
  {I.~E.}\ \bibnamefont {Castelli}}, \bibinfo {author} {\bibfnamefont {S.~J.}\
  \bibnamefont {Clark}}, \bibinfo {author} {\bibfnamefont {A.~D.}\ \bibnamefont
  {Corso}}, \bibinfo {author} {\bibfnamefont {S.}~\bibnamefont {de~Gironcoli}},
  \bibinfo {author} {\bibfnamefont {T.}~\bibnamefont {Deutsch}}, \bibinfo
  {author} {\bibfnamefont {J.~K.}\ \bibnamefont {Dewhurst}}, \bibinfo {author}
  {\bibfnamefont {I.~D.}\ \bibnamefont {Marco}}, \bibinfo {author}
  {\bibfnamefont {C.}~\bibnamefont {Draxl}}, \bibinfo {author} {\bibfnamefont
  {M.}~\bibnamefont {Dułak}}, \bibinfo {author} {\bibfnamefont
  {O.}~\bibnamefont {Eriksson}}, \bibinfo {author} {\bibfnamefont {J.~A.}\
  \bibnamefont {Flores-Livas}}, \bibinfo {author} {\bibfnamefont {K.~F.}\
  \bibnamefont {Garrity}}, \bibinfo {author} {\bibfnamefont {L.}~\bibnamefont
  {Genovese}}, \bibinfo {author} {\bibfnamefont {P.}~\bibnamefont {Giannozzi}},
  \bibinfo {author} {\bibfnamefont {M.}~\bibnamefont {Giantomassi}}, \bibinfo
  {author} {\bibfnamefont {S.}~\bibnamefont {Goedecker}}, \bibinfo {author}
  {\bibfnamefont {X.}~\bibnamefont {Gonze}}, \bibinfo {author} {\bibfnamefont
  {O.}~\bibnamefont {Grånäs}}, \bibinfo {author} {\bibfnamefont {E.~K.~U.}\
  \bibnamefont {Gross}}, \bibinfo {author} {\bibfnamefont {A.}~\bibnamefont
  {Gulans}}, \bibinfo {author} {\bibfnamefont {F.}~\bibnamefont {Gygi}},
  \bibinfo {author} {\bibfnamefont {D.~R.}\ \bibnamefont {Hamann}}, \bibinfo
  {author} {\bibfnamefont {P.~J.}\ \bibnamefont {Hasnip}}, \bibinfo {author}
  {\bibfnamefont {N.~A.~W.}\ \bibnamefont {Holzwarth}}, \bibinfo {author}
  {\bibfnamefont {D.}~\bibnamefont {Iuşan}}, \bibinfo {author} {\bibfnamefont
  {D.~B.}\ \bibnamefont {Jochym}}, \bibinfo {author} {\bibfnamefont
  {F.}~\bibnamefont {Jollet}}, \bibinfo {author} {\bibfnamefont
  {D.}~\bibnamefont {Jones}}, \bibinfo {author} {\bibfnamefont
  {G.}~\bibnamefont {Kresse}}, \bibinfo {author} {\bibfnamefont
  {K.}~\bibnamefont {Koepernik}}, \bibinfo {author} {\bibfnamefont
  {E.}~\bibnamefont {Küçükbenli}}, \bibinfo {author} {\bibfnamefont {Y.~O.}\
  \bibnamefont {Kvashnin}}, \bibinfo {author} {\bibfnamefont {I.~L.~M.}\
  \bibnamefont {Locht}}, \bibinfo {author} {\bibfnamefont {S.}~\bibnamefont
  {Lubeck}}, \bibinfo {author} {\bibfnamefont {M.}~\bibnamefont {Marsman}},
  \bibinfo {author} {\bibfnamefont {N.}~\bibnamefont {Marzari}}, \bibinfo
  {author} {\bibfnamefont {U.}~\bibnamefont {Nitzsche}}, \bibinfo {author}
  {\bibfnamefont {L.}~\bibnamefont {Nordström}}, \bibinfo {author}
  {\bibfnamefont {T.}~\bibnamefont {Ozaki}}, \bibinfo {author} {\bibfnamefont
  {L.}~\bibnamefont {Paulatto}}, \bibinfo {author} {\bibfnamefont {C.~J.}\
  \bibnamefont {Pickard}}, \bibinfo {author} {\bibfnamefont {W.}~\bibnamefont
  {Poelmans}}, \bibinfo {author} {\bibfnamefont {M.~I.~J.}\ \bibnamefont
  {Probert}}, \bibinfo {author} {\bibfnamefont {K.}~\bibnamefont {Refson}},
  \bibinfo {author} {\bibfnamefont {M.}~\bibnamefont {Richter}}, \bibinfo
  {author} {\bibfnamefont {G.-M.}\ \bibnamefont {Rignanese}}, \bibinfo {author}
  {\bibfnamefont {S.}~\bibnamefont {Saha}}, \bibinfo {author} {\bibfnamefont
  {M.}~\bibnamefont {Scheffler}}, \bibinfo {author} {\bibfnamefont
  {M.}~\bibnamefont {Schlipf}}, \bibinfo {author} {\bibfnamefont
  {K.}~\bibnamefont {Schwarz}}, \bibinfo {author} {\bibfnamefont
  {S.}~\bibnamefont {Sharma}}, \bibinfo {author} {\bibfnamefont
  {F.}~\bibnamefont {Tavazza}}, \bibinfo {author} {\bibfnamefont
  {P.}~\bibnamefont {Thunström}}, \bibinfo {author} {\bibfnamefont
  {A.}~\bibnamefont {Tkatchenko}}, \bibinfo {author} {\bibfnamefont
  {M.}~\bibnamefont {Torrent}}, \bibinfo {author} {\bibfnamefont
  {D.}~\bibnamefont {Vanderbilt}}, \bibinfo {author} {\bibfnamefont {M.~J.}\
  \bibnamefont {van Setten}}, \bibinfo {author} {\bibfnamefont {V.~V.}\
  \bibnamefont {Speybroeck}}, \bibinfo {author} {\bibfnamefont {J.~M.}\
  \bibnamefont {Wills}}, \bibinfo {author} {\bibfnamefont {J.~R.}\ \bibnamefont
  {Yates}}, \bibinfo {author} {\bibfnamefont {G.-X.}\ \bibnamefont {Zhang}}, \
  and\ \bibinfo {author} {\bibfnamefont {S.}~\bibnamefont {Cottenier}},\ }\href
  {\doibase 10.1126/science.aad3000} {\bibfield  {journal} {\bibinfo  {journal}
  {Science}\ }\textbf {\bibinfo {volume} {351}},\ \bibinfo {pages} {aad3000}
  (\bibinfo {year} {2016})}\BibitemShut {NoStop}%
\bibitem [{\citenamefont {Ku}\ and\ \citenamefont {Eguiluz}(2002)}]{Ku2002}%
  \BibitemOpen
  \bibfield  {author} {\bibinfo {author} {\bibfnamefont {W.}~\bibnamefont
  {Ku}}\ and\ \bibinfo {author} {\bibfnamefont {A.~G.}\ \bibnamefont
  {Eguiluz}},\ }\href {\doibase 10.1103/PhysRevLett.89.126401} {\bibfield
  {journal} {\bibinfo  {journal} {Phys. Rev. Lett.}\ }\textbf {\bibinfo
  {volume} {89}},\ \bibinfo {pages} {126401} (\bibinfo {year}
  {2002})}\BibitemShut {NoStop}%
\bibitem [{\citenamefont {Delaney}\ \emph {et~al.}(2004)\citenamefont
  {Delaney}, \citenamefont {Garc\'{\i}a-Gonz\'alez}, \citenamefont {Rubio},
  \citenamefont {Rinke},\ and\ \citenamefont {Godby}}]{Delaney2004}%
  \BibitemOpen
  \bibfield  {author} {\bibinfo {author} {\bibfnamefont {K.}~\bibnamefont
  {Delaney}}, \bibinfo {author} {\bibfnamefont {P.}~\bibnamefont
  {Garc\'{\i}a-Gonz\'alez}}, \bibinfo {author} {\bibfnamefont {A.}~\bibnamefont
  {Rubio}}, \bibinfo {author} {\bibfnamefont {P.}~\bibnamefont {Rinke}}, \ and\
  \bibinfo {author} {\bibfnamefont {R.~W.}\ \bibnamefont {Godby}},\ }\href
  {\doibase 10.1103/PhysRevLett.93.249701} {\bibfield  {journal} {\bibinfo
  {journal} {Phys. Rev. Lett.}\ }\textbf {\bibinfo {volume} {93}},\ \bibinfo
  {pages} {249701} (\bibinfo {year} {2004})}\BibitemShut {NoStop}%
\bibitem [{\citenamefont {Tiago}\ \emph {et~al.}(2004)\citenamefont {Tiago},
  \citenamefont {Ismail-Beigi},\ and\ \citenamefont {Louie}}]{Tiago_2004}%
  \BibitemOpen
  \bibfield  {author} {\bibinfo {author} {\bibfnamefont {M.~L.}\ \bibnamefont
  {Tiago}}, \bibinfo {author} {\bibfnamefont {S.}~\bibnamefont {Ismail-Beigi}},
  \ and\ \bibinfo {author} {\bibfnamefont {S.~G.}\ \bibnamefont {Louie}},\
  }\href {\doibase 10.1103/PhysRevB.69.125212} {\bibfield  {journal} {\bibinfo
  {journal} {Phys. Rev. B}\ }\textbf {\bibinfo {volume} {69}},\ \bibinfo
  {pages} {125212} (\bibinfo {year} {2004})}\BibitemShut {NoStop}%
\bibitem [{\citenamefont {van Schilfgaarde}\ \emph
  {et~al.}(2006{\natexlab{a}})\citenamefont {van Schilfgaarde}, \citenamefont
  {Kotani},\ and\ \citenamefont {Faleev}}]{vanSchilfgaarde2006}%
  \BibitemOpen
  \bibfield  {author} {\bibinfo {author} {\bibfnamefont {M.}~\bibnamefont {van
  Schilfgaarde}}, \bibinfo {author} {\bibfnamefont {T.}~\bibnamefont {Kotani}},
  \ and\ \bibinfo {author} {\bibfnamefont {S.~V.}\ \bibnamefont {Faleev}},\
  }\href {\doibase 10.1103/PhysRevB.74.245125} {\bibfield  {journal} {\bibinfo
  {journal} {Phys. Rev. B}\ }\textbf {\bibinfo {volume} {74}},\ \bibinfo
  {pages} {245125} (\bibinfo {year} {2006}{\natexlab{a}})}\BibitemShut
  {NoStop}%
\bibitem [{\citenamefont {Friedrich}\ \emph {et~al.}(2006)\citenamefont
  {Friedrich}, \citenamefont {Schindlmayr}, \citenamefont {Bl\"ugel},\ and\
  \citenamefont {Kotani}}]{Friedrich2006}%
  \BibitemOpen
  \bibfield  {author} {\bibinfo {author} {\bibfnamefont {C.}~\bibnamefont
  {Friedrich}}, \bibinfo {author} {\bibfnamefont {A.}~\bibnamefont
  {Schindlmayr}}, \bibinfo {author} {\bibfnamefont {S.}~\bibnamefont
  {Bl\"ugel}}, \ and\ \bibinfo {author} {\bibfnamefont {T.}~\bibnamefont
  {Kotani}},\ }\href {\doibase 10.1103/PhysRevB.74.045104} {\bibfield
  {journal} {\bibinfo  {journal} {Phys. Rev. B}\ }\textbf {\bibinfo {volume}
  {74}},\ \bibinfo {pages} {045104} (\bibinfo {year} {2006})}\BibitemShut
  {NoStop}%
\bibitem [{\citenamefont {G\'omez-Abal}\ \emph {et~al.}(2008)\citenamefont
  {G\'omez-Abal}, \citenamefont {Li}, \citenamefont {Scheffler},\ and\
  \citenamefont {Ambrosch-Draxl}}]{Gomez_2008}%
  \BibitemOpen
  \bibfield  {author} {\bibinfo {author} {\bibfnamefont {R.}~\bibnamefont
  {G\'omez-Abal}}, \bibinfo {author} {\bibfnamefont {X.}~\bibnamefont {Li}},
  \bibinfo {author} {\bibfnamefont {M.}~\bibnamefont {Scheffler}}, \ and\
  \bibinfo {author} {\bibfnamefont {C.}~\bibnamefont {Ambrosch-Draxl}},\ }\href
  {\doibase 10.1103/PhysRevLett.101.106404} {\bibfield  {journal} {\bibinfo
  {journal} {Phys. Rev. Lett.}\ }\textbf {\bibinfo {volume} {101}},\ \bibinfo
  {pages} {106404} (\bibinfo {year} {2008})}\BibitemShut {NoStop}%
\bibitem [{\citenamefont {Luppi}\ \emph {et~al.}(2008)\citenamefont {Luppi},
  \citenamefont {Weissker}, \citenamefont {Bottaro}, \citenamefont {Sottile},
  \citenamefont {Veniard}, \citenamefont {Reining},\ and\ \citenamefont
  {Onida}}]{Luppi_2008}%
  \BibitemOpen
  \bibfield  {author} {\bibinfo {author} {\bibfnamefont {E.}~\bibnamefont
  {Luppi}}, \bibinfo {author} {\bibfnamefont {H.-C.}\ \bibnamefont {Weissker}},
  \bibinfo {author} {\bibfnamefont {S.}~\bibnamefont {Bottaro}}, \bibinfo
  {author} {\bibfnamefont {F.}~\bibnamefont {Sottile}}, \bibinfo {author}
  {\bibfnamefont {V.}~\bibnamefont {Veniard}}, \bibinfo {author} {\bibfnamefont
  {L.}~\bibnamefont {Reining}}, \ and\ \bibinfo {author} {\bibfnamefont
  {G.}~\bibnamefont {Onida}},\ }\href {\doibase 10.1103/PhysRevB.78.245124}
  {\bibfield  {journal} {\bibinfo  {journal} {Phys. Rev. B}\ }\textbf {\bibinfo
  {volume} {78}},\ \bibinfo {pages} {245124} (\bibinfo {year}
  {2008})}\BibitemShut {NoStop}%
\bibitem [{\citenamefont {Klime\ifmmode~\check{s}\else \v{s}\fi{}}\ \emph
  {et~al.}(2014)\citenamefont {Klime\ifmmode~\check{s}\else \v{s}\fi{}},
  \citenamefont {Kaltak},\ and\ \citenamefont {Kresse}}]{Klimes2014}%
  \BibitemOpen
  \bibfield  {author} {\bibinfo {author} {\bibfnamefont {J.~c.~v.}\
  \bibnamefont {Klime\ifmmode~\check{s}\else \v{s}\fi{}}}, \bibinfo {author}
  {\bibfnamefont {M.}~\bibnamefont {Kaltak}}, \ and\ \bibinfo {author}
  {\bibfnamefont {G.}~\bibnamefont {Kresse}},\ }\href {\doibase
  10.1103/PhysRevB.90.075125} {\bibfield  {journal} {\bibinfo  {journal} {Phys.
  Rev. B}\ }\textbf {\bibinfo {volume} {90}},\ \bibinfo {pages} {075125}
  (\bibinfo {year} {2014})}\BibitemShut {NoStop}%
\bibitem [{\citenamefont {Friedrich}\ \emph
  {et~al.}(2011{\natexlab{a}})\citenamefont {Friedrich}, \citenamefont
  {M\"uller},\ and\ \citenamefont {Bl\"ugel}}]{Friedrich2011}%
  \BibitemOpen
  \bibfield  {author} {\bibinfo {author} {\bibfnamefont {C.}~\bibnamefont
  {Friedrich}}, \bibinfo {author} {\bibfnamefont {M.~C.}\ \bibnamefont
  {M\"uller}}, \ and\ \bibinfo {author} {\bibfnamefont {S.}~\bibnamefont
  {Bl\"ugel}},\ }\href {\doibase 10.1103/PhysRevB.83.081101} {\bibfield
  {journal} {\bibinfo  {journal} {Phys. Rev. B}\ }\textbf {\bibinfo {volume}
  {83}},\ \bibinfo {pages} {081101} (\bibinfo {year}
  {2011}{\natexlab{a}})}\BibitemShut {NoStop}%
\bibitem [{\citenamefont {Friedrich}\ \emph
  {et~al.}(2011{\natexlab{b}})\citenamefont {Friedrich}, \citenamefont
  {M\"uller},\ and\ \citenamefont {Bl\"ugel}}]{Friedrich2011_Erratum}%
  \BibitemOpen
  \bibfield  {author} {\bibinfo {author} {\bibfnamefont {C.}~\bibnamefont
  {Friedrich}}, \bibinfo {author} {\bibfnamefont {M.~C.}\ \bibnamefont
  {M\"uller}}, \ and\ \bibinfo {author} {\bibfnamefont {S.}~\bibnamefont
  {Bl\"ugel}},\ }\href {\doibase 10.1103/PhysRevB.84.039906} {\bibfield
  {journal} {\bibinfo  {journal} {Phys. Rev. B}\ }\textbf {\bibinfo {volume}
  {84}},\ \bibinfo {pages} {039906} (\bibinfo {year}
  {2011}{\natexlab{b}})}\BibitemShut {NoStop}%
\bibitem [{\citenamefont {Jiang}\ and\ \citenamefont
  {Blaha}(2016)}]{Jiang2016}%
  \BibitemOpen
  \bibfield  {author} {\bibinfo {author} {\bibfnamefont {H.}~\bibnamefont
  {Jiang}}\ and\ \bibinfo {author} {\bibfnamefont {P.}~\bibnamefont {Blaha}},\
  }\href {\doibase 10.1103/PhysRevB.93.115203} {\bibfield  {journal} {\bibinfo
  {journal} {Phys. Rev. B}\ }\textbf {\bibinfo {volume} {93}},\ \bibinfo
  {pages} {115203} (\bibinfo {year} {2016})}\BibitemShut {NoStop}%
\bibitem [{\citenamefont {Jiang}(2018)}]{Jiang2018}%
  \BibitemOpen
  \bibfield  {author} {\bibinfo {author} {\bibfnamefont {H.}~\bibnamefont
  {Jiang}},\ }\href {\doibase 10.1103/PhysRevB.97.245132} {\bibfield  {journal}
  {\bibinfo  {journal} {Phys. Rev. B}\ }\textbf {\bibinfo {volume} {97}},\
  \bibinfo {pages} {245132} (\bibinfo {year} {2018})}\BibitemShut {NoStop}%
\bibitem [{\citenamefont {Hamann}(2013)}]{Hamann_2013}%
  \BibitemOpen
  \bibfield  {author} {\bibinfo {author} {\bibfnamefont {D.~R.}\ \bibnamefont
  {Hamann}},\ }\href {\doibase 10.1103/PhysRevB.88.085117} {\bibfield
  {journal} {\bibinfo  {journal} {Phys. Rev. B}\ }\textbf {\bibinfo {volume}
  {88}},\ \bibinfo {pages} {085117} (\bibinfo {year} {2013})}\BibitemShut
  {NoStop}%
\bibitem [{Note1()}]{Note1}%
  \BibitemOpen
  \bibinfo {note} {The same hypothesis is made when the core orbitals are
  obtained from a calculation of the isolated atom\cite
  {Shirley2004,Bloechl1994,Unzog2022}.}\BibitemShut {Stop}%
\bibitem [{\citenamefont {French}(1990)}]{French_1990}%
  \BibitemOpen
  \bibfield  {author} {\bibinfo {author} {\bibfnamefont {R.~H.}\ \bibnamefont
  {French}},\ }\href {\doibase
  https://doi.org/10.1111/j.1151-2916.1990.tb06541.x} {\bibfield  {journal}
  {\bibinfo  {journal} {Journal of the American Ceramic Society}\ }\textbf
  {\bibinfo {volume} {73}},\ \bibinfo {pages} {477} (\bibinfo {year}
  {1990})}\BibitemShut {NoStop}%
\bibitem [{\citenamefont {French}\ \emph {et~al.}(1994)\citenamefont {French},
  \citenamefont {Jones},\ and\ \citenamefont {Loughin}}]{French_1994}%
  \BibitemOpen
  \bibfield  {author} {\bibinfo {author} {\bibfnamefont {R.~H.}\ \bibnamefont
  {French}}, \bibinfo {author} {\bibfnamefont {D.~J.}\ \bibnamefont {Jones}}, \
  and\ \bibinfo {author} {\bibfnamefont {S.}~\bibnamefont {Loughin}},\ }\href
  {\doibase https://doi.org/10.1111/j.1151-2916.1994.tb07009.x} {\bibfield
  {journal} {\bibinfo  {journal} {Journal of the American Ceramic Society}\
  }\textbf {\bibinfo {volume} {77}},\ \bibinfo {pages} {412} (\bibinfo {year}
  {1994})}\BibitemShut {NoStop}%
\bibitem [{\citenamefont {Tanaka}\ and\ \citenamefont
  {Adachi}(1996)}]{Tanaka_1996}%
  \BibitemOpen
  \bibfield  {author} {\bibinfo {author} {\bibfnamefont {I.}~\bibnamefont
  {Tanaka}}\ and\ \bibinfo {author} {\bibfnamefont {H.}~\bibnamefont
  {Adachi}},\ }\href {\doibase 10.1103/PhysRevB.54.4604} {\bibfield  {journal}
  {\bibinfo  {journal} {Phys. Rev. B}\ }\textbf {\bibinfo {volume} {54}},\
  \bibinfo {pages} {4604} (\bibinfo {year} {1996})}\BibitemShut {NoStop}%
\bibitem [{\citenamefont {Cabaret}\ \emph {et~al.}(1996)\citenamefont
  {Cabaret}, \citenamefont {Sainctavit}, \citenamefont {Ildefonse},\ and\
  \citenamefont {Flank}}]{Cabaret_1996}%
  \BibitemOpen
  \bibfield  {author} {\bibinfo {author} {\bibfnamefont {D.}~\bibnamefont
  {Cabaret}}, \bibinfo {author} {\bibfnamefont {P.}~\bibnamefont {Sainctavit}},
  \bibinfo {author} {\bibfnamefont {P.}~\bibnamefont {Ildefonse}}, \ and\
  \bibinfo {author} {\bibfnamefont {A.-M.}\ \bibnamefont {Flank}},\ }\href
  {\doibase 10.1088/0953-8984/8/20/015} {\bibfield  {journal} {\bibinfo
  {journal} {Journal of Physics: Condensed Matter}\ }\textbf {\bibinfo {volume}
  {8}},\ \bibinfo {pages} {3691} (\bibinfo {year} {1996})}\BibitemShut
  {NoStop}%
\bibitem [{\citenamefont {Ildefonse}\ \emph {et~al.}(1998)\citenamefont
  {Ildefonse}, \citenamefont {Cabaret}, \citenamefont {Sainctavit},
  \citenamefont {Calas}, \citenamefont {Flank},\ and\ \citenamefont
  {Lagarde}}]{Ildefonse_1998}%
  \BibitemOpen
  \bibfield  {author} {\bibinfo {author} {\bibfnamefont {P.}~\bibnamefont
  {Ildefonse}}, \bibinfo {author} {\bibfnamefont {D.}~\bibnamefont {Cabaret}},
  \bibinfo {author} {\bibfnamefont {P.}~\bibnamefont {Sainctavit}}, \bibinfo
  {author} {\bibfnamefont {G.}~\bibnamefont {Calas}}, \bibinfo {author}
  {\bibfnamefont {A.-M.}\ \bibnamefont {Flank}}, \ and\ \bibinfo {author}
  {\bibfnamefont {P.}~\bibnamefont {Lagarde}},\ }\href
  {https://doi.org/10.1007/s002690050093} {\bibfield  {journal} {\bibinfo
  {journal} {Physics and Chemistry of Minerals}\ }\textbf {\bibinfo {volume}
  {25}},\ \bibinfo {pages} {112} (\bibinfo {year} {1998})}\BibitemShut
  {NoStop}%
\bibitem [{\citenamefont {van Bokhoven}\ \emph {et~al.}(2001)\citenamefont {van
  Bokhoven}, \citenamefont {Nabi}, \citenamefont {Sambe}, \citenamefont
  {Ramaker},\ and\ \citenamefont {Koningsberger}}]{vanBokhoven_2001}%
  \BibitemOpen
  \bibfield  {author} {\bibinfo {author} {\bibfnamefont {J.~A.}\ \bibnamefont
  {van Bokhoven}}, \bibinfo {author} {\bibfnamefont {T.}~\bibnamefont {Nabi}},
  \bibinfo {author} {\bibfnamefont {H.}~\bibnamefont {Sambe}}, \bibinfo
  {author} {\bibfnamefont {D.~E.}\ \bibnamefont {Ramaker}}, \ and\ \bibinfo
  {author} {\bibfnamefont {D.~C.}\ \bibnamefont {Koningsberger}},\ }\href
  {\doibase 10.1088/0953-8984/13/45/311} {\bibfield  {journal} {\bibinfo
  {journal} {Journal of Physics: Condensed Matter}\ }\textbf {\bibinfo {volume}
  {13}},\ \bibinfo {pages} {10247} (\bibinfo {year} {2001})}\BibitemShut
  {NoStop}%
\bibitem [{\citenamefont {Strinati}(1988)}]{Strinati1988}%
  \BibitemOpen
  \bibfield  {author} {\bibinfo {author} {\bibfnamefont {G.}~\bibnamefont
  {Strinati}},\ }\href {http://dx.doi.org/10.1007/BF0272596} {\bibfield
  {journal} {\bibinfo  {journal} {Rivista del Nuovo Cimento}\ }\textbf
  {\bibinfo {volume} {11}},\ \bibinfo {pages} {1} (\bibinfo {year}
  {1988})}\BibitemShut {NoStop}%
\bibitem [{\citenamefont {Hedin}(1965)}]{Hedin1965}%
  \BibitemOpen
  \bibfield  {author} {\bibinfo {author} {\bibfnamefont {L.}~\bibnamefont
  {Hedin}},\ }\href {\doibase 10.1103/PhysRev.139.A796} {\bibfield  {journal}
  {\bibinfo  {journal} {Phys. Rev.}\ }\textbf {\bibinfo {volume} {139}},\
  \bibinfo {pages} {A796} (\bibinfo {year} {1965})}\BibitemShut {NoStop}%
\bibitem [{\citenamefont {Albrecht}\ \emph {et~al.}(1998)\citenamefont
  {Albrecht}, \citenamefont {Reining}, \citenamefont {Del~Sole},\ and\
  \citenamefont {Onida}}]{Albrecht1998}%
  \BibitemOpen
  \bibfield  {author} {\bibinfo {author} {\bibfnamefont {S.}~\bibnamefont
  {Albrecht}}, \bibinfo {author} {\bibfnamefont {L.}~\bibnamefont {Reining}},
  \bibinfo {author} {\bibfnamefont {R.}~\bibnamefont {Del~Sole}}, \ and\
  \bibinfo {author} {\bibfnamefont {G.}~\bibnamefont {Onida}},\ }\href@noop {}
  {\bibfield  {journal} {\bibinfo  {journal} {Phys. Rev. Lett.}\ }\textbf
  {\bibinfo {volume} {80}},\ \bibinfo {pages} {4510} (\bibinfo {year}
  {1998})}\BibitemShut {NoStop}%
\bibitem [{\citenamefont {Benedict}\ \emph {et~al.}(1998)\citenamefont
  {Benedict}, \citenamefont {Shirley},\ and\ \citenamefont
  {Bohn}}]{Benedict1998}%
  \BibitemOpen
  \bibfield  {author} {\bibinfo {author} {\bibfnamefont {L.~X.}\ \bibnamefont
  {Benedict}}, \bibinfo {author} {\bibfnamefont {E.~L.}\ \bibnamefont
  {Shirley}}, \ and\ \bibinfo {author} {\bibfnamefont {R.~B.}\ \bibnamefont
  {Bohn}},\ }\href {\doibase 10.1103/PhysRevLett.80.4514} {\bibfield  {journal}
  {\bibinfo  {journal} {Phys. Rev. Lett.}\ }\textbf {\bibinfo {volume} {80}},\
  \bibinfo {pages} {4514} (\bibinfo {year} {1998})}\BibitemShut {NoStop}%
\bibitem [{\citenamefont {Rohlfing}\ and\ \citenamefont
  {Louie}(2000)}]{Rohlfing2000}%
  \BibitemOpen
  \bibfield  {author} {\bibinfo {author} {\bibfnamefont {M.}~\bibnamefont
  {Rohlfing}}\ and\ \bibinfo {author} {\bibfnamefont {S.~G.}\ \bibnamefont
  {Louie}},\ }\href {\doibase 10.1103/PhysRevB.62.4927} {\bibfield  {journal}
  {\bibinfo  {journal} {Phys. Rev. B}\ }\textbf {\bibinfo {volume} {62}},\
  \bibinfo {pages} {4927} (\bibinfo {year} {2000})}\BibitemShut {NoStop}%
\bibitem [{\citenamefont {Vinson}\ \emph {et~al.}(2011)\citenamefont {Vinson},
  \citenamefont {Rehr}, \citenamefont {Kas},\ and\ \citenamefont
  {Shirley}}]{Vinson2011}%
  \BibitemOpen
  \bibfield  {author} {\bibinfo {author} {\bibfnamefont {J.}~\bibnamefont
  {Vinson}}, \bibinfo {author} {\bibfnamefont {J.~J.}\ \bibnamefont {Rehr}},
  \bibinfo {author} {\bibfnamefont {J.~J.}\ \bibnamefont {Kas}}, \ and\
  \bibinfo {author} {\bibfnamefont {E.~L.}\ \bibnamefont {Shirley}},\ }\href
  {\doibase 10.1103/PhysRevB.83.115106} {\bibfield  {journal} {\bibinfo
  {journal} {Phys. Rev. B}\ }\textbf {\bibinfo {volume} {83}},\ \bibinfo
  {pages} {115106} (\bibinfo {year} {2011})}\BibitemShut {NoStop}%
\bibitem [{\citenamefont {Vinson}\ and\ \citenamefont
  {Rehr}(2012)}]{Vinson2012}%
  \BibitemOpen
  \bibfield  {author} {\bibinfo {author} {\bibfnamefont {J.}~\bibnamefont
  {Vinson}}\ and\ \bibinfo {author} {\bibfnamefont {J.~J.}\ \bibnamefont
  {Rehr}},\ }\href {\doibase 10.1103/PhysRevB.86.195135} {\bibfield  {journal}
  {\bibinfo  {journal} {Phys. Rev. B}\ }\textbf {\bibinfo {volume} {86}},\
  \bibinfo {pages} {195135} (\bibinfo {year} {2012})}\BibitemShut {NoStop}%
\bibitem [{\citenamefont {Gilmore}\ \emph {et~al.}(2015)\citenamefont
  {Gilmore}, \citenamefont {Vinson}, \citenamefont {Shirley}, \citenamefont
  {Prendergast}, \citenamefont {Pemmaraju}, \citenamefont {Kas}, \citenamefont
  {Vila},\ and\ \citenamefont {Rehr}}]{Gilmore2015}%
  \BibitemOpen
  \bibfield  {author} {\bibinfo {author} {\bibfnamefont {K.}~\bibnamefont
  {Gilmore}}, \bibinfo {author} {\bibfnamefont {J.}~\bibnamefont {Vinson}},
  \bibinfo {author} {\bibfnamefont {E.}~\bibnamefont {Shirley}}, \bibinfo
  {author} {\bibfnamefont {D.}~\bibnamefont {Prendergast}}, \bibinfo {author}
  {\bibfnamefont {C.}~\bibnamefont {Pemmaraju}}, \bibinfo {author}
  {\bibfnamefont {J.}~\bibnamefont {Kas}}, \bibinfo {author} {\bibfnamefont
  {F.}~\bibnamefont {Vila}}, \ and\ \bibinfo {author} {\bibfnamefont
  {J.}~\bibnamefont {Rehr}},\ }\href {\doibase
  https://doi.org/10.1016/j.cpc.2015.08.014} {\bibfield  {journal} {\bibinfo
  {journal} {Computer Physics Communications}\ }\textbf {\bibinfo {volume}
  {197}},\ \bibinfo {pages} {109} (\bibinfo {year} {2015})}\BibitemShut
  {NoStop}%
\bibitem [{\citenamefont {Gilmore}\ \emph {et~al.}(2021)\citenamefont
  {Gilmore}, \citenamefont {Pelliciari}, \citenamefont {Huang}, \citenamefont
  {Kas}, \citenamefont {Dantz}, \citenamefont {Strocov}, \citenamefont
  {Kasahara}, \citenamefont {Matsuda}, \citenamefont {Das}, \citenamefont
  {Shibauchi},\ and\ \citenamefont {Schmitt}}]{Gilmore2021}%
  \BibitemOpen
  \bibfield  {author} {\bibinfo {author} {\bibfnamefont {K.}~\bibnamefont
  {Gilmore}}, \bibinfo {author} {\bibfnamefont {J.}~\bibnamefont {Pelliciari}},
  \bibinfo {author} {\bibfnamefont {Y.}~\bibnamefont {Huang}}, \bibinfo
  {author} {\bibfnamefont {J.~J.}\ \bibnamefont {Kas}}, \bibinfo {author}
  {\bibfnamefont {M.}~\bibnamefont {Dantz}}, \bibinfo {author} {\bibfnamefont
  {V.~N.}\ \bibnamefont {Strocov}}, \bibinfo {author} {\bibfnamefont
  {S.}~\bibnamefont {Kasahara}}, \bibinfo {author} {\bibfnamefont
  {Y.}~\bibnamefont {Matsuda}}, \bibinfo {author} {\bibfnamefont
  {T.}~\bibnamefont {Das}}, \bibinfo {author} {\bibfnamefont {T.}~\bibnamefont
  {Shibauchi}}, \ and\ \bibinfo {author} {\bibfnamefont {T.}~\bibnamefont
  {Schmitt}},\ }\href {\doibase 10.1103/PhysRevX.11.031013} {\bibfield
  {journal} {\bibinfo  {journal} {Phys. Rev. X}\ }\textbf {\bibinfo {volume}
  {11}},\ \bibinfo {pages} {031013} (\bibinfo {year} {2021})}\BibitemShut
  {NoStop}%
\bibitem [{\citenamefont {Geondzhian}\ and\ \citenamefont
  {Gilmore}(2018)}]{Geondzhian2015}%
  \BibitemOpen
  \bibfield  {author} {\bibinfo {author} {\bibfnamefont {A.}~\bibnamefont
  {Geondzhian}}\ and\ \bibinfo {author} {\bibfnamefont {K.}~\bibnamefont
  {Gilmore}},\ }\href {\doibase 10.1103/PhysRevB.98.214305} {\bibfield
  {journal} {\bibinfo  {journal} {Phys. Rev. B}\ }\textbf {\bibinfo {volume}
  {98}},\ \bibinfo {pages} {214305} (\bibinfo {year} {2018})}\BibitemShut
  {NoStop}%
\bibitem [{\citenamefont {Dashwood}\ \emph {et~al.}(2021)\citenamefont
  {Dashwood}, \citenamefont {Geondzhian}, \citenamefont {Vale}, \citenamefont
  {Pakpour-Tabrizi}, \citenamefont {Howard}, \citenamefont {Faure},
  \citenamefont {Veiga}, \citenamefont {Meyers}, \citenamefont
  {Chiuzb\ifmmode~\u{a}\else \u{a}\fi{}ian}, \citenamefont {Nicolaou},
  \citenamefont {Jaouen}, \citenamefont {Jackman}, \citenamefont {Nag},
  \citenamefont {Garc\'{\i}a-Fern\'andez}, \citenamefont {Zhou}, \citenamefont
  {Walters}, \citenamefont {Gilmore}, \citenamefont {McMorrow},\ and\
  \citenamefont {Dean}}]{Dashwood2021}%
  \BibitemOpen
  \bibfield  {author} {\bibinfo {author} {\bibfnamefont {C.~D.}\ \bibnamefont
  {Dashwood}}, \bibinfo {author} {\bibfnamefont {A.}~\bibnamefont
  {Geondzhian}}, \bibinfo {author} {\bibfnamefont {J.~G.}\ \bibnamefont
  {Vale}}, \bibinfo {author} {\bibfnamefont {A.~C.}\ \bibnamefont
  {Pakpour-Tabrizi}}, \bibinfo {author} {\bibfnamefont {C.~A.}\ \bibnamefont
  {Howard}}, \bibinfo {author} {\bibfnamefont {Q.}~\bibnamefont {Faure}},
  \bibinfo {author} {\bibfnamefont {L.~S.~I.}\ \bibnamefont {Veiga}}, \bibinfo
  {author} {\bibfnamefont {D.}~\bibnamefont {Meyers}}, \bibinfo {author}
  {\bibfnamefont {S.~G.}\ \bibnamefont {Chiuzb\ifmmode~\u{a}\else
  \u{a}\fi{}ian}}, \bibinfo {author} {\bibfnamefont {A.}~\bibnamefont
  {Nicolaou}}, \bibinfo {author} {\bibfnamefont {N.}~\bibnamefont {Jaouen}},
  \bibinfo {author} {\bibfnamefont {R.~B.}\ \bibnamefont {Jackman}}, \bibinfo
  {author} {\bibfnamefont {A.}~\bibnamefont {Nag}}, \bibinfo {author}
  {\bibfnamefont {M.}~\bibnamefont {Garc\'{\i}a-Fern\'andez}}, \bibinfo
  {author} {\bibfnamefont {K.-J.}\ \bibnamefont {Zhou}}, \bibinfo {author}
  {\bibfnamefont {A.~C.}\ \bibnamefont {Walters}}, \bibinfo {author}
  {\bibfnamefont {K.}~\bibnamefont {Gilmore}}, \bibinfo {author} {\bibfnamefont
  {D.~F.}\ \bibnamefont {McMorrow}}, \ and\ \bibinfo {author} {\bibfnamefont
  {M.~P.~M.}\ \bibnamefont {Dean}},\ }\href {\doibase
  10.1103/PhysRevX.11.041052} {\bibfield  {journal} {\bibinfo  {journal} {Phys.
  Rev. X}\ }\textbf {\bibinfo {volume} {11}},\ \bibinfo {pages} {041052}
  (\bibinfo {year} {2021})}\BibitemShut {NoStop}%
\bibitem [{\citenamefont {Vinson}(2022)}]{Vinson2022}%
  \BibitemOpen
  \bibfield  {author} {\bibinfo {author} {\bibfnamefont {J.}~\bibnamefont
  {Vinson}},\ }\href {\doibase 10.1039/D2CP01030E} {\bibfield  {journal}
  {\bibinfo  {journal} {Phys. Chem. Chem. Phys.}\ }\textbf {\bibinfo {volume}
  {24}},\ \bibinfo {pages} {12787} (\bibinfo {year} {2022})}\BibitemShut
  {NoStop}%
\bibitem [{\citenamefont {Olovsson}\ \emph
  {et~al.}(2009{\natexlab{a}})\citenamefont {Olovsson}, \citenamefont {Tanaka},
  \citenamefont {Puschnig},\ and\ \citenamefont
  {Ambrosch-Draxl}}]{Olovsson2009}%
  \BibitemOpen
  \bibfield  {author} {\bibinfo {author} {\bibfnamefont {W.}~\bibnamefont
  {Olovsson}}, \bibinfo {author} {\bibfnamefont {I.}~\bibnamefont {Tanaka}},
  \bibinfo {author} {\bibfnamefont {P.}~\bibnamefont {Puschnig}}, \ and\
  \bibinfo {author} {\bibfnamefont {C.}~\bibnamefont {Ambrosch-Draxl}},\ }\href
  {\doibase 10.1088/0953-8984/21/10/104205} {\bibfield  {journal} {\bibinfo
  {journal} {Journal of Physics: Condensed Matter}\ }\textbf {\bibinfo {volume}
  {21}},\ \bibinfo {pages} {104205} (\bibinfo {year}
  {2009}{\natexlab{a}})}\BibitemShut {NoStop}%
\bibitem [{\citenamefont {Olovsson}\ \emph
  {et~al.}(2009{\natexlab{b}})\citenamefont {Olovsson}, \citenamefont {Tanaka},
  \citenamefont {Mizoguchi}, \citenamefont {Puschnig},\ and\ \citenamefont
  {Ambrosch-Draxl}}]{Olovsson2009b}%
  \BibitemOpen
  \bibfield  {author} {\bibinfo {author} {\bibfnamefont {W.}~\bibnamefont
  {Olovsson}}, \bibinfo {author} {\bibfnamefont {I.}~\bibnamefont {Tanaka}},
  \bibinfo {author} {\bibfnamefont {T.}~\bibnamefont {Mizoguchi}}, \bibinfo
  {author} {\bibfnamefont {P.}~\bibnamefont {Puschnig}}, \ and\ \bibinfo
  {author} {\bibfnamefont {C.}~\bibnamefont {Ambrosch-Draxl}},\ }\href
  {\doibase 10.1103/PhysRevB.79.041102} {\bibfield  {journal} {\bibinfo
  {journal} {Phys. Rev. B}\ }\textbf {\bibinfo {volume} {79}},\ \bibinfo
  {pages} {041102} (\bibinfo {year} {2009}{\natexlab{b}})}\BibitemShut
  {NoStop}%
\bibitem [{\citenamefont {Olovsson}\ \emph {et~al.}(2011)\citenamefont
  {Olovsson}, \citenamefont {Tanaka}, \citenamefont {Mizoguchi}, \citenamefont
  {Radtke}, \citenamefont {Puschnig},\ and\ \citenamefont
  {Ambrosch-Draxl}}]{Olovsson2011}%
  \BibitemOpen
  \bibfield  {author} {\bibinfo {author} {\bibfnamefont {W.}~\bibnamefont
  {Olovsson}}, \bibinfo {author} {\bibfnamefont {I.}~\bibnamefont {Tanaka}},
  \bibinfo {author} {\bibfnamefont {T.}~\bibnamefont {Mizoguchi}}, \bibinfo
  {author} {\bibfnamefont {G.}~\bibnamefont {Radtke}}, \bibinfo {author}
  {\bibfnamefont {P.}~\bibnamefont {Puschnig}}, \ and\ \bibinfo {author}
  {\bibfnamefont {C.}~\bibnamefont {Ambrosch-Draxl}},\ }\href {\doibase
  10.1103/PhysRevB.83.195206} {\bibfield  {journal} {\bibinfo  {journal} {Phys.
  Rev. B}\ }\textbf {\bibinfo {volume} {83}},\ \bibinfo {pages} {195206}
  (\bibinfo {year} {2011})}\BibitemShut {NoStop}%
\bibitem [{\citenamefont {Vorwerk}\ \emph {et~al.}(2017)\citenamefont
  {Vorwerk}, \citenamefont {Cocchi},\ and\ \citenamefont
  {Draxl}}]{Vorwerk2017}%
  \BibitemOpen
  \bibfield  {author} {\bibinfo {author} {\bibfnamefont {C.}~\bibnamefont
  {Vorwerk}}, \bibinfo {author} {\bibfnamefont {C.}~\bibnamefont {Cocchi}}, \
  and\ \bibinfo {author} {\bibfnamefont {C.}~\bibnamefont {Draxl}},\ }\href
  {\doibase 10.1103/PhysRevB.95.155121} {\bibfield  {journal} {\bibinfo
  {journal} {Phys. Rev. B}\ }\textbf {\bibinfo {volume} {95}},\ \bibinfo
  {pages} {155121} (\bibinfo {year} {2017})}\BibitemShut {NoStop}%
\bibitem [{\citenamefont {Vorwerk}\ \emph {et~al.}(2019)\citenamefont
  {Vorwerk}, \citenamefont {Aurich}, \citenamefont {Cocchi},\ and\
  \citenamefont {Draxl}}]{Vorwerk2019}%
  \BibitemOpen
  \bibfield  {author} {\bibinfo {author} {\bibfnamefont {C.}~\bibnamefont
  {Vorwerk}}, \bibinfo {author} {\bibfnamefont {B.}~\bibnamefont {Aurich}},
  \bibinfo {author} {\bibfnamefont {C.}~\bibnamefont {Cocchi}}, \ and\ \bibinfo
  {author} {\bibfnamefont {C.}~\bibnamefont {Draxl}},\ }\href {\doibase
  10.1088/2516-1075/ab3123} {\bibfield  {journal} {\bibinfo  {journal}
  {Electronic Structure}\ }\textbf {\bibinfo {volume} {1}},\ \bibinfo {pages}
  {037001} (\bibinfo {year} {2019})}\BibitemShut {NoStop}%
\bibitem [{\citenamefont {Vorwerk}\ \emph {et~al.}(2020)\citenamefont
  {Vorwerk}, \citenamefont {Sottile},\ and\ \citenamefont
  {Draxl}}]{Vorwerk2020}%
  \BibitemOpen
  \bibfield  {author} {\bibinfo {author} {\bibfnamefont {C.}~\bibnamefont
  {Vorwerk}}, \bibinfo {author} {\bibfnamefont {F.}~\bibnamefont {Sottile}}, \
  and\ \bibinfo {author} {\bibfnamefont {C.}~\bibnamefont {Draxl}},\ }\href
  {\doibase 10.1103/PhysRevResearch.2.042003} {\bibfield  {journal} {\bibinfo
  {journal} {Phys. Rev. Research}\ }\textbf {\bibinfo {volume} {2}},\ \bibinfo
  {pages} {042003} (\bibinfo {year} {2020})}\BibitemShut {NoStop}%
\bibitem [{\citenamefont {Laskowski}\ and\ \citenamefont
  {Blaha}(2010)}]{Laskowski2010}%
  \BibitemOpen
  \bibfield  {author} {\bibinfo {author} {\bibfnamefont {R.}~\bibnamefont
  {Laskowski}}\ and\ \bibinfo {author} {\bibfnamefont {P.}~\bibnamefont
  {Blaha}},\ }\href {\doibase 10.1103/PhysRevB.82.205104} {\bibfield  {journal}
  {\bibinfo  {journal} {Phys. Rev. B}\ }\textbf {\bibinfo {volume} {82}},\
  \bibinfo {pages} {205104} (\bibinfo {year} {2010})}\BibitemShut {NoStop}%
\bibitem [{\citenamefont {Yao}\ \emph {et~al.}(2022)\citenamefont {Yao},
  \citenamefont {Golze}, \citenamefont {Rinke}, \citenamefont {Blum},\ and\
  \citenamefont {Kanai}}]{Yao2022}%
  \BibitemOpen
  \bibfield  {author} {\bibinfo {author} {\bibfnamefont {Y.}~\bibnamefont
  {Yao}}, \bibinfo {author} {\bibfnamefont {D.}~\bibnamefont {Golze}}, \bibinfo
  {author} {\bibfnamefont {P.}~\bibnamefont {Rinke}}, \bibinfo {author}
  {\bibfnamefont {V.}~\bibnamefont {Blum}}, \ and\ \bibinfo {author}
  {\bibfnamefont {Y.}~\bibnamefont {Kanai}},\ }\href {\doibase
  10.1021/acs.jctc.1c01180} {\bibfield  {journal} {\bibinfo  {journal} {Journal
  of Chemical Theory and Computation}\ }\textbf {\bibinfo {volume} {18}},\
  \bibinfo {pages} {1569} (\bibinfo {year} {2022})}\BibitemShut {NoStop}%
\bibitem [{\citenamefont {Vorwerk}\ \emph {et~al.}(2022)\citenamefont
  {Vorwerk}, \citenamefont {Sottile},\ and\ \citenamefont
  {Draxl}}]{Vorwerk_2022}%
  \BibitemOpen
  \bibfield  {author} {\bibinfo {author} {\bibfnamefont {C.}~\bibnamefont
  {Vorwerk}}, \bibinfo {author} {\bibfnamefont {F.}~\bibnamefont {Sottile}}, \
  and\ \bibinfo {author} {\bibfnamefont {C.}~\bibnamefont {Draxl}},\ }\href
  {\doibase 10.1039/D2CP00994C} {\bibfield  {journal} {\bibinfo  {journal}
  {Phys. Chem. Chem. Phys.}\ }\textbf {\bibinfo {volume} {24}},\ \bibinfo
  {pages} {17439} (\bibinfo {year} {2022})}\BibitemShut {NoStop}%
\bibitem [{\citenamefont {Unzog}\ \emph {et~al.}(2022)\citenamefont {Unzog},
  \citenamefont {Tal},\ and\ \citenamefont {Kresse}}]{Unzog2022}%
  \BibitemOpen
  \bibfield  {author} {\bibinfo {author} {\bibfnamefont {M.}~\bibnamefont
  {Unzog}}, \bibinfo {author} {\bibfnamefont {A.}~\bibnamefont {Tal}}, \ and\
  \bibinfo {author} {\bibfnamefont {G.}~\bibnamefont {Kresse}},\ }\href
  {\doibase 10.1103/PhysRevB.106.155133} {\bibfield  {journal} {\bibinfo
  {journal} {Phys. Rev. B}\ }\textbf {\bibinfo {volume} {106}},\ \bibinfo
  {pages} {155133} (\bibinfo {year} {2022})}\BibitemShut {NoStop}%
\bibitem [{\citenamefont {Adler}(1962)}]{Adler1962}%
  \BibitemOpen
  \bibfield  {author} {\bibinfo {author} {\bibfnamefont {S.~L.}\ \bibnamefont
  {Adler}},\ }\href {\doibase 10.1103/PhysRev.126.413} {\bibfield  {journal}
  {\bibinfo  {journal} {Phys. Rev.}\ }\textbf {\bibinfo {volume} {126}},\
  \bibinfo {pages} {413} (\bibinfo {year} {1962})}\BibitemShut {NoStop}%
\bibitem [{\citenamefont {Wiser}(1963)}]{Wiser1963}%
  \BibitemOpen
  \bibfield  {author} {\bibinfo {author} {\bibfnamefont {N.}~\bibnamefont
  {Wiser}},\ }\href {\doibase 10.1103/PhysRev.129.62} {\bibfield  {journal}
  {\bibinfo  {journal} {Phys. Rev.}\ }\textbf {\bibinfo {volume} {129}},\
  \bibinfo {pages} {62} (\bibinfo {year} {1963})}\BibitemShut {NoStop}%
\bibitem [{Note2()}]{Note2}%
  \BibitemOpen
  \bibinfo {note} {There is also the possibility to include $-W$ and exclude
  $\protect \bar v_c$, which corresponds to the description of triplet
  excitations.}\BibitemShut {Stop}%
\bibitem [{\citenamefont {Gonze}\ \emph {et~al.}(2016)\citenamefont {Gonze},
  \citenamefont {Jollet}, \citenamefont {Abreu~Araujo}, \citenamefont {Adams},
  \citenamefont {Amadon}, \citenamefont {Applencourt}, \citenamefont {Audouze},
  \citenamefont {Beuken}, \citenamefont {Bieder}, \citenamefont {Bokhanchuk},
  \citenamefont {Bousquet}, \citenamefont {Bruneval}, \citenamefont {Caliste},
  \citenamefont {Côté}, \citenamefont {Dahm}, \citenamefont {Da~Pieve},
  \citenamefont {Delaveau}, \citenamefont {Di~Gennaro}, \citenamefont {Dorado},
  \citenamefont {Espejo}, \citenamefont {Geneste}, \citenamefont {Genovese},
  \citenamefont {Gerossier}, \citenamefont {Giantomassi}, \citenamefont
  {Gillet}, \citenamefont {Hamann}, \citenamefont {He}, \citenamefont {Jomard},
  \citenamefont {Laflamme~Janssen}, \citenamefont {Le~Roux}, \citenamefont
  {Levitt}, \citenamefont {Lherbier}, \citenamefont {Liu}, \citenamefont
  {Lukačević}, \citenamefont {Martin}, \citenamefont {Martins}, \citenamefont
  {Oliveira}, \citenamefont {Poncé}, \citenamefont {Pouillon}, \citenamefont
  {Rangel}, \citenamefont {Rignanese}, \citenamefont {Romero}, \citenamefont
  {Rousseau}, \citenamefont {Rubel}, \citenamefont {Shukri}, \citenamefont
  {Stankovski}, \citenamefont {Torrent}, \citenamefont {Van~Setten},
  \citenamefont {Van~Troeye}, \citenamefont {Verstraete}, \citenamefont
  {Waroquiers}, \citenamefont {Wiktor}, \citenamefont {Xu}, \citenamefont
  {Zhou},\ and\ \citenamefont {Zwanziger}}]{Gonze2016}%
  \BibitemOpen
  \bibfield  {author} {\bibinfo {author} {\bibfnamefont {X.}~\bibnamefont
  {Gonze}}, \bibinfo {author} {\bibfnamefont {F.}~\bibnamefont {Jollet}},
  \bibinfo {author} {\bibfnamefont {F.}~\bibnamefont {Abreu~Araujo}}, \bibinfo
  {author} {\bibfnamefont {D.}~\bibnamefont {Adams}}, \bibinfo {author}
  {\bibfnamefont {B.}~\bibnamefont {Amadon}}, \bibinfo {author} {\bibfnamefont
  {T.}~\bibnamefont {Applencourt}}, \bibinfo {author} {\bibfnamefont
  {C.}~\bibnamefont {Audouze}}, \bibinfo {author} {\bibfnamefont {J.-M.}\
  \bibnamefont {Beuken}}, \bibinfo {author} {\bibfnamefont {J.}~\bibnamefont
  {Bieder}}, \bibinfo {author} {\bibfnamefont {A.}~\bibnamefont {Bokhanchuk}},
  \bibinfo {author} {\bibfnamefont {E.}~\bibnamefont {Bousquet}}, \bibinfo
  {author} {\bibfnamefont {F.}~\bibnamefont {Bruneval}}, \bibinfo {author}
  {\bibfnamefont {D.}~\bibnamefont {Caliste}}, \bibinfo {author} {\bibfnamefont
  {M.}~\bibnamefont {Côté}}, \bibinfo {author} {\bibfnamefont
  {F.}~\bibnamefont {Dahm}}, \bibinfo {author} {\bibfnamefont {F.}~\bibnamefont
  {Da~Pieve}}, \bibinfo {author} {\bibfnamefont {M.}~\bibnamefont {Delaveau}},
  \bibinfo {author} {\bibfnamefont {M.}~\bibnamefont {Di~Gennaro}}, \bibinfo
  {author} {\bibfnamefont {B.}~\bibnamefont {Dorado}}, \bibinfo {author}
  {\bibfnamefont {C.}~\bibnamefont {Espejo}}, \bibinfo {author} {\bibfnamefont
  {G.}~\bibnamefont {Geneste}}, \bibinfo {author} {\bibfnamefont
  {L.}~\bibnamefont {Genovese}}, \bibinfo {author} {\bibfnamefont
  {A.}~\bibnamefont {Gerossier}}, \bibinfo {author} {\bibfnamefont
  {M.}~\bibnamefont {Giantomassi}}, \bibinfo {author} {\bibfnamefont
  {Y.}~\bibnamefont {Gillet}}, \bibinfo {author} {\bibfnamefont
  {D.}~\bibnamefont {Hamann}}, \bibinfo {author} {\bibfnamefont
  {L.}~\bibnamefont {He}}, \bibinfo {author} {\bibfnamefont {G.}~\bibnamefont
  {Jomard}}, \bibinfo {author} {\bibfnamefont {J.}~\bibnamefont
  {Laflamme~Janssen}}, \bibinfo {author} {\bibfnamefont {S.}~\bibnamefont
  {Le~Roux}}, \bibinfo {author} {\bibfnamefont {A.}~\bibnamefont {Levitt}},
  \bibinfo {author} {\bibfnamefont {A.}~\bibnamefont {Lherbier}}, \bibinfo
  {author} {\bibfnamefont {F.}~\bibnamefont {Liu}}, \bibinfo {author}
  {\bibfnamefont {I.}~\bibnamefont {Lukačević}}, \bibinfo {author}
  {\bibfnamefont {A.}~\bibnamefont {Martin}}, \bibinfo {author} {\bibfnamefont
  {C.}~\bibnamefont {Martins}}, \bibinfo {author} {\bibfnamefont
  {M.}~\bibnamefont {Oliveira}}, \bibinfo {author} {\bibfnamefont
  {S.}~\bibnamefont {Poncé}}, \bibinfo {author} {\bibfnamefont
  {Y.}~\bibnamefont {Pouillon}}, \bibinfo {author} {\bibfnamefont
  {T.}~\bibnamefont {Rangel}}, \bibinfo {author} {\bibfnamefont {G.-M.}\
  \bibnamefont {Rignanese}}, \bibinfo {author} {\bibfnamefont {A.}~\bibnamefont
  {Romero}}, \bibinfo {author} {\bibfnamefont {B.}~\bibnamefont {Rousseau}},
  \bibinfo {author} {\bibfnamefont {O.}~\bibnamefont {Rubel}}, \bibinfo
  {author} {\bibfnamefont {A.}~\bibnamefont {Shukri}}, \bibinfo {author}
  {\bibfnamefont {M.}~\bibnamefont {Stankovski}}, \bibinfo {author}
  {\bibfnamefont {M.}~\bibnamefont {Torrent}}, \bibinfo {author} {\bibfnamefont
  {M.}~\bibnamefont {Van~Setten}}, \bibinfo {author} {\bibfnamefont
  {B.}~\bibnamefont {Van~Troeye}}, \bibinfo {author} {\bibfnamefont
  {M.}~\bibnamefont {Verstraete}}, \bibinfo {author} {\bibfnamefont
  {D.}~\bibnamefont {Waroquiers}}, \bibinfo {author} {\bibfnamefont
  {J.}~\bibnamefont {Wiktor}}, \bibinfo {author} {\bibfnamefont
  {B.}~\bibnamefont {Xu}}, \bibinfo {author} {\bibfnamefont {A.}~\bibnamefont
  {Zhou}}, \ and\ \bibinfo {author} {\bibfnamefont {J.}~\bibnamefont
  {Zwanziger}},\ }\href {\doibase 10.1016/j.cpc.2016.04.003} {\bibfield
  {journal} {\bibinfo  {journal} {Comput. Phys. Commun.}\ }\textbf {\bibinfo
  {volume} {205}},\ \bibinfo {pages} {106} (\bibinfo {year}
  {2016})}\BibitemShut {NoStop}%
\bibitem [{\citenamefont {Reining}\ \emph {et~al.}()\citenamefont {Reining},
  \citenamefont {Olevano}, \citenamefont {Sottile}, \citenamefont {Albrecht},\
  and\ \citenamefont {Onida}}]{EXCcode}%
  \BibitemOpen
  \bibfield  {author} {\bibinfo {author} {\bibfnamefont {L.}~\bibnamefont
  {Reining}}, \bibinfo {author} {\bibfnamefont {V.}~\bibnamefont {Olevano}},
  \bibinfo {author} {\bibfnamefont {F.}~\bibnamefont {Sottile}}, \bibinfo
  {author} {\bibfnamefont {S.}~\bibnamefont {Albrecht}}, \ and\ \bibinfo
  {author} {\bibfnamefont {G.}~\bibnamefont {Onida}},\ }\href
  {http://www.bethe-salpeter.org/} {\enquote {\bibinfo {title} {The exc
  code},}\ }\bibinfo {howpublished}
  {\url{https://etsf.polytechnique.fr/software/Ab_Initio/}},\ \bibinfo {note}
  {unpublished}\BibitemShut {NoStop}%
\bibitem [{\citenamefont {Gulans}\ \emph {et~al.}(2014)\citenamefont {Gulans},
  \citenamefont {Kontur}, \citenamefont {Meisenbichler}, \citenamefont {Nabok},
  \citenamefont {Pavone}, \citenamefont {Rigamonti}, \citenamefont
  {Sagmeister}, \citenamefont {Werner},\ and\ \citenamefont
  {Draxl}}]{Gulans_2014}%
  \BibitemOpen
  \bibfield  {author} {\bibinfo {author} {\bibfnamefont {A.}~\bibnamefont
  {Gulans}}, \bibinfo {author} {\bibfnamefont {S.}~\bibnamefont {Kontur}},
  \bibinfo {author} {\bibfnamefont {C.}~\bibnamefont {Meisenbichler}}, \bibinfo
  {author} {\bibfnamefont {D.}~\bibnamefont {Nabok}}, \bibinfo {author}
  {\bibfnamefont {P.}~\bibnamefont {Pavone}}, \bibinfo {author} {\bibfnamefont
  {S.}~\bibnamefont {Rigamonti}}, \bibinfo {author} {\bibfnamefont
  {S.}~\bibnamefont {Sagmeister}}, \bibinfo {author} {\bibfnamefont
  {U.}~\bibnamefont {Werner}}, \ and\ \bibinfo {author} {\bibfnamefont
  {C.}~\bibnamefont {Draxl}},\ }\href {\doibase 10.1088/0953-8984/26/36/363202}
  {\bibfield  {journal} {\bibinfo  {journal} {Journal of Physics: Condensed
  Matter}\ }\textbf {\bibinfo {volume} {26}},\ \bibinfo {pages} {363202}
  (\bibinfo {year} {2014})}\BibitemShut {NoStop}%
\bibitem [{\citenamefont {Kohn}\ and\ \citenamefont {Sham}(1965)}]{Kohn1965}%
  \BibitemOpen
  \bibfield  {author} {\bibinfo {author} {\bibfnamefont {W.}~\bibnamefont
  {Kohn}}\ and\ \bibinfo {author} {\bibfnamefont {L.~J.}\ \bibnamefont
  {Sham}},\ }\href {\doibase 10.1103/PhysRev.140.A1133} {\bibfield  {journal}
  {\bibinfo  {journal} {Phys. Rev.}\ }\textbf {\bibinfo {volume} {140}},\
  \bibinfo {pages} {A1133} (\bibinfo {year} {1965})}\BibitemShut {NoStop}%
\bibitem [{\citenamefont {Troullier}\ and\ \citenamefont
  {Martins}(1991)}]{Troullier_1991}%
  \BibitemOpen
  \bibfield  {author} {\bibinfo {author} {\bibfnamefont {N.}~\bibnamefont
  {Troullier}}\ and\ \bibinfo {author} {\bibfnamefont {J.~L.}\ \bibnamefont
  {Martins}},\ }\href {\doibase 10.1103/PhysRevB.43.1993} {\bibfield  {journal}
  {\bibinfo  {journal} {Phys. Rev. B}\ }\textbf {\bibinfo {volume} {43}},\
  \bibinfo {pages} {1993} (\bibinfo {year} {1991})}\BibitemShut {NoStop}%
\bibitem [{\citenamefont {{van Setten}}\ \emph {et~al.}(2018)\citenamefont
  {{van Setten}}, \citenamefont {Giantomassi}, \citenamefont {Bousquet},
  \citenamefont {Verstraete}, \citenamefont {Hamann}, \citenamefont {Gonze},\
  and\ \citenamefont {Rignanese}}]{vanSetten_2018}%
  \BibitemOpen
  \bibfield  {author} {\bibinfo {author} {\bibfnamefont {M.}~\bibnamefont {{van
  Setten}}}, \bibinfo {author} {\bibfnamefont {M.}~\bibnamefont {Giantomassi}},
  \bibinfo {author} {\bibfnamefont {E.}~\bibnamefont {Bousquet}}, \bibinfo
  {author} {\bibfnamefont {M.}~\bibnamefont {Verstraete}}, \bibinfo {author}
  {\bibfnamefont {D.}~\bibnamefont {Hamann}}, \bibinfo {author} {\bibfnamefont
  {X.}~\bibnamefont {Gonze}}, \ and\ \bibinfo {author} {\bibfnamefont {G.-M.}\
  \bibnamefont {Rignanese}},\ }\href {\doibase
  https://doi.org/10.1016/j.cpc.2018.01.012} {\bibfield  {journal} {\bibinfo
  {journal} {Computer Physics Communications}\ }\textbf {\bibinfo {volume}
  {226}},\ \bibinfo {pages} {39} (\bibinfo {year} {2018})}\BibitemShut
  {NoStop}%
\bibitem [{\citenamefont {Zhou}\ \emph {et~al.}(2020)\citenamefont {Zhou},
  \citenamefont {Reining}, \citenamefont {Nicolaou}, \citenamefont {Bendounan},
  \citenamefont {Ruotsalainen}, \citenamefont {Vanzini}, \citenamefont {Kas},
  \citenamefont {Rehr}, \citenamefont {Muntwiler}, \citenamefont {Strocov},
  \citenamefont {Sirotti},\ and\ \citenamefont {Gatti}}]{Zhou_2020}%
  \BibitemOpen
  \bibfield  {author} {\bibinfo {author} {\bibfnamefont {J.~S.}\ \bibnamefont
  {Zhou}}, \bibinfo {author} {\bibfnamefont {L.}~\bibnamefont {Reining}},
  \bibinfo {author} {\bibfnamefont {A.}~\bibnamefont {Nicolaou}}, \bibinfo
  {author} {\bibfnamefont {A.}~\bibnamefont {Bendounan}}, \bibinfo {author}
  {\bibfnamefont {K.}~\bibnamefont {Ruotsalainen}}, \bibinfo {author}
  {\bibfnamefont {M.}~\bibnamefont {Vanzini}}, \bibinfo {author} {\bibfnamefont
  {J.~J.}\ \bibnamefont {Kas}}, \bibinfo {author} {\bibfnamefont {J.~J.}\
  \bibnamefont {Rehr}}, \bibinfo {author} {\bibfnamefont {M.}~\bibnamefont
  {Muntwiler}}, \bibinfo {author} {\bibfnamefont {V.~N.}\ \bibnamefont
  {Strocov}}, \bibinfo {author} {\bibfnamefont {F.}~\bibnamefont {Sirotti}}, \
  and\ \bibinfo {author} {\bibfnamefont {M.}~\bibnamefont {Gatti}},\ }\href
  {\doibase 10.1073/pnas.2012625117} {\bibfield  {journal} {\bibinfo  {journal}
  {Proceedings of the National Academy of Sciences}\ }\textbf {\bibinfo
  {volume} {117}},\ \bibinfo {pages} {28596} (\bibinfo {year}
  {2020})}\BibitemShut {NoStop}%
\bibitem [{\citenamefont {Sturm}\ \emph {et~al.}(1990)\citenamefont {Sturm},
  \citenamefont {Zaremba},\ and\ \citenamefont {Nuroh}}]{Sturm_1990}%
  \BibitemOpen
  \bibfield  {author} {\bibinfo {author} {\bibfnamefont {K.}~\bibnamefont
  {Sturm}}, \bibinfo {author} {\bibfnamefont {E.}~\bibnamefont {Zaremba}}, \
  and\ \bibinfo {author} {\bibfnamefont {K.}~\bibnamefont {Nuroh}},\ }\href
  {\doibase 10.1103/PhysRevB.42.6973} {\bibfield  {journal} {\bibinfo
  {journal} {Phys. Rev. B}\ }\textbf {\bibinfo {volume} {42}},\ \bibinfo
  {pages} {6973} (\bibinfo {year} {1990})}\BibitemShut {NoStop}%
\bibitem [{\citenamefont {Quong}\ and\ \citenamefont
  {Eguiluz}(1993)}]{Quong_1993}%
  \BibitemOpen
  \bibfield  {author} {\bibinfo {author} {\bibfnamefont {A.~A.}\ \bibnamefont
  {Quong}}\ and\ \bibinfo {author} {\bibfnamefont {A.~G.}\ \bibnamefont
  {Eguiluz}},\ }\href {\doibase 10.1103/PhysRevLett.70.3955} {\bibfield
  {journal} {\bibinfo  {journal} {Phys. Rev. Lett.}\ }\textbf {\bibinfo
  {volume} {70}},\ \bibinfo {pages} {3955} (\bibinfo {year}
  {1993})}\BibitemShut {NoStop}%
\bibitem [{\citenamefont {Zhou}\ \emph {et~al.}(2018)\citenamefont {Zhou},
  \citenamefont {Gatti}, \citenamefont {Kas}, \citenamefont {Rehr},\ and\
  \citenamefont {Reining}}]{Zhou_2018}%
  \BibitemOpen
  \bibfield  {author} {\bibinfo {author} {\bibfnamefont {J.~S.}\ \bibnamefont
  {Zhou}}, \bibinfo {author} {\bibfnamefont {M.}~\bibnamefont {Gatti}},
  \bibinfo {author} {\bibfnamefont {J.~J.}\ \bibnamefont {Kas}}, \bibinfo
  {author} {\bibfnamefont {J.~J.}\ \bibnamefont {Rehr}}, \ and\ \bibinfo
  {author} {\bibfnamefont {L.}~\bibnamefont {Reining}},\ }\href {\doibase
  10.1103/PhysRevB.97.035137} {\bibfield  {journal} {\bibinfo  {journal} {Phys.
  Rev. B}\ }\textbf {\bibinfo {volume} {97}},\ \bibinfo {pages} {035137}
  (\bibinfo {year} {2018})}\BibitemShut {NoStop}%
\bibitem [{\citenamefont {Marinopoulos}\ and\ \citenamefont
  {Gr\"uning}(2011)}]{Myrta_2011}%
  \BibitemOpen
  \bibfield  {author} {\bibinfo {author} {\bibfnamefont {A.~G.}\ \bibnamefont
  {Marinopoulos}}\ and\ \bibinfo {author} {\bibfnamefont {M.}~\bibnamefont
  {Gr\"uning}},\ }\href {\doibase 10.1103/PhysRevB.83.195129} {\bibfield
  {journal} {\bibinfo  {journal} {Phys. Rev. B}\ }\textbf {\bibinfo {volume}
  {83}},\ \bibinfo {pages} {195129} (\bibinfo {year} {2011})}\BibitemShut
  {NoStop}%
\bibitem [{\citenamefont {Lorin}\ \emph {et~al.}(2021)\citenamefont {Lorin},
  \citenamefont {Gatti}, \citenamefont {Reining},\ and\ \citenamefont
  {Sottile}}]{Lorin2021}%
  \BibitemOpen
  \bibfield  {author} {\bibinfo {author} {\bibfnamefont {A.}~\bibnamefont
  {Lorin}}, \bibinfo {author} {\bibfnamefont {M.}~\bibnamefont {Gatti}},
  \bibinfo {author} {\bibfnamefont {L.}~\bibnamefont {Reining}}, \ and\
  \bibinfo {author} {\bibfnamefont {F.}~\bibnamefont {Sottile}},\ }\href
  {\doibase 10.1103/PhysRevB.104.235149} {\bibfield  {journal} {\bibinfo
  {journal} {Phys. Rev. B}\ }\textbf {\bibinfo {volume} {104}},\ \bibinfo
  {pages} {235149} (\bibinfo {year} {2021})}\BibitemShut {NoStop}%
\bibitem [{\citenamefont {Newnham}\ and\ \citenamefont
  {Haan}(1962)}]{Newnham_1962}%
  \BibitemOpen
  \bibfield  {author} {\bibinfo {author} {\bibfnamefont {E.~E.}\ \bibnamefont
  {Newnham}}\ and\ \bibinfo {author} {\bibfnamefont {Y.~M.}\ \bibnamefont
  {Haan}},\ }\href {\doibase doi:10.1524/zkri.1962.117.16.235} {\bibfield
  {journal} {\bibinfo  {journal} {Zeitschrift fur Kristallographie -
  Crystalline Materials}\ }\textbf {\bibinfo {volume} {117}},\ \bibinfo {pages}
  {235} (\bibinfo {year} {1962})}\BibitemShut {NoStop}%
\bibitem [{\citenamefont {Mackrodt}\ \emph {et~al.}(2019)\citenamefont
  {Mackrodt}, \citenamefont {Rérat}, \citenamefont {Gentile},\ and\
  \citenamefont {Dovesi}}]{Mackrodt_2020}%
  \BibitemOpen
  \bibfield  {author} {\bibinfo {author} {\bibfnamefont {W.~C.}\ \bibnamefont
  {Mackrodt}}, \bibinfo {author} {\bibfnamefont {M.}~\bibnamefont {Rérat}},
  \bibinfo {author} {\bibfnamefont {F.~S.}\ \bibnamefont {Gentile}}, \ and\
  \bibinfo {author} {\bibfnamefont {R.}~\bibnamefont {Dovesi}},\ }\href
  {\doibase 10.1088/1361-648X/ab4c0e} {\bibfield  {journal} {\bibinfo
  {journal} {Journal of Physics: Condensed Matter}\ }\textbf {\bibinfo {volume}
  {32}},\ \bibinfo {pages} {085901} (\bibinfo {year} {2019})}\BibitemShut
  {NoStop}%
\bibitem [{\citenamefont {Ahuja}\ \emph {et~al.}(2004)\citenamefont {Ahuja},
  \citenamefont {Osorio-Guillen}, \citenamefont {de~Almeida}, \citenamefont
  {Holm}, \citenamefont {Ching},\ and\ \citenamefont {Johansson}}]{Ahuja_2004}%
  \BibitemOpen
  \bibfield  {author} {\bibinfo {author} {\bibfnamefont {R.}~\bibnamefont
  {Ahuja}}, \bibinfo {author} {\bibfnamefont {J.~M.}\ \bibnamefont
  {Osorio-Guillen}}, \bibinfo {author} {\bibfnamefont {J.~S.}\ \bibnamefont
  {de~Almeida}}, \bibinfo {author} {\bibfnamefont {B.}~\bibnamefont {Holm}},
  \bibinfo {author} {\bibfnamefont {W.~Y.}\ \bibnamefont {Ching}}, \ and\
  \bibinfo {author} {\bibfnamefont {B.}~\bibnamefont {Johansson}},\ }\href
  {\doibase 10.1088/0953-8984/16/16/013} {\bibfield  {journal} {\bibinfo
  {journal} {Journal of Physics: Condensed Matter}\ }\textbf {\bibinfo {volume}
  {16}},\ \bibinfo {pages} {2891} (\bibinfo {year} {2004})}\BibitemShut
  {NoStop}%
\bibitem [{\citenamefont {Santos}\ \emph {et~al.}(2015)\citenamefont {Santos},
  \citenamefont {Longhinotti}, \citenamefont {Freire}, \citenamefont
  {Reimberg},\ and\ \citenamefont {Caetano}}]{Santos_2015}%
  \BibitemOpen
  \bibfield  {author} {\bibinfo {author} {\bibfnamefont {R.}~\bibnamefont
  {Santos}}, \bibinfo {author} {\bibfnamefont {E.}~\bibnamefont {Longhinotti}},
  \bibinfo {author} {\bibfnamefont {V.}~\bibnamefont {Freire}}, \bibinfo
  {author} {\bibfnamefont {R.}~\bibnamefont {Reimberg}}, \ and\ \bibinfo
  {author} {\bibfnamefont {E.}~\bibnamefont {Caetano}},\ }\href {\doibase
  https://doi.org/10.1016/j.cplett.2015.08.004} {\bibfield  {journal} {\bibinfo
   {journal} {Chemical Physics Letters}\ }\textbf {\bibinfo {volume} {637}},\
  \bibinfo {pages} {172} (\bibinfo {year} {2015})}\BibitemShut {NoStop}%
\bibitem [{\citenamefont {Will}\ \emph {et~al.}(1992)\citenamefont {Will},
  \citenamefont {DeLorenzi},\ and\ \citenamefont {Janora}}]{Will_1992}%
  \BibitemOpen
  \bibfield  {author} {\bibinfo {author} {\bibfnamefont {F.~G.}\ \bibnamefont
  {Will}}, \bibinfo {author} {\bibfnamefont {H.~G.}\ \bibnamefont {DeLorenzi}},
  \ and\ \bibinfo {author} {\bibfnamefont {K.~H.}\ \bibnamefont {Janora}},\
  }\href {\doibase https://doi.org/10.1111/j.1151-2916.1992.tb08179.x}
  {\bibfield  {journal} {\bibinfo  {journal} {Journal of the American Ceramic
  Society}\ }\textbf {\bibinfo {volume} {75}},\ \bibinfo {pages} {295}
  (\bibinfo {year} {1992})}\BibitemShut {NoStop}%
\bibitem [{\citenamefont {Crist}(2004)}]{Crist_2004}%
  \BibitemOpen
  \bibfield  {author} {\bibinfo {author} {\bibfnamefont {B.}~\bibnamefont
  {Crist}},\ }\href@noop {} {\emph {\bibinfo {title} {Handbooks of
  Monochromatic XPS Spectra: Volume 2 : Commercially Pure Binary Oxides}}}\
  (\bibinfo  {publisher} {XPS International LLC},\ \bibinfo {year}
  {2004})\BibitemShut {NoStop}%
\bibitem [{\citenamefont {Gonze}\ \emph {et~al.}(1991)\citenamefont {Gonze},
  \citenamefont {Stumpf},\ and\ \citenamefont {Scheffler}}]{Gonze1991}%
  \BibitemOpen
  \bibfield  {author} {\bibinfo {author} {\bibfnamefont {X.}~\bibnamefont
  {Gonze}}, \bibinfo {author} {\bibfnamefont {R.}~\bibnamefont {Stumpf}}, \
  and\ \bibinfo {author} {\bibfnamefont {M.}~\bibnamefont {Scheffler}},\ }\href
  {\doibase 10.1103/PhysRevB.44.8503} {\bibfield  {journal} {\bibinfo
  {journal} {Phys. Rev. B}\ }\textbf {\bibinfo {volume} {44}},\ \bibinfo
  {pages} {8503} (\bibinfo {year} {1991})}\BibitemShut {NoStop}%
\bibitem [{\citenamefont {Puschnig}\ and\ \citenamefont
  {Ambrosch-Draxl}(2002)}]{Draxl-Puschnig_2002}%
  \BibitemOpen
  \bibfield  {author} {\bibinfo {author} {\bibfnamefont {P.}~\bibnamefont
  {Puschnig}}\ and\ \bibinfo {author} {\bibfnamefont {C.}~\bibnamefont
  {Ambrosch-Draxl}},\ }\href {\doibase 10.1103/PhysRevB.66.165105} {\bibfield
  {journal} {\bibinfo  {journal} {Phys. Rev. B}\ }\textbf {\bibinfo {volume}
  {66}},\ \bibinfo {pages} {165105} (\bibinfo {year} {2002})}\BibitemShut
  {NoStop}%
\bibitem [{\citenamefont {Rangel}\ \emph {et~al.}(2020)\citenamefont {Rangel},
  \citenamefont {Ben}, \citenamefont {Varsano}, \citenamefont {Antonius},
  \citenamefont {Bruneval}, \citenamefont {da~Jornada}, \citenamefont {van
  Setten}, \citenamefont {Orhan}, \citenamefont {O'Regan}, \citenamefont
  {Canning}, \citenamefont {Ferretti}, \citenamefont {Marini}, \citenamefont
  {Rignanese}, \citenamefont {Deslippe}, \citenamefont {Louie},\ and\
  \citenamefont {Neaton}}]{Rangel_2020}%
  \BibitemOpen
  \bibfield  {author} {\bibinfo {author} {\bibfnamefont {T.}~\bibnamefont
  {Rangel}}, \bibinfo {author} {\bibfnamefont {M.~D.}\ \bibnamefont {Ben}},
  \bibinfo {author} {\bibfnamefont {D.}~\bibnamefont {Varsano}}, \bibinfo
  {author} {\bibfnamefont {G.}~\bibnamefont {Antonius}}, \bibinfo {author}
  {\bibfnamefont {F.}~\bibnamefont {Bruneval}}, \bibinfo {author}
  {\bibfnamefont {F.~H.}\ \bibnamefont {da~Jornada}}, \bibinfo {author}
  {\bibfnamefont {M.~J.}\ \bibnamefont {van Setten}}, \bibinfo {author}
  {\bibfnamefont {O.~K.}\ \bibnamefont {Orhan}}, \bibinfo {author}
  {\bibfnamefont {D.~D.}\ \bibnamefont {O'Regan}}, \bibinfo {author}
  {\bibfnamefont {A.}~\bibnamefont {Canning}}, \bibinfo {author} {\bibfnamefont
  {A.}~\bibnamefont {Ferretti}}, \bibinfo {author} {\bibfnamefont
  {A.}~\bibnamefont {Marini}}, \bibinfo {author} {\bibfnamefont {G.-M.}\
  \bibnamefont {Rignanese}}, \bibinfo {author} {\bibfnamefont {J.}~\bibnamefont
  {Deslippe}}, \bibinfo {author} {\bibfnamefont {S.~G.}\ \bibnamefont {Louie}},
  \ and\ \bibinfo {author} {\bibfnamefont {J.~B.}\ \bibnamefont {Neaton}},\
  }\href@noop {} {\bibfield  {journal} {\bibinfo  {journal} {Computer Physics
  Communications}\ }\textbf {\bibinfo {volume} {255}},\ \bibinfo {pages}
  {107242} (\bibinfo {year} {2020})}\BibitemShut {NoStop}%
\bibitem [{\citenamefont {Albrecht}(1999)}]{Albrecht1999}%
  \BibitemOpen
  \bibfield  {author} {\bibinfo {author} {\bibfnamefont {S.}~\bibnamefont
  {Albrecht}},\ }\emph {\bibinfo {title} {Optical Absorption Spectra of
  Semiconductors and Insulators: ab initio calculations of many-body
  effects}},\ \href
  {https://etsf.polytechnique.fr/system/files/thesis_stefan_albrecht.pdf}
  {Ph.D. thesis},\ \bibinfo  {school} {Ecole Polytechnique, Palaiseau}
  (\bibinfo {year} {1999})\BibitemShut {NoStop}%
\bibitem [{\citenamefont {Puschnig}(2002)}]{Puschnig2002}%
  \BibitemOpen
  \bibfield  {author} {\bibinfo {author} {\bibfnamefont {P.}~\bibnamefont
  {Puschnig}},\ }\emph {\bibinfo {title} {Excitonic Effects in Organic
  Semi-Conductors - An Ab-initio Study within the LAPW Method}},\ \href
  {https://www.researchgate.net/profile/Peter-Puschnig/publication/235762448_Excitonic_Effects_in_Organic_Semi-Conductors_-_An_Ab-initio_Study_within_the_LAPW_Method/links/004635208ec44e3018000000/Excitonic-Effects-in-Organic-Semi-Conductors-An-Ab-initio-Study-within-the-LAPW-Method.pdf}
  {Ph.D. thesis} (\bibinfo {year} {2002})\BibitemShut {NoStop}%
\bibitem [{\citenamefont {Fuchs}\ \emph {et~al.}(2008)\citenamefont {Fuchs},
  \citenamefont {R\"odl}, \citenamefont {Schleife},\ and\ \citenamefont
  {Bechstedt}}]{Fuchs2008}%
  \BibitemOpen
  \bibfield  {author} {\bibinfo {author} {\bibfnamefont {F.}~\bibnamefont
  {Fuchs}}, \bibinfo {author} {\bibfnamefont {C.}~\bibnamefont {R\"odl}},
  \bibinfo {author} {\bibfnamefont {A.}~\bibnamefont {Schleife}}, \ and\
  \bibinfo {author} {\bibfnamefont {F.}~\bibnamefont {Bechstedt}},\ }\href
  {\doibase 10.1103/PhysRevB.78.085103} {\bibfield  {journal} {\bibinfo
  {journal} {Phys. Rev. B}\ }\textbf {\bibinfo {volume} {78}},\ \bibinfo
  {pages} {085103} (\bibinfo {year} {2008})}\BibitemShut {NoStop}%
\bibitem [{\citenamefont {Gorelov}\ \emph {et~al.}(2022)\citenamefont
  {Gorelov}, \citenamefont {Reining}, \citenamefont {Lambrecht},\ and\
  \citenamefont {Gatti}}]{Gorelov2023}%
  \BibitemOpen
  \bibfield  {author} {\bibinfo {author} {\bibfnamefont {V.}~\bibnamefont
  {Gorelov}}, \bibinfo {author} {\bibfnamefont {L.}~\bibnamefont {Reining}},
  \bibinfo {author} {\bibfnamefont {W.~R.~L.}\ \bibnamefont {Lambrecht}}, \
  and\ \bibinfo {author} {\bibfnamefont {M.}~\bibnamefont {Gatti}},\ }\href
  {\doibase 10.48550/ARXIV.2211.06285} {\enquote {\bibinfo {title} {Robustness
  of electronic screening effects in electron spectroscopies: example of
  v$_2$o$_5$},}\ } (\bibinfo {year} {2022}),\ \bibinfo {note}
  {https://arxiv.org/abs/2211.06285}\BibitemShut {NoStop}%
\bibitem [{\citenamefont {Goedecker}(1993)}]{muffin-tin_1993}%
  \BibitemOpen
  \bibfield  {author} {\bibinfo {author} {\bibfnamefont {S.}~\bibnamefont
  {Goedecker}},\ }\href {\doibase 10.1103/PhysRevB.47.9881} {\bibfield
  {journal} {\bibinfo  {journal} {Phys. Rev. B}\ }\textbf {\bibinfo {volume}
  {47}},\ \bibinfo {pages} {9881} (\bibinfo {year} {1993})}\BibitemShut
  {NoStop}%
\bibitem [{\citenamefont {Singh}(1991)}]{local-orbital_1991}%
  \BibitemOpen
  \bibfield  {author} {\bibinfo {author} {\bibfnamefont {D.}~\bibnamefont
  {Singh}},\ }\href {\doibase 10.1103/PhysRevB.43.6388} {\bibfield  {journal}
  {\bibinfo  {journal} {Phys. Rev. B}\ }\textbf {\bibinfo {volume} {43}},\
  \bibinfo {pages} {6388} (\bibinfo {year} {1991})}\BibitemShut {NoStop}%
\bibitem [{\citenamefont {Tomiki}\ \emph {et~al.}(1993)\citenamefont {Tomiki},
  \citenamefont {Ganaha}, \citenamefont {Shikenbaru}, \citenamefont {Futemma},
  \citenamefont {Yuri}, \citenamefont {Aiura}, \citenamefont {Sato},
  \citenamefont {Fukutani}, \citenamefont {Kato}, \citenamefont {Miyahara},
  \citenamefont {Yonesu},\ and\ \citenamefont {Tamashiro}}]{Tomiki_1993}%
  \BibitemOpen
  \bibfield  {author} {\bibinfo {author} {\bibfnamefont {T.}~\bibnamefont
  {Tomiki}}, \bibinfo {author} {\bibfnamefont {Y.}~\bibnamefont {Ganaha}},
  \bibinfo {author} {\bibfnamefont {T.}~\bibnamefont {Shikenbaru}}, \bibinfo
  {author} {\bibfnamefont {T.}~\bibnamefont {Futemma}}, \bibinfo {author}
  {\bibfnamefont {M.}~\bibnamefont {Yuri}}, \bibinfo {author} {\bibfnamefont
  {Y.}~\bibnamefont {Aiura}}, \bibinfo {author} {\bibfnamefont
  {S.}~\bibnamefont {Sato}}, \bibinfo {author} {\bibfnamefont {H.}~\bibnamefont
  {Fukutani}}, \bibinfo {author} {\bibfnamefont {H.}~\bibnamefont {Kato}},
  \bibinfo {author} {\bibfnamefont {T.}~\bibnamefont {Miyahara}}, \bibinfo
  {author} {\bibfnamefont {A.}~\bibnamefont {Yonesu}}, \ and\ \bibinfo {author}
  {\bibfnamefont {J.}~\bibnamefont {Tamashiro}},\ }\href {\doibase
  10.1143/JPSJ.62.573} {\bibfield  {journal} {\bibinfo  {journal} {Journal of
  the Physical Society of Japan}\ }\textbf {\bibinfo {volume} {62}},\ \bibinfo
  {pages} {573} (\bibinfo {year} {1993})}\BibitemShut {NoStop}%
\bibitem [{\citenamefont {French}\ \emph {et~al.}(1998)\citenamefont {French},
  \citenamefont {Müllejans},\ and\ \citenamefont {Jones}}]{French_1998}%
  \BibitemOpen
  \bibfield  {author} {\bibinfo {author} {\bibfnamefont {R.~H.}\ \bibnamefont
  {French}}, \bibinfo {author} {\bibfnamefont {H.}~\bibnamefont {Müllejans}},
  \ and\ \bibinfo {author} {\bibfnamefont {D.~J.}\ \bibnamefont {Jones}},\
  }\href {\doibase https://doi.org/10.1111/j.1151-2916.1998.tb02660.x}
  {\bibfield  {journal} {\bibinfo  {journal} {Journal of the American Ceramic
  Society}\ }\textbf {\bibinfo {volume} {81}},\ \bibinfo {pages} {2549}
  (\bibinfo {year} {1998})}\BibitemShut {NoStop}%
\bibitem [{\citenamefont {Weigel}\ \emph {et~al.}(2008)\citenamefont {Weigel},
  \citenamefont {Calas}, \citenamefont {Cormier}, \citenamefont {Galoisy},\
  and\ \citenamefont {Henderson}}]{Weigel_2008}%
  \BibitemOpen
  \bibfield  {author} {\bibinfo {author} {\bibfnamefont {C.}~\bibnamefont
  {Weigel}}, \bibinfo {author} {\bibfnamefont {G.}~\bibnamefont {Calas}},
  \bibinfo {author} {\bibfnamefont {L.}~\bibnamefont {Cormier}}, \bibinfo
  {author} {\bibfnamefont {L.}~\bibnamefont {Galoisy}}, \ and\ \bibinfo
  {author} {\bibfnamefont {G.~S.}\ \bibnamefont {Henderson}},\ }\href {\doibase
  10.1088/0953-8984/20/13/135219} {\bibfield  {journal} {\bibinfo  {journal}
  {Journal of Physics: Condensed Matter}\ }\textbf {\bibinfo {volume} {20}},\
  \bibinfo {pages} {135219} (\bibinfo {year} {2008})}\BibitemShut {NoStop}%
\bibitem [{\citenamefont {van Bokhoven}\ \emph {et~al.}(1999)\citenamefont {van
  Bokhoven}, \citenamefont {Sambe}, \citenamefont {Ramaker},\ and\
  \citenamefont {Koningsberger}}]{vanBokhoven_1999}%
  \BibitemOpen
  \bibfield  {author} {\bibinfo {author} {\bibfnamefont {J.~A.}\ \bibnamefont
  {van Bokhoven}}, \bibinfo {author} {\bibfnamefont {H.}~\bibnamefont {Sambe}},
  \bibinfo {author} {\bibfnamefont {D.~E.}\ \bibnamefont {Ramaker}}, \ and\
  \bibinfo {author} {\bibfnamefont {D.~C.}\ \bibnamefont {Koningsberger}},\
  }\href {\doibase 10.1021/jp990478t} {\bibfield  {journal} {\bibinfo
  {journal} {The Journal of Physical Chemistry B}\ }\textbf {\bibinfo {volume}
  {103}},\ \bibinfo {pages} {7557} (\bibinfo {year} {1999})}\BibitemShut
  {NoStop}%
\bibitem [{\citenamefont {Mizoguchi}\ \emph {et~al.}(2009)\citenamefont
  {Mizoguchi}, \citenamefont {Tanaka}, \citenamefont {Gao},\ and\ \citenamefont
  {Pickard}}]{Mizoguchi_2009}%
  \BibitemOpen
  \bibfield  {author} {\bibinfo {author} {\bibfnamefont {T.}~\bibnamefont
  {Mizoguchi}}, \bibinfo {author} {\bibfnamefont {I.}~\bibnamefont {Tanaka}},
  \bibinfo {author} {\bibfnamefont {S.-P.}\ \bibnamefont {Gao}}, \ and\
  \bibinfo {author} {\bibfnamefont {C.~J.}\ \bibnamefont {Pickard}},\ }\href
  {\doibase 10.1088/0953-8984/21/10/104204} {\bibfield  {journal} {\bibinfo
  {journal} {Journal of Physics: Condensed Matter}\ }\textbf {\bibinfo {volume}
  {21}},\ \bibinfo {pages} {104204} (\bibinfo {year} {2009})}\BibitemShut
  {NoStop}%
\bibitem [{\citenamefont {van Schilfgaarde}\ \emph
  {et~al.}(2006{\natexlab{b}})\citenamefont {van Schilfgaarde}, \citenamefont
  {Kotani},\ and\ \citenamefont {Faleev}}]{vanSchilfgaarde_2006}%
  \BibitemOpen
  \bibfield  {author} {\bibinfo {author} {\bibfnamefont {M.}~\bibnamefont {van
  Schilfgaarde}}, \bibinfo {author} {\bibfnamefont {T.}~\bibnamefont {Kotani}},
  \ and\ \bibinfo {author} {\bibfnamefont {S.}~\bibnamefont {Faleev}},\ }\href
  {\doibase 10.1103/PhysRevLett.96.226402} {\bibfield  {journal} {\bibinfo
  {journal} {Phys. Rev. Lett.}\ }\textbf {\bibinfo {volume} {96}},\ \bibinfo
  {pages} {226402} (\bibinfo {year} {2006}{\natexlab{b}})}\BibitemShut
  {NoStop}%
\bibitem [{\citenamefont {O'Brien}\ \emph {et~al.}(1991)\citenamefont
  {O'Brien}, \citenamefont {Jia}, \citenamefont {Dong}, \citenamefont
  {Callcott}, \citenamefont {Rubensson}, \citenamefont {Mueller},\ and\
  \citenamefont {Ederer}}]{OBrien-Al2O3-L23_1991}%
  \BibitemOpen
  \bibfield  {author} {\bibinfo {author} {\bibfnamefont {W.~L.}\ \bibnamefont
  {O'Brien}}, \bibinfo {author} {\bibfnamefont {J.}~\bibnamefont {Jia}},
  \bibinfo {author} {\bibfnamefont {Q.-Y.}\ \bibnamefont {Dong}}, \bibinfo
  {author} {\bibfnamefont {T.~A.}\ \bibnamefont {Callcott}}, \bibinfo {author}
  {\bibfnamefont {J.-E.}\ \bibnamefont {Rubensson}}, \bibinfo {author}
  {\bibfnamefont {D.~L.}\ \bibnamefont {Mueller}}, \ and\ \bibinfo {author}
  {\bibfnamefont {D.~L.}\ \bibnamefont {Ederer}},\ }\href {\doibase
  10.1103/PhysRevB.44.1013} {\bibfield  {journal} {\bibinfo  {journal} {Phys.
  Rev. B}\ }\textbf {\bibinfo {volume} {44}},\ \bibinfo {pages} {1013}
  (\bibinfo {year} {1991})}\BibitemShut {NoStop}%
\bibitem [{\citenamefont {O'Brien}\ \emph {et~al.}(1993)\citenamefont
  {O'Brien}, \citenamefont {Jia}, \citenamefont {Dong}, \citenamefont
  {Callcott}, \citenamefont {Mueller}, \citenamefont {Ederer},\ and\
  \citenamefont {Kao}}]{OBrien_1993}%
  \BibitemOpen
  \bibfield  {author} {\bibinfo {author} {\bibfnamefont {W.~L.}\ \bibnamefont
  {O'Brien}}, \bibinfo {author} {\bibfnamefont {J.}~\bibnamefont {Jia}},
  \bibinfo {author} {\bibfnamefont {Q.-Y.}\ \bibnamefont {Dong}}, \bibinfo
  {author} {\bibfnamefont {T.~A.}\ \bibnamefont {Callcott}}, \bibinfo {author}
  {\bibfnamefont {D.~R.}\ \bibnamefont {Mueller}}, \bibinfo {author}
  {\bibfnamefont {D.~L.}\ \bibnamefont {Ederer}}, \ and\ \bibinfo {author}
  {\bibfnamefont {C.-C.}\ \bibnamefont {Kao}},\ }\href {\doibase
  10.1103/PhysRevB.47.15482} {\bibfield  {journal} {\bibinfo  {journal} {Phys.
  Rev. B}\ }\textbf {\bibinfo {volume} {47}},\ \bibinfo {pages} {15482}
  (\bibinfo {year} {1993})}\BibitemShut {NoStop}%
\bibitem [{\citenamefont {Cabaret}\ \emph {et~al.}(2005)\citenamefont
  {Cabaret}, \citenamefont {Gaudry}, \citenamefont {Taillefumier},
  \citenamefont {Sainctavit},\ and\ \citenamefont {Mauri}}]{Cabaret_2005}%
  \BibitemOpen
  \bibfield  {author} {\bibinfo {author} {\bibfnamefont {D.}~\bibnamefont
  {Cabaret}}, \bibinfo {author} {\bibfnamefont {E.}~\bibnamefont {Gaudry}},
  \bibinfo {author} {\bibfnamefont {M.}~\bibnamefont {Taillefumier}}, \bibinfo
  {author} {\bibfnamefont {P.}~\bibnamefont {Sainctavit}}, \ and\ \bibinfo
  {author} {\bibfnamefont {F.}~\bibnamefont {Mauri}},\ }\href {\doibase
  10.1238/Physica.Topical.115a00131} {\bibfield  {journal} {\bibinfo  {journal}
  {Physica Scripta}\ }\textbf {\bibinfo {volume} {2005}},\ \bibinfo {pages}
  {131} (\bibinfo {year} {2005})}\BibitemShut {NoStop}%
\bibitem [{\citenamefont {Cabaret}\ and\ \citenamefont
  {Brouder}(2009)}]{Cabaret_2009}%
  \BibitemOpen
  \bibfield  {author} {\bibinfo {author} {\bibfnamefont {D.}~\bibnamefont
  {Cabaret}}\ and\ \bibinfo {author} {\bibfnamefont {C.}~\bibnamefont
  {Brouder}},\ }\href {\doibase 10.1088/1742-6596/190/1/012003} {\bibfield
  {journal} {\bibinfo  {journal} {Journal of Physics: Conference Series}\
  }\textbf {\bibinfo {volume} {190}},\ \bibinfo {pages} {012003} (\bibinfo
  {year} {2009})}\BibitemShut {NoStop}%
\bibitem [{\citenamefont {Brouder}\ \emph {et~al.}(2010)\citenamefont
  {Brouder}, \citenamefont {Cabaret}, \citenamefont {Juhin},\ and\
  \citenamefont {Sainctavit}}]{Brouder2010}%
  \BibitemOpen
  \bibfield  {author} {\bibinfo {author} {\bibfnamefont {C.}~\bibnamefont
  {Brouder}}, \bibinfo {author} {\bibfnamefont {D.}~\bibnamefont {Cabaret}},
  \bibinfo {author} {\bibfnamefont {A.}~\bibnamefont {Juhin}}, \ and\ \bibinfo
  {author} {\bibfnamefont {P.}~\bibnamefont {Sainctavit}},\ }\href {\doibase
  10.1103/PhysRevB.81.115125} {\bibfield  {journal} {\bibinfo  {journal} {Phys.
  Rev. B}\ }\textbf {\bibinfo {volume} {81}},\ \bibinfo {pages} {115125}
  (\bibinfo {year} {2010})}\BibitemShut {NoStop}%
\bibitem [{\citenamefont {Manuel}\ \emph {et~al.}(2012)\citenamefont {Manuel},
  \citenamefont {Cabaret}, \citenamefont {Brouder}, \citenamefont {Sainctavit},
  \citenamefont {Bordage},\ and\ \citenamefont {Trcera}}]{Manuel_2012}%
  \BibitemOpen
  \bibfield  {author} {\bibinfo {author} {\bibfnamefont {D.}~\bibnamefont
  {Manuel}}, \bibinfo {author} {\bibfnamefont {D.}~\bibnamefont {Cabaret}},
  \bibinfo {author} {\bibfnamefont {C.}~\bibnamefont {Brouder}}, \bibinfo
  {author} {\bibfnamefont {P.}~\bibnamefont {Sainctavit}}, \bibinfo {author}
  {\bibfnamefont {A.}~\bibnamefont {Bordage}}, \ and\ \bibinfo {author}
  {\bibfnamefont {N.}~\bibnamefont {Trcera}},\ }\href {\doibase
  10.1103/PhysRevB.85.224108} {\bibfield  {journal} {\bibinfo  {journal} {Phys.
  Rev. B}\ }\textbf {\bibinfo {volume} {85}},\ \bibinfo {pages} {224108}
  (\bibinfo {year} {2012})}\BibitemShut {NoStop}%
\bibitem [{\citenamefont {Nemausat}\ \emph {et~al.}(2016)\citenamefont
  {Nemausat}, \citenamefont {Brouder}, \citenamefont {Gervais},\ and\
  \citenamefont {Cabaret}}]{Nemausat_2016}%
  \BibitemOpen
  \bibfield  {author} {\bibinfo {author} {\bibfnamefont {R.}~\bibnamefont
  {Nemausat}}, \bibinfo {author} {\bibfnamefont {C.}~\bibnamefont {Brouder}},
  \bibinfo {author} {\bibfnamefont {C.}~\bibnamefont {Gervais}}, \ and\
  \bibinfo {author} {\bibfnamefont {D.}~\bibnamefont {Cabaret}},\ }\href
  {\doibase 10.1088/1742-6596/712/1/012006} {\bibfield  {journal} {\bibinfo
  {journal} {Journal of Physics: Conference Series}\ }\textbf {\bibinfo
  {volume} {712}},\ \bibinfo {pages} {012006} (\bibinfo {year}
  {2016})}\BibitemShut {NoStop}%
\bibitem [{\citenamefont {Delhommaye}\ \emph {et~al.}(2021)\citenamefont
  {Delhommaye}, \citenamefont {Radtke}, \citenamefont {Brouder}, \citenamefont
  {Collins}, \citenamefont {Huotari}, \citenamefont {Sahle}, \citenamefont
  {Lazzeri}, \citenamefont {Paulatto},\ and\ \citenamefont
  {Cabaret}}]{Delhommaye_2021}%
  \BibitemOpen
  \bibfield  {author} {\bibinfo {author} {\bibfnamefont {S.}~\bibnamefont
  {Delhommaye}}, \bibinfo {author} {\bibfnamefont {G.}~\bibnamefont {Radtke}},
  \bibinfo {author} {\bibfnamefont {C.}~\bibnamefont {Brouder}}, \bibinfo
  {author} {\bibfnamefont {S.~P.}\ \bibnamefont {Collins}}, \bibinfo {author}
  {\bibfnamefont {S.}~\bibnamefont {Huotari}}, \bibinfo {author} {\bibfnamefont
  {C.}~\bibnamefont {Sahle}}, \bibinfo {author} {\bibfnamefont
  {M.}~\bibnamefont {Lazzeri}}, \bibinfo {author} {\bibfnamefont
  {L.}~\bibnamefont {Paulatto}}, \ and\ \bibinfo {author} {\bibfnamefont
  {D.}~\bibnamefont {Cabaret}},\ }\href {\doibase 10.1103/PhysRevB.104.024302}
  {\bibfield  {journal} {\bibinfo  {journal} {Phys. Rev. B}\ }\textbf {\bibinfo
  {volume} {104}},\ \bibinfo {pages} {024302} (\bibinfo {year}
  {2021})}\BibitemShut {NoStop}%
\bibitem [{\citenamefont {Marinopoulos}\ \emph {et~al.}(2002)\citenamefont
  {Marinopoulos}, \citenamefont {Reining}, \citenamefont {Olevano},
  \citenamefont {Rubio}, \citenamefont {Pichler}, \citenamefont {Liu},
  \citenamefont {Knupfer},\ and\ \citenamefont {Fink}}]{Marinopoulos_2002}%
  \BibitemOpen
  \bibfield  {author} {\bibinfo {author} {\bibfnamefont {A.~G.}\ \bibnamefont
  {Marinopoulos}}, \bibinfo {author} {\bibfnamefont {L.}~\bibnamefont
  {Reining}}, \bibinfo {author} {\bibfnamefont {V.}~\bibnamefont {Olevano}},
  \bibinfo {author} {\bibfnamefont {A.}~\bibnamefont {Rubio}}, \bibinfo
  {author} {\bibfnamefont {T.}~\bibnamefont {Pichler}}, \bibinfo {author}
  {\bibfnamefont {X.}~\bibnamefont {Liu}}, \bibinfo {author} {\bibfnamefont
  {M.}~\bibnamefont {Knupfer}}, \ and\ \bibinfo {author} {\bibfnamefont
  {J.}~\bibnamefont {Fink}},\ }\href {\doibase 10.1103/PhysRevLett.89.076402}
  {\bibfield  {journal} {\bibinfo  {journal} {Phys. Rev. Lett.}\ }\textbf
  {\bibinfo {volume} {89}},\ \bibinfo {pages} {076402} (\bibinfo {year}
  {2002})}\BibitemShut {NoStop}%
\bibitem [{\citenamefont {Vast}\ \emph {et~al.}(2002)\citenamefont {Vast},
  \citenamefont {Reining}, \citenamefont {Olevano}, \citenamefont
  {Schattschneider},\ and\ \citenamefont {Jouffrey}}]{Vast_2002}%
  \BibitemOpen
  \bibfield  {author} {\bibinfo {author} {\bibfnamefont {N.}~\bibnamefont
  {Vast}}, \bibinfo {author} {\bibfnamefont {L.}~\bibnamefont {Reining}},
  \bibinfo {author} {\bibfnamefont {V.}~\bibnamefont {Olevano}}, \bibinfo
  {author} {\bibfnamefont {P.}~\bibnamefont {Schattschneider}}, \ and\ \bibinfo
  {author} {\bibfnamefont {B.}~\bibnamefont {Jouffrey}},\ }\href {\doibase
  10.1103/PhysRevLett.88.037601} {\bibfield  {journal} {\bibinfo  {journal}
  {Phys. Rev. Lett.}\ }\textbf {\bibinfo {volume} {88}},\ \bibinfo {pages}
  {037601} (\bibinfo {year} {2002})}\BibitemShut {NoStop}%
\bibitem [{\citenamefont {Dash}\ \emph {et~al.}(2007)\citenamefont {Dash},
  \citenamefont {Bruneval}, \citenamefont {Trinité}, \citenamefont {Vast},\
  and\ \citenamefont {Reining}}]{Dash_2007}%
  \BibitemOpen
  \bibfield  {author} {\bibinfo {author} {\bibfnamefont {L.}~\bibnamefont
  {Dash}}, \bibinfo {author} {\bibfnamefont {F.}~\bibnamefont {Bruneval}},
  \bibinfo {author} {\bibfnamefont {V.}~\bibnamefont {Trinité}}, \bibinfo
  {author} {\bibfnamefont {N.}~\bibnamefont {Vast}}, \ and\ \bibinfo {author}
  {\bibfnamefont {L.}~\bibnamefont {Reining}},\ }\href {\doibase
  https://doi.org/10.1016/j.commatsci.2005.09.010} {\bibfield  {journal}
  {\bibinfo  {journal} {Computational Materials Science}\ }\textbf {\bibinfo
  {volume} {38}},\ \bibinfo {pages} {482} (\bibinfo {year} {2007})},\ \bibinfo
  {note} {selected papers from the International Conference on Computational
  Methods in Sciences and Engineering 2004}\BibitemShut {NoStop}%
\bibitem [{\citenamefont {Huotari}\ \emph {et~al.}(2010)\citenamefont
  {Huotari}, \citenamefont {Soininen}, \citenamefont {Vank\'o}, \citenamefont
  {Monaco},\ and\ \citenamefont {Olevano}}]{Huotari_2010}%
  \BibitemOpen
  \bibfield  {author} {\bibinfo {author} {\bibfnamefont {S.}~\bibnamefont
  {Huotari}}, \bibinfo {author} {\bibfnamefont {J.~A.}\ \bibnamefont
  {Soininen}}, \bibinfo {author} {\bibfnamefont {G.}~\bibnamefont {Vank\'o}},
  \bibinfo {author} {\bibfnamefont {G.}~\bibnamefont {Monaco}}, \ and\ \bibinfo
  {author} {\bibfnamefont {V.}~\bibnamefont {Olevano}},\ }\href {\doibase
  10.1103/PhysRevB.82.064514} {\bibfield  {journal} {\bibinfo  {journal} {Phys.
  Rev. B}\ }\textbf {\bibinfo {volume} {82}},\ \bibinfo {pages} {064514}
  (\bibinfo {year} {2010})}\BibitemShut {NoStop}%
\bibitem [{\citenamefont {Cudazzo}\ \emph {et~al.}(2014)\citenamefont
  {Cudazzo}, \citenamefont {Ruotsalainen}, \citenamefont {Sahle}, \citenamefont
  {Al-Zein}, \citenamefont {Berger}, \citenamefont {Navarro-Moratalla},
  \citenamefont {Huotari}, \citenamefont {Gatti},\ and\ \citenamefont
  {Rubio}}]{Cudazzo_2014}%
  \BibitemOpen
  \bibfield  {author} {\bibinfo {author} {\bibfnamefont {P.}~\bibnamefont
  {Cudazzo}}, \bibinfo {author} {\bibfnamefont {K.~O.}\ \bibnamefont
  {Ruotsalainen}}, \bibinfo {author} {\bibfnamefont {C.~J.}\ \bibnamefont
  {Sahle}}, \bibinfo {author} {\bibfnamefont {A.}~\bibnamefont {Al-Zein}},
  \bibinfo {author} {\bibfnamefont {H.}~\bibnamefont {Berger}}, \bibinfo
  {author} {\bibfnamefont {E.}~\bibnamefont {Navarro-Moratalla}}, \bibinfo
  {author} {\bibfnamefont {S.}~\bibnamefont {Huotari}}, \bibinfo {author}
  {\bibfnamefont {M.}~\bibnamefont {Gatti}}, \ and\ \bibinfo {author}
  {\bibfnamefont {A.}~\bibnamefont {Rubio}},\ }\href {\doibase
  10.1103/PhysRevB.90.125125} {\bibfield  {journal} {\bibinfo  {journal} {Phys.
  Rev. B}\ }\textbf {\bibinfo {volume} {90}},\ \bibinfo {pages} {125125}
  (\bibinfo {year} {2014})}\BibitemShut {NoStop}%
\bibitem [{\citenamefont {Ruotsalainen}\ \emph {et~al.}(2021)\citenamefont
  {Ruotsalainen}, \citenamefont {Nicolaou}, \citenamefont {Sahle},
  \citenamefont {Efimenko}, \citenamefont {Ablett}, \citenamefont {Rueff},
  \citenamefont {Prabhakaran},\ and\ \citenamefont
  {Gatti}}]{Ruotsalainen_2021}%
  \BibitemOpen
  \bibfield  {author} {\bibinfo {author} {\bibfnamefont {K.}~\bibnamefont
  {Ruotsalainen}}, \bibinfo {author} {\bibfnamefont {A.}~\bibnamefont
  {Nicolaou}}, \bibinfo {author} {\bibfnamefont {C.~J.}\ \bibnamefont {Sahle}},
  \bibinfo {author} {\bibfnamefont {A.}~\bibnamefont {Efimenko}}, \bibinfo
  {author} {\bibfnamefont {J.~M.}\ \bibnamefont {Ablett}}, \bibinfo {author}
  {\bibfnamefont {J.-P.}\ \bibnamefont {Rueff}}, \bibinfo {author}
  {\bibfnamefont {D.}~\bibnamefont {Prabhakaran}}, \ and\ \bibinfo {author}
  {\bibfnamefont {M.}~\bibnamefont {Gatti}},\ }\href {\doibase
  10.1103/PhysRevB.103.235136} {\bibfield  {journal} {\bibinfo  {journal}
  {Phys. Rev. B}\ }\textbf {\bibinfo {volume} {103}},\ \bibinfo {pages}
  {235136} (\bibinfo {year} {2021})}\BibitemShut {NoStop}%
\bibitem [{Note3()}]{Note3}%
  \BibitemOpen
  \bibinfo {note} {It is well known that local field effects, expressed as
  electron-hole exchange interaction in the BSE framework, are essential to get
  the correct branching ratios between L$_{2}$ and L$_{3}$ components, see e.g.
  \cite {Gilmore2015,Vinson2012,Ankudinov_2005}. However, in the present case
  the neglect of spin-orbit coupling does not allow us to resolve the two
  components. For {$\alpha $-Al$_2$O$_3$} an electron–hole exchange energy of
  0.3 eV has been estimated \cite
  {Weigel_2008,OBrien-Al2O3-L23_1991}.}\BibitemShut {Stop}%
\bibitem [{\citenamefont {Rohlfing}\ and\ \citenamefont
  {Louie}(1998)}]{Rohlfing1998}%
  \BibitemOpen
  \bibfield  {author} {\bibinfo {author} {\bibfnamefont {M.}~\bibnamefont
  {Rohlfing}}\ and\ \bibinfo {author} {\bibfnamefont {S.~G.}\ \bibnamefont
  {Louie}},\ }\href {\doibase 10.1103/PhysRevLett.81.2312} {\bibfield
  {journal} {\bibinfo  {journal} {Phys. Rev. Lett.}\ }\textbf {\bibinfo
  {volume} {81}},\ \bibinfo {pages} {2312} (\bibinfo {year}
  {1998})}\BibitemShut {NoStop}%
\bibitem [{\citenamefont {Gatti}\ and\ \citenamefont
  {Sottile}(2013)}]{Gatti2013}%
  \BibitemOpen
  \bibfield  {author} {\bibinfo {author} {\bibfnamefont {M.}~\bibnamefont
  {Gatti}}\ and\ \bibinfo {author} {\bibfnamefont {F.}~\bibnamefont
  {Sottile}},\ }\href {\doibase 10.1103/PhysRevB.88.155113} {\bibfield
  {journal} {\bibinfo  {journal} {Phys. Rev. B}\ }\textbf {\bibinfo {volume}
  {88}},\ \bibinfo {pages} {155113} (\bibinfo {year} {2013})}\BibitemShut
  {NoStop}%
\bibitem [{\citenamefont {Shirley}(2004)}]{Shirley2004}%
  \BibitemOpen
  \bibfield  {author} {\bibinfo {author} {\bibfnamefont {E.~L.}\ \bibnamefont
  {Shirley}},\ }\href {\doibase https://doi.org/10.1016/j.elspec.2004.02.134}
  {\bibfield  {journal} {\bibinfo  {journal} {Journal of Electron Spectroscopy
  and Related Phenomena}\ }\textbf {\bibinfo {volume} {136}},\ \bibinfo {pages}
  {77} (\bibinfo {year} {2004})},\ \bibinfo {note} {progress in Core-Level
  Spectroscopy of Condensed Systems.}\BibitemShut {Stop}%
\bibitem [{\citenamefont {Bl\"ochl}(1994)}]{Bloechl1994}%
  \BibitemOpen
  \bibfield  {author} {\bibinfo {author} {\bibfnamefont {P.~E.}\ \bibnamefont
  {Bl\"ochl}},\ }\href {\doibase 10.1103/PhysRevB.50.17953} {\bibfield
  {journal} {\bibinfo  {journal} {Phys. Rev. B}\ }\textbf {\bibinfo {volume}
  {50}},\ \bibinfo {pages} {17953} (\bibinfo {year} {1994})}\BibitemShut
  {NoStop}%
\bibitem [{\citenamefont {Ankudinov}\ \emph {et~al.}(2005)\citenamefont
  {Ankudinov}, \citenamefont {Takimoto},\ and\ \citenamefont
  {Rehr}}]{Ankudinov_2005}%
  \BibitemOpen
  \bibfield  {author} {\bibinfo {author} {\bibfnamefont {A.~L.}\ \bibnamefont
  {Ankudinov}}, \bibinfo {author} {\bibfnamefont {Y.}~\bibnamefont {Takimoto}},
  \ and\ \bibinfo {author} {\bibfnamefont {J.~J.}\ \bibnamefont {Rehr}},\
  }\href {\doibase 10.1103/PhysRevB.71.165110} {\bibfield  {journal} {\bibinfo
  {journal} {Phys. Rev. B}\ }\textbf {\bibinfo {volume} {71}},\ \bibinfo
  {pages} {165110} (\bibinfo {year} {2005})}\BibitemShut {NoStop}%
\end{thebibliography}%

\end{document}